\documentclass[letterpaper,11pt]{article}
\pdfoutput=1

\usepackage{jheppub}
\newcommand{\setmainskip}{\setlength\baselineskip{18pt}}

\usepackage{subfig}

\def\beq{\begin{equation}}
\def\eeq{\end{equation}}
\def\bsp#1\esp{\begin{split}#1\end{split}}
\newcommand{\be}{\begin{equation}}
\newcommand{\ee}{\end{equation}}

\def\Fig#1{Fig.~{\ref{#1}}}

\def\Figs#1#2{Figs.~{\ref{#1}} and {\ref{#2}}}

\def\App#1{Appendix~\ref{#1}}

\def\to{\rightarrow}

\DeclareRobustCommand{\Sec}[1]{Sec.~\ref{#1}}

\DeclareRobustCommand{\App}[1]{App.~\ref{#1}}

\DeclareRobustCommand{\Fig}[1]{Fig.~\ref{#1}}
\DeclareRobustCommand{\Figs}[2]{Figs.~\ref{#1} and \ref{#2}}
\DeclareRobustCommand{\Eq}[1]{Eq.~(\ref{#1})}
\DeclareRobustCommand{\Eqs}[2]{Eqs.~(\ref{#1}) and (\ref{#2})}
\DeclareRobustCommand{\Ref}[1]{Ref.~\cite{#1}}
\DeclareRobustCommand{\Refs}[1]{Refs.~\cite{#1}}

\def\cE{\mathcal{E}}

\def\cN{\mathcal{N}}

\def \as {\relax\ifmmode\alpha_s\else{$\alpha_s${ }}\fi}

\def\ksl{\not{\hbox{\kern-2.3pt $k$}}}

\def\cE{\mathcal{E}}

\def\spa#1.#2{\left\langle#1\,#2\right\rangle}
\def\spb#1.#2{\left[#1\,#2\right]}
\def\lor#1.#2{\left(#1\,#2\right)}
\def\sand#1.#2.#3{%
\left\langle\smash{#1}{\vphantom1}^{-}\right|{#2}%
\left|\smash{#3}{\vphantom1}^{-}\right\rangle}

\def\beqa{\begin{eqnarray}}
\def\eeqa{\end{eqnarray}}

\begin{document}

\preprint{MIT-CTP 5430}

\title{Non-Gaussianities in Collider Energy Flux}

\author[1]{Hao Chen,}
\author[2]{Ian Moult,}
\author[3]{Jesse Thaler,}
\author[1]{and Hua Xing Zhu}
\affiliation[1]{Zhejiang Institute of Modern Physics, Department of Physics, Zhejiang University, Hangzhou, Zhejiang 310027, China}
\affiliation[2]{Department of Physics, Yale University, New Haven, CT 06511, USA\vspace{0.5ex}}
\affiliation[3]{Center for Theoretical Physics, Massachusetts Institute of Technology, Cambridge, MA 02139, USA\vspace{0.5ex}}

\emailAdd{chenhao201224@zju.edu.cn}
\emailAdd{ian.moult@yale.edu}
\emailAdd{jthaler@mit.edu}
\emailAdd{zhuhx@zju.edu.cn}

\abstract{
The microscopic dynamics of particle collisions is imprinted into the statistical properties of asymptotic energy flux, much like the dynamics of inflation is imprinted into the cosmic microwave background.
This energy flux is characterized by correlation functions $\langle \cE(\vec n_1)\cdots \cE(\vec n_k) \rangle$ of energy flow operators $ \cE(\vec n)$.
There has been significant recent progress in studying energy flux, including the calculation of multi-point correlation functions and their direct measurement inside high-energy jets at the Large Hadron Collider (LHC).
In this paper, we build on these advances by defining a notion of ``celestial non-gaussianity" as a ratio of the three-point function to a product of two-point functions.
We show that this celestial non-gaussianity is under perturbative control within jets at the LHC, allowing us to cleanly access the non-gaussian interactions of quarks and gluons.
We find good agreement between perturbative calculations of the non-gaussianity and a charged-particle-based analysis using CMS Open Data, and we observe a strong non-gaussianity peaked in the ``flattened triangle" regime. 
The ability to robustly study three-point correlations is a significant step in advancing our understanding of jet substructure at the LHC.
We anticipate that the celestial non-gaussianity, and its generalizations, will play an important role in the development of higher-order parton showers simulations and in the hunt for ever more subtle signals of potential new physics within jets.
}

\maketitle
\setmainskip

%%%%%%%%%%%%%%%%%%%%%%
\section{Introduction}\label{sec:intro}
%%%%%%%%%%%%%%%%%%%%%%

Collimated sprays of hadrons called jets are the manifestation of Quantum Chromodynamics (QCD) at high-energy colliders \cite{Hanson:1975fe,Sterman:1977wj}.
The seminal introduction of experimentally robust infrared-and-collinear-safe jet algorithms \cite{Cacciari:2005hq,Cacciari:2008gp,Cacciari:2011ma}, combined with the remarkable resolution of the Large Hadron Collider (LHC) detectors \cite{CMSPF,ATLAS:2017ghe}, has enabled the precision study of the detailed structure of energy flow within jets, a field referred to as jet substructure \cite{Larkoski:2017jix,Marzani:2019hun}.

The ability to exploit the detailed internal structure of jets has opened up numerous new avenues to search for new physics signals \cite{Butterworth:2008iy,Kaplan:2008ie,Krohn:2009th}, and it continues to provide some of the most innovative searches for new physics \cite{CMS:2022jed,CMS:2021yqw}.
It has also provided numerous new ways to probe QCD, both in vacuum and in the medium \cite{Andrews:2018jcm,Cunqueiro:2021wls}.
These new approaches necessitate the theoretical understanding of the detailed structure of energy flow \emph{within} jets.
While there has been extensive progress in the theoretical understanding of jet cross sections, with remarkable calculations including $W+9$-jet cross sections at tree level \cite{Hoche:2019flt}, 5-jet cross sections at next-to-leading order (NLO) \cite{Badger:2013yda}, and 3-jet cross sections at NNLO \cite{Czakon:2021mjy}, much less is understood about the structure of energy flow within jets themselves.

\begin{figure}[t]
\begin{center}
\subfloat[]{
\includegraphics[scale=0.22]{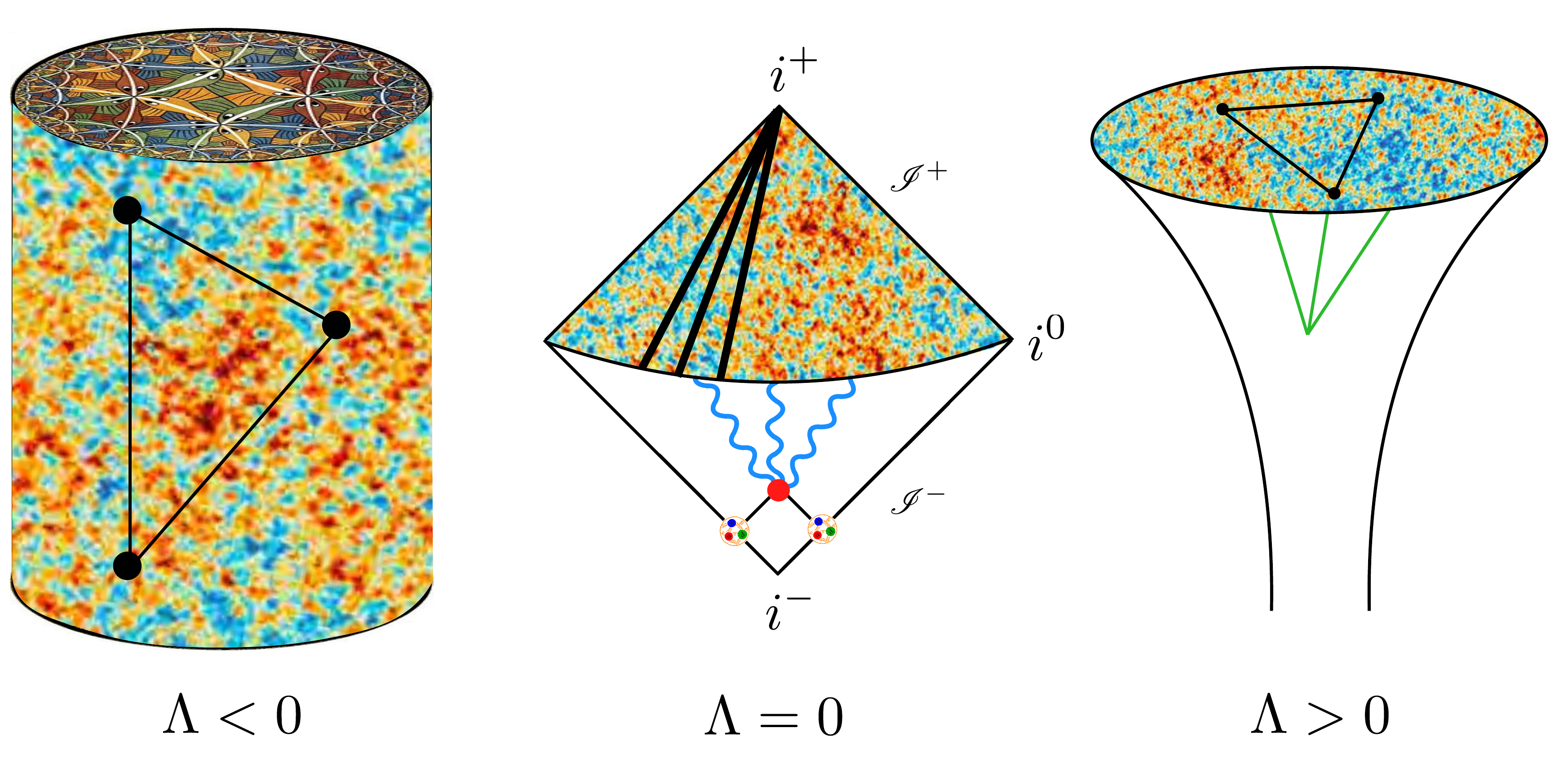}\label{fig:intro_a}
}\qquad\qquad
\subfloat[]{
\includegraphics[scale=0.22]{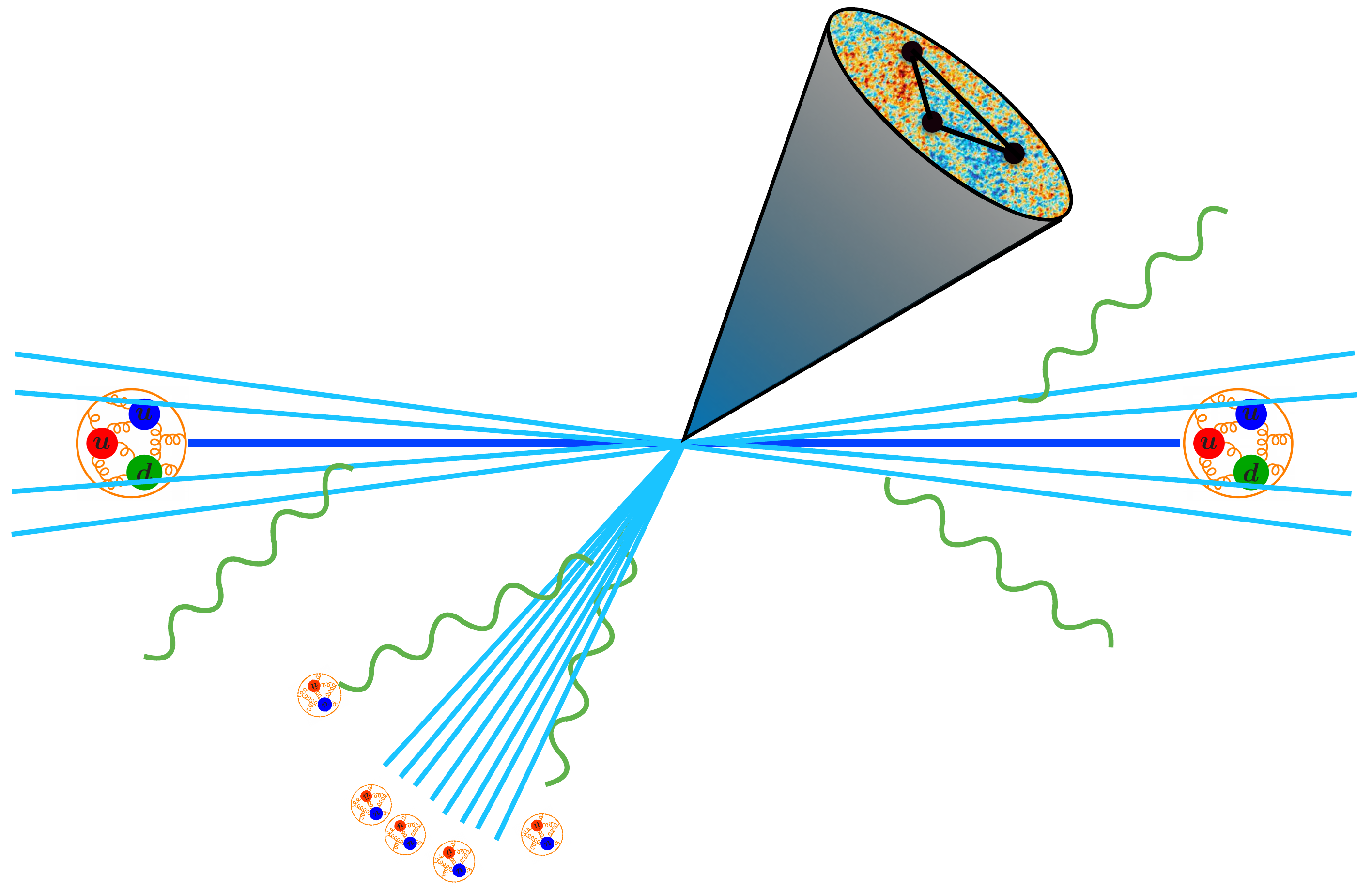}\label{fig:intro_b}
}
\end{center}
\caption{
(a) Correlation functions measured on the asymptotic boundaries of different spacetimes.
(b) In the flat space case, a wealth of data exists inside high-energy jets at the LHC, allowing for the direct measurement of higher-point correlators.}
\label{fig:intro}
\end{figure}

Asymptotic energy flow in collider experiments is characterized by correlation functions
$\langle \mathcal{E}(\vec n_1) \mathcal{E}(\vec n_2) \cdots \mathcal{E}(\vec n_k) \rangle$ of the energy flow operator~\cite{Sveshnikov:1995vi,Tkachov:1995kk,Korchemsky:1999kt,Bauer:2008dt,Hofman:2008ar,Belitsky:2013xxa,Belitsky:2013bja,Kravchuk:2018htv}:
\begin{align}
\label{energy_flow_operator}
\mathcal{E}(\vec n) = \lim_{r\to \infty} \int\limits_0^\infty dt \, r^2 \, n^i \, T_{0i}(t,r \vec n)\,.
\end{align}
These energy flow operators are illustrated in the Penrose diagram in \Fig{fig:intro_a}, which shows their non-local spacetime structure, and inside a high-energy jet at the LHC in \Fig{fig:intro_b}, where they are shown as points, illustrating their locality on the celestial sphere.
The underlying microscopic details of the collision are imprinted in the detailed structure of these correlation functions.
This is in analogy with how the details of inflation are imprinted in cosmological correlation functions of scalar $\langle \zeta_{\vec k_1}  \zeta_{\vec k_2} \cdots  \zeta_{\vec k_n} \rangle $ or tensor $\langle \gamma_{\vec k_1}  \gamma_{\vec k_2} \cdots   \gamma_{\vec k_n} \rangle$ fluctuations, illustrated schematically in \Fig{fig:intro_a}.
Despite their direct relation to experiments of interest, correlation functions that live on the boundary of flat or de-Sitter space are much less understood than the correlation functions of local operators in a conformal field theory on the boundary of AdS \cite{Maldacena:1997re,Gubser:1998bc,Witten:1998qj}, but they have recently received significant interest.

\begin{figure}
\begin{center}
\subfloat[]{
\includegraphics[width=0.2\textwidth]{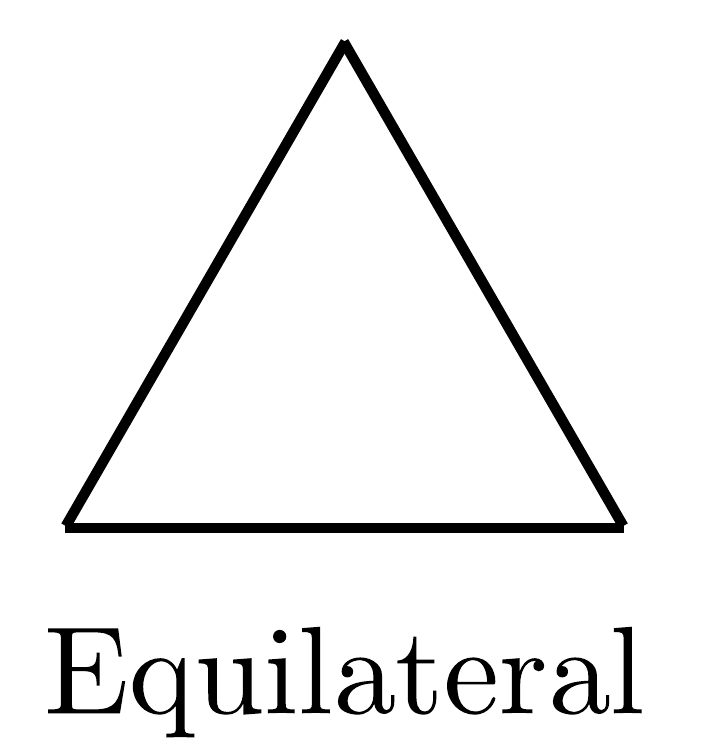}\label{fig:shapes_a}
}\qquad
\subfloat[]{
\includegraphics[width=0.30\textwidth]{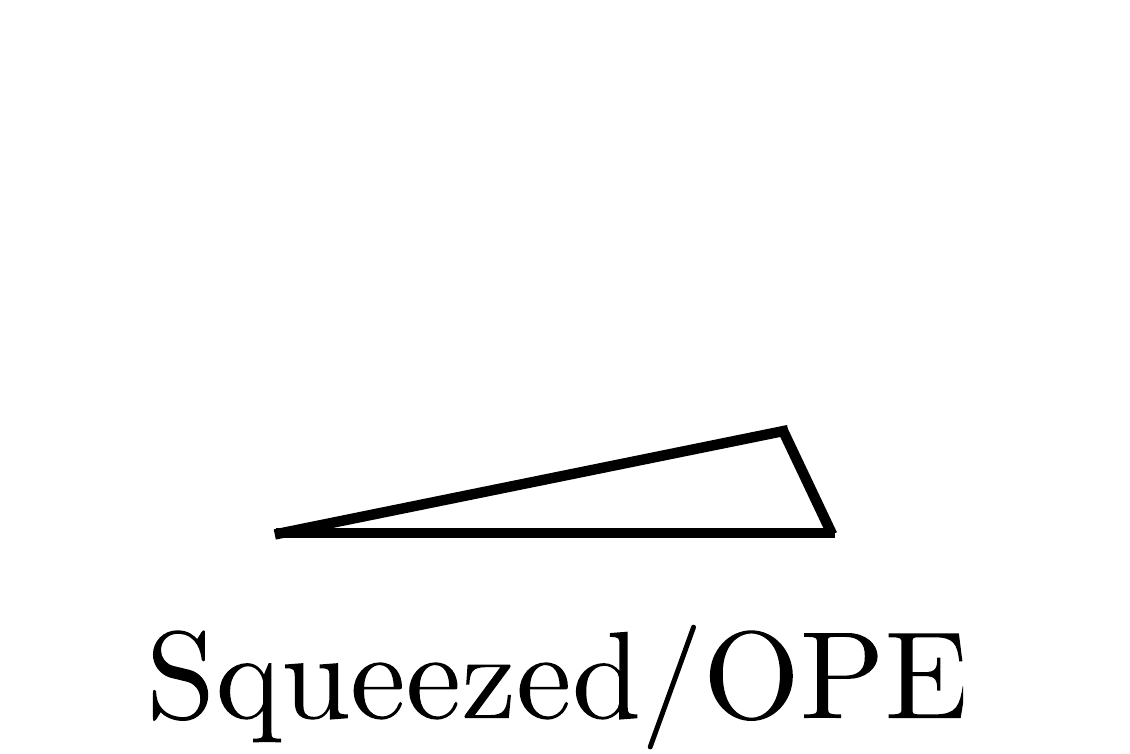}\label{fig:shapes_b}
}\qquad
\subfloat[]{
\includegraphics[width=0.30\textwidth]{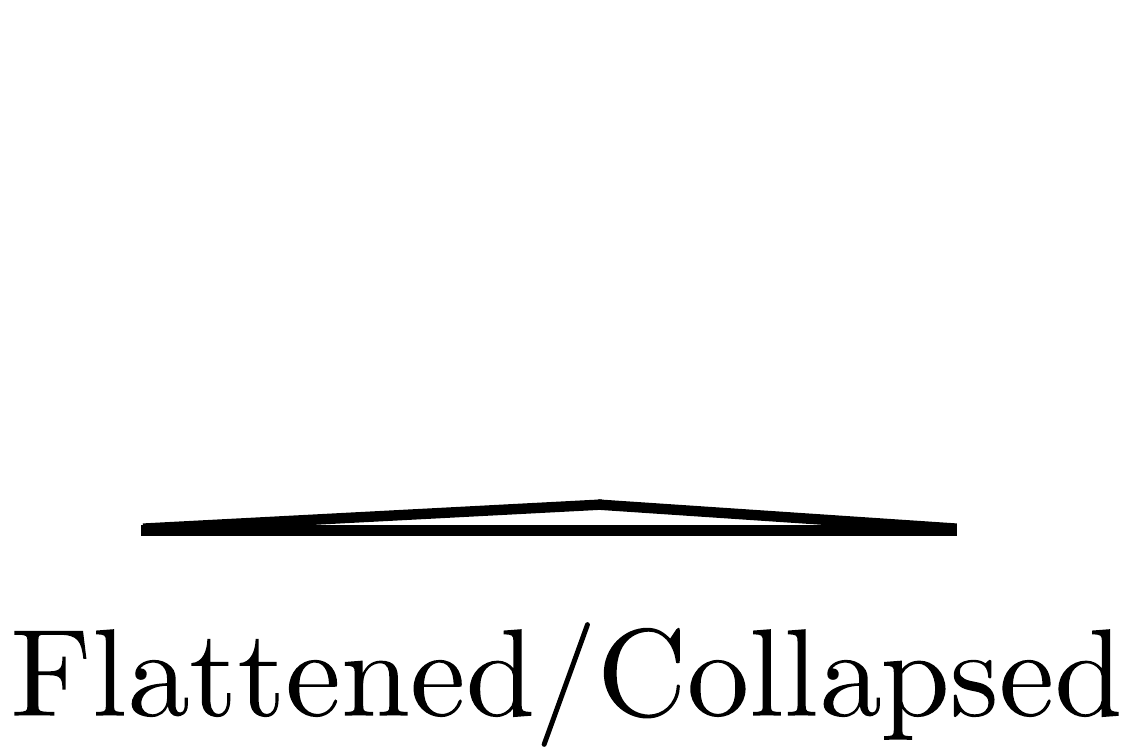}\label{fig:shapes_c}
}
\end{center}
\caption{
Limiting shapes for the three-point correlation function that will play a central role in our discussion: (a) equilateral, (b) squeezed, and (c) flattened.
The behavior of the correlation function in these different limits is determined by distinct physics.
The naming conventions used depend on the particular community.
}
\label{fig:shapes}
\end{figure}

The cosmological three-point function  $\langle \zeta_{\vec k_1}  \zeta_{\vec k_2}  \zeta_{\vec k_3} \rangle $ was first computed in single-field inflation in the seminal work of Maldacena \cite{Maldacena:2011nz}, and it has since been studied extensively \cite{Babich:2004gb,Chen:2006nt,Cheung:2007sv,Cheung:2007st,Weinberg:2008hq} and extended to an understanding of three-point functions of tensor fluctuations \cite{Maldacena:2011nz}. 
These three-point functions encode a wealth of information about the underlying dynamics through their shape dependence.
Limiting shapes are shown in \Fig{fig:shapes}, each of which encode different physics, see e.g.~\cite{Bartolo:2004if,Babich:2004gb,Chen:2006nt,Baumann:2009ds,Chen:2010xka,Meerburg:2019qqi,Planck:2015zfm,Planck:2019kim}.
Phenomenological applications of the shapes of four-point correlators have also been studied~\cite{Arroja:2009pd,Chen:2009bc,Hindmarsh:2009es,Senatore:2010jy,Bartolo:2010di,Lewis:2011au}.
There has recently been significant advances in the understanding of cosmological correlation functions, driven by advances from the amplitudes and conformal bootstrap programs; see e.g.~\cite{DiPietro:2021sjt,Cabass:2021fnw,Goodhew:2021oqg,Melville:2021lst,Benincasa:2020aoj,Arkani-Hamed:2018bjr,Arkani-Hamed:2017fdk,Baumann:2021fxj,Baumann:2020dch,Baumann:2019oyu,Arkani-Hamed:2018kmz,Lee:2016vti}.

By contrast, despite the wealth of collider data, the understanding of the asymptotic structure of energy flow operators lags behind its cosmological counterparts.
The two-point correlator was studied early on in the history of QCD~\cite{Basham:1978bw,Basham:1977iq,Basham:1979gh,Basham:1978zq,Konishi:1979cb}, measured at $e^+e^-$ colliders~\cite{SLD:1994idb,L3:1992btq,OPAL:1991uui,TOPAZ:1989yod,TASSO:1987mcs,JADE:1984taa,Fernandez:1984db,Wood:1987uf,CELLO:1982rca,PLUTO:1985yzc}, and more recently computed analytically to higher perturbative orders~\cite{Belitsky:2013ofa,Dixon:2018qgp,Henn:2019gkr,Luo:2019nig}.
However, the first calculation of multi-point correlators of energy flow operators was the seminal work of Hofman and Maldacena \cite{Hofman:2008ar} as an expansion about strong coupling.
This motivated significant theoretical study of these correlators, particularly in the context of conformal field theories~\cite{Belitsky:2013xxa,Belitsky:2013bja,Belitsky:2013ofa,Belitsky:2014zha,Korchemsky:2015ssa} and the development of the light-ray-operator product expansion (OPE)~\cite{Hofman:2008ar,Kravchuk:2018htv,Kologlu:2019bco,Kologlu:2019mfz,Chang:2020qpj}.

Recently, the three-point correlator was computed in the collinear limit at weak coupling in both QCD and $\cN=4$ super Yang-Mills~\cite{Chang:2022ryc,Chen:2022jhb}, where it was analyzed in detail and expressed as a sum over celestial blocks incorporating the symmetries of the Lorentz group.
(This has since been extended to a calculation of the full angular dependence~\cite{Yan:2022cye}.)
It was also shown that it can be directly analyzed inside jets at the LHC using CMS Open Data~\cite{Komiske:2022enw}.
This is part of a broader program to reformulate the study of jet substructure in terms of energy correlators~\cite{Dixon:2019uzg,Chen:2019bpb,Chen:2020adz,Chen:2020vvp,Chen:2021gdk,Chen:2022jhb,Holguin:2022epo}.%
\footnote{The study of energy correlators in the back-to-back (Sudakov) region has also seen significant progress; see e.g.~\cite{Gao:2019ojf,Moult:2018jzp,Moult:2019vou,Ebert:2020sfi,Li:2021txc}.}
An important part of this program includes the development of techniques to compute energy correlators on charged particles (tracks) \cite{Chang:2013rca,Chang:2013iba,Chen:2020vvp,Li:2021zcf,Jaarsma:2022kdd} to enable the experimental measurement of higher-point correlators.
This availability of higher-point correlators in the collider context allows us to begin to start asking a similar class of questions to those studied in cosmology, namely about the shape dependence of correlations in the QCD energy flux.

\begin{figure}[t]
\begin{center}
\includegraphics[scale=0.5]{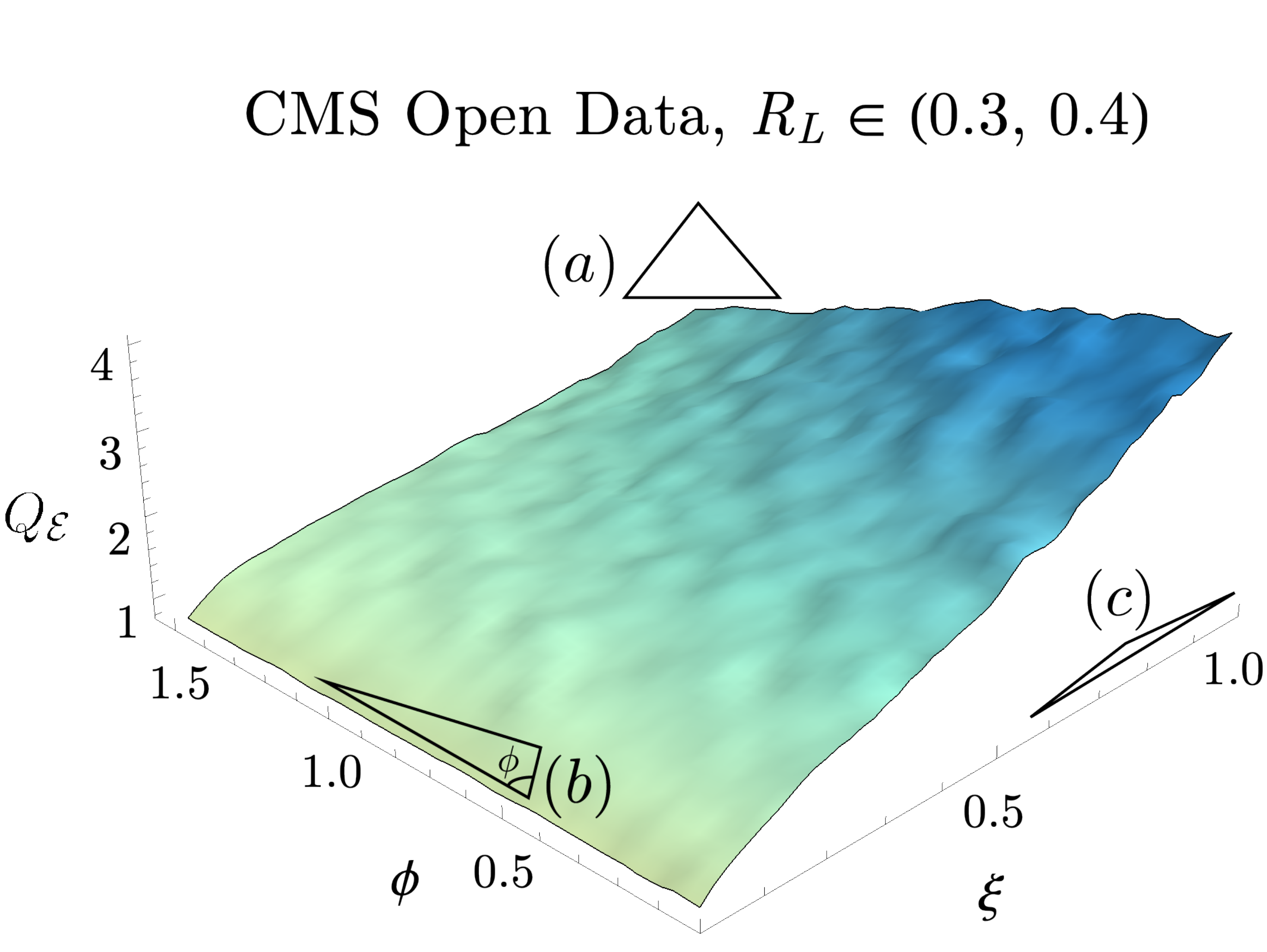}
\caption{
The full shape dependence of the celestial non-gaussianity $Q_\mathcal{E}$ (whose complete definition is given in \Eq{eq:NG_def}) in CMS Open Data, showing a strong peak in the  ``flattened triangle" region.
The $(\xi, \phi)$ coordinates parametrize the shape of the three-point correlator and are defined in \Eq{eq:transf}, but the representative shapes from \Fig{fig:shapes} are drawn to guide the reader.
The squeezed limit is characterized by $\xi\to 0$, while the flattened triangle is characterized by $\phi\to 0$. To our knowledge, this is the first study of non-gaussianities in QCD energy flux.} 
\label{fig:shape_intro}
\end{center}
\end{figure}

A first step in the use of the three-point correlator in collider physics was taken in \Refs{Chen:2020adz,Chen:2021gdk}, which focused on the squeezed or operator product expansion (OPE) limit, as illustrated in \Fig{fig:shapes_b}.
There, it was shown that interference effects associated with the spin of the gluon are encoded in the azimuthal structure as two squeezed correlators are rotated with respect to a third. 
This is in analogy to similar effects in the squeezed limit of the cosmological three-point correlator~\cite{Arkani-Hamed:2015bza}.
The analytic resummation of these effects \cite{Chen:2020adz,Chen:2021gdk} was then used to verify their incorporation into parton showers \cite{Karlberg:2021kwr}.
Note that in the squeezed limit, the three-point correlator factorizes into a product of two-point correlators, much in analogy with the consistency relations in the cosmological case \cite{Maldacena:2002vr,Creminelli:2004yq,Cheung:2007sv,Goldberger:2013rsa}.
The full shape dependence, and the wealth of physics incorporated into it, has not yet been exploited.

In this paper, we introduce ``celestial non-gaussianities", which are a particular ratio of the three-point correlator to a product of two-point correlators, in analogy to non-gaussianities for correlation functions of local operators in conformal field theories and for cosmological correlation functions.
We show that this observable is robust to hadronization effects, allowing it to be computed perturbatively and compared with data from high-energy jets at the LHC, and we study it in detail using perturbative results and parton showers.
We then plot the celestial non-gaussianities using publicly available data from the CMS experiment at the LHC~\cite{CMS:JetPrimary2011A}, finding good agreement with our theoretical calculations for a track-based analysis.

A plot of the shape dependence of the celestial non-gaussianity is shown in \Fig{fig:shape_intro}. The coordinates $(\xi, \phi)$ will be defined in \Sec{sec:NG}, however, we have drawn representative shapes of the three-point correlator to guide the reader.
Interestingly, we find that the non-gaussianity is highly peaked in the flattened triangle region, from which we can draw an analogy with the case of cosmological correlation functions. 
Our results provide the first study of the non-gaussianities of QCD radiation, and they show for the first time control over three-point correlations within jets at the LHC.
This provides a significant step in our understanding of the structure of energy flow inside jets, which we believe will be useful in a number of different directions for improving our understanding of QCD in the high-energy regime.
First, and most obviously, our results provide new detailed probes of the perturbative interactions of quarks and gluons in QCD.
Second, increasingly sophisticated properties of energy flow are being exploited by machine learning techniques to search for increasingly subtle imprints of new physics within jets (see e.g.~\cite{Komiske:2018cqr,Qu:2019gqs,CMS:2020poo}), greatly extending the reach of previous observables~\cite{Thaler:2010tr,Thaler:2011gf,Larkoski:2013eya,Larkoski:2014zma,Larkoski:2014gra,Larkoski:2015kga,Moult:2016cvt,Larkoski:2017cqq,Larkoski:2017iuy,Komiske:2017aww,Komiske:2018cqr,Komiske:2019fks}.
Supervised machine learning relies on the accurate description of this energy flow using parton shower Monte Carlo generators.
There is currently a push to extend these to incorporate higher-order effects into shower generators, from a variety of different directions~\cite{Li:2016yez,Hoche:2017hno,Hoche:2017iem,Dulat:2018vuy,Gellersen:2021eci,Hamilton:2020rcu,Dasgupta:2020fwr,Hamilton:2021dyz,Karlberg:2021kwr}.
Having an analytic understanding of properties of the radiation flow will prove important for this goal.
Indeed, the understanding of the three-point correlator in the collinear limit \cite{Chen:2020adz} has already been useful for verifying the incorporation of spin effects \cite{Hamilton:2021dyz,Karlberg:2021kwr}.
Finally, jet substructure has begun to play an important role in the study of heavy-ion collisions \cite{Andrews:2018jcm,Cunqueiro:2021wls}.
Much like in the case of inflation, the three-point correlator provides a wealth of information potentially allowing one to distinguish different mechanisms of medium modification.
All of these motivate pushing our understanding of radiation patterns within jets to the level of three-point correlations.

An outline of this paper is as follows.
In \Sec{sec:NG}, we introduce our definition of ``celestial non-gaussianity" and describe its basic features and theoretical motivation.
We study its basic theoretical properties in \Sec{sec:NG_analytic}, using both perturbative fixed-order results and parton shower generators.
In \Sec{sec:open_data}, we analyze the celestial non-gaussianity using CMS Open Data, and compare with our analytic results.
We conclude in \Sec{sec:conc}.
Additional technical details related to both the theory and data analysis are provided in the Appendices.

%%%%%%%%%%%%%%%%%%%%%%
\section{Celestial Non-Gaussianities}\label{sec:NG}
%%%%%%%%%%%%%%%%%%%%%%

A main point of this paper is to introduce a robust definition of non-gaussianity for asymptotic energy flow, which we term ``celestial non-gaussianity".
This is inspired by the definition of non-gaussianities often used in the study of condensed matter systems, conformal field theory, and cosmology.
Taking as an example the four-point correlator of a local operator $\sigma$ in a CFT, one traditionally defines the non-gaussianity by dividing the four-point correlator by the sum of Wick contractions:
\begin{align} \label{eq:Ising_NG_def}
Q_\sigma(1,2,3,4)=\frac{\langle  \sigma_1 \sigma_2  \sigma_3 \sigma_4  \rangle}{\langle \sigma_1 \sigma_2 \rangle \langle \sigma_3 \sigma_4 \rangle   +   \langle \sigma_1 \sigma_3 \rangle \langle \sigma_2 \sigma_4 \rangle   +   \langle \sigma_1 \sigma_4 \rangle \langle \sigma_2 \sigma_3 \rangle}\,.
\end{align}
In the case of a Gaussian (free) theory, $Q_\sigma=1$.
The non-gaussianity $Q_\sigma$ is known exactly in the 2d Ising model~\cite{Belavin:1984vu}, and it has recently been studied in the 3d Ising model in \Ref{Rychkov:2016mrc} using state of the art data \cite{Kos:2016ysd,Komargodski:2016auf,Simmons-Duffin:2016wlq} from the conformal bootstrap program \cite{Rattazzi:2008pe,El-Showk:2012cjh,El-Showk:2014dwa,Poland:2018epd}.
There, it was found that the non-gaussianity of the 3d Ising model is strongly peaked for an equilateral configuration.
Similar measures are used in cosmology \cite{Baumann:2009ds}.
Here, we will introduce a similar quantity for the asymptotic energy flux, which is both experimentally robust and insensitive to hadronization effects.

%%%%%%%%%%%%%%%%%%%%%%
\subsection{Definition}\label{sec:def}
%%%%%%%%%%%%%%%%%%%%%%

In the collinear limit, the first correlator with a non-trivial shape dependence is the three-point correlator (EEEC):
\begin{equation}
\langle\mathcal{E}(\vec{n}_1)\, \mathcal{E}(\vec{n}_2)\, \mathcal{E}(\vec{n}_3)\rangle,
\end{equation}
as illustrated in \Fig{fig:intro_b}.
Much like the four-point correlator of local operators discussed above, the three-point correlator contains in it iterations of two-point correlations.
In the case of asymptotic energy flux, these are interpreted physically as iterated $1\to 2$ splittings, as are implemented in standard parton shower generators.
For a recent discussion in terms of the $1\to 3$ splitting function, see \Ref{Braun-White:2022rtg}.
These iterations give rise to the leading singular behavior in the squeezed/OPE limit, where the three-point correlator factorizes into a product of two-point correlators~\cite{Chen:2021gdk,Chen:2020adz}.
To derive an appropriate non-gaussianity for asymptotic energy flux $Q_\cE$, one therefore wants to remove these contributions from the three-point correlator, defining an observable such that  $Q_\cE\to \text{const}$ in the squeezed limits.

%%%%%%%%%%%%%%%%%%%%%%
\begin{figure}[t]
\begin{center}
\includegraphics[scale=0.45]{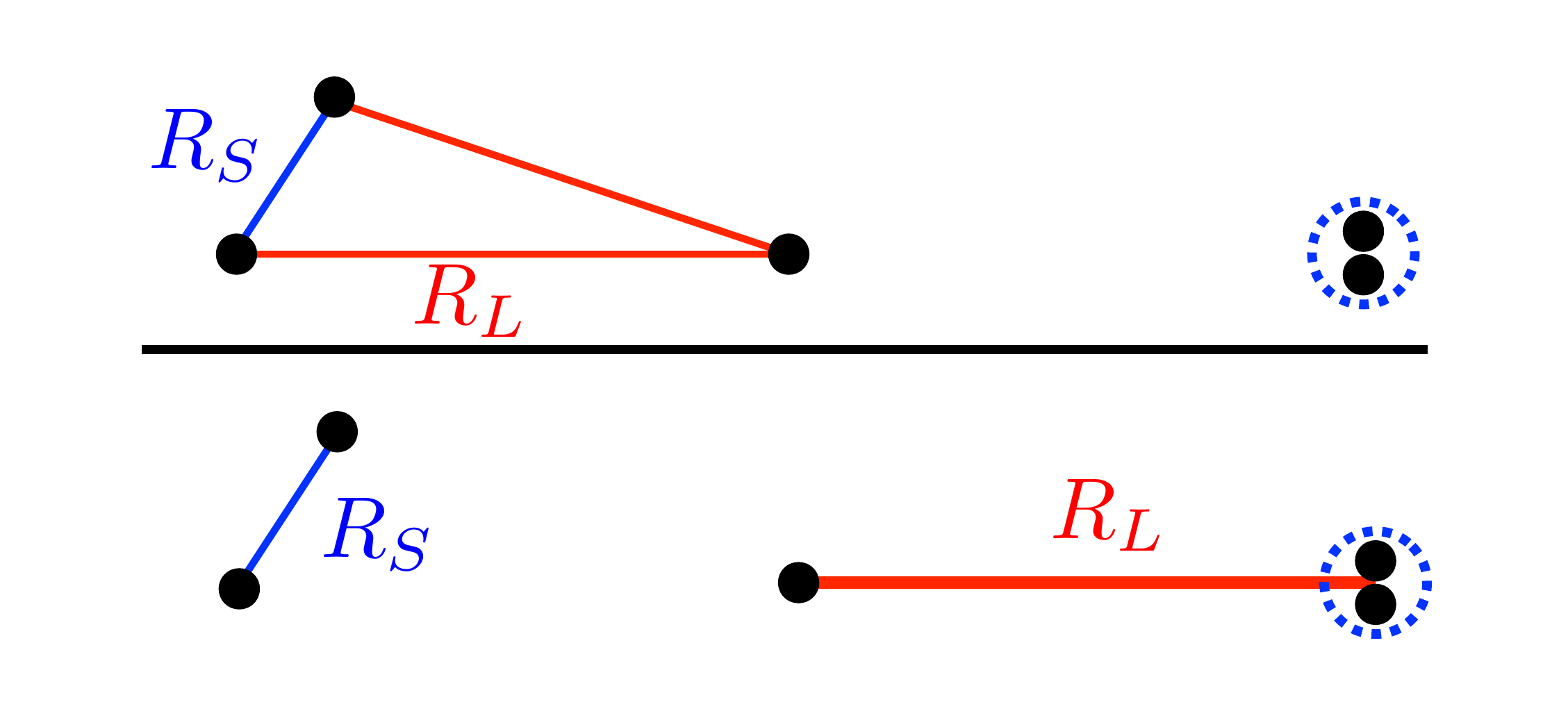}
\end{center}
\caption{The celestial non-gaussianity $Q_{\mathcal{E}}(\vec n_1, \vec n_2, \vec n_3)$, as defined in \Eq{eq:NG_def} as a ratio of correlators. }
\label{fig:NG_def}
\end{figure}
%%%%%%%%%%%%%%%%%%%%%%

At a hadron collider, the three-particle correlator is expressed in terms of the boost invariant angles $R_{ij}=\sqrt{\Delta y_{ij}^2 + \Delta \phi_{ij}^2}$ on the rapidity-azimuth cylinder.
Without loss of generality, we assume the angles are ordered as $R_{12} < R_{23} < R_{13}$. We then define the ``celestial non-gaussianity"%
\footnote{We use the word ``non-gaussianity'', since in the squeezed limit, the three-point correlator reduces to a product of two-point correlators. More generally in squeezed limits of high-point functions, they reduce to products of two-point functions. A non-standard feature is that these two-point functions involve higher powers, $\cE^n$ of the energy flow operator. In this sense, it is different from a standard non-gaussianity as in \Eq{eq:Ising_NG_def}, where the fields are strictly gaussian (free). However, we believe that \Eq{eq:NG_def} is one of the appropriate generalizations for the case of asymptotic energy flow. 
Note that unlike \Eq{eq:Ising_NG_def}, the denominator of the celestial non-gaussianity in \Eq{eq:NG_def} is not symmetric due to the ordering of the angles. In \App{sec:new_ratio}, we present numerical results for a symmetric celestial non-gaussianity $\widetilde{Q}_{\mathcal{E}}$. Although we focus primarily on the asymmetric definition in this paper, we believe that both definitions are worth exploring since they may make manifest different features of the three-point correlation.
} as the ratio
\begin{align}\label{eq:NG_def}
\boxed{Q_{\mathcal{E}}(\vec n_1, \vec n_2, \vec n_3)=\frac{\langle\mathcal{E}(\vec{n}_1)\, \mathcal{E}(\vec{n}_2)\, \mathcal{E}(\vec{n}_3)\rangle_{\Psi} ~ \langle\mathcal{E}^2(\vec{n}_1)\rangle_{\Psi} }
{\langle\mathcal{E}(\vec{n}_1)\, \mathcal{E}(\vec{n}_2)\rangle_{\Psi}  ~ \langle\mathcal{E}^2(\vec{n}_1)\, \mathcal{E}(\vec{n}_3)\rangle_{\Psi} }\,. }
\end{align}
Here, we have divided the three-point correlator by a product of two two-point correlators, one involving $\mathcal{E}^m(\vec{n})$, which is used to denote the measurement of $m$-th power weighting along direction $\vec{n}$.
Precise definitions of all observables appearing in the definition of the celestial non-gaussianity are provided in \App{sec:def_app}, and for our later studies, individual energy correlators are computed using \Ref{EEC_github}.%
\footnote{We are indebted to Patrick Komiske for creating and maintaining this software package.}
We have made the state dependence explicit, where $\Psi$ denotes a generic state in which the energy correlators are evaluated.
This ratio is illustrated in  \Fig{fig:NG_def}.
One can think of the terms in the denominator, which reproduce the squeezed limit through a product of two-point correlators, as being a form of Wick contraction.
Although the motivation for introducing $Q_\cE$ was to isolate the non-gaussianity, we find that it also turns out to be quite robust to hadronization, capturing in a clean way the perturbative structure of the three-point correlator.

For simplicity, in this paper, we consider the simplest case where the state $\Psi$ is unpolarized, which we will denote by $\Phi$ to distinguish it from the more general case.
(For a detailed discussion of energy correlators in polarized states, see e.g.~\cite{Chang:2020qpj}).
In this case, the energy correlators only depend on the relative angles between the $n_i$, namely $R_{12} < R_{23} < R_{13}$.
To study the celestial non-gaussianity experimentally, it is convenient to map the region over which it is defined into a square so that it can be binned in a simple manner.
A mapping allowing this was introduced in \Ref{Komiske:2022enw}, which we follow here.
To simplify the notation, we define the  long, medium, and small sides of the correlator by $(R_L, R_M, R_S)$.
We then change to the coordinates $(R_L, \xi, \phi)$, where 
\begin{equation}
\xi=\frac{R_S}{R_M} \,, \qquad \phi=\mathrm{sgn} (\tau) \, \arcsin \sqrt{1 - \frac{(R_L-R_M)^2}{R_S^2}}\,.
\label{eq:transf}
\end{equation}
Here $\mathrm{sgn}(\tau)$ is the sign of the determinant $\tau = \mathrm{det}(\vec{n}_3, \vec{n}_2, \vec{n}_1)$, characterizing whether the ordering of $(R_S, R_M, R_L)$ is  clockwise or counter-clockwise on the celestial sphere. With this choice of coordinates, $R_L$ is used to characterize the overall size of the correlator, and $(\xi, \phi)$ are used to characterize its shape. Since the primary focus of this paper is on the shape dependence, plots in this paper have a fixed $R_L$. Detailed studies of the scaling behavior in $R_L$ were presented in \Ref{Komiske:2022enw}.

The $(\xi, \phi)$ coordinates in \Eq{eq:transf} blow up the OPE region into a line, with $\xi \in (0,1)$ acting as a radial coordinate about the OPE limit and $\phi \in (-\pi/2, \pi/2)$ as an azimuthal coordinate.
In QCD, this observable is $\mathbb{Z}_2$-symmetric under $\phi \to -\phi$, and hence we restrict ourselves in the region $\phi \in (0, \pi/2)$.
A detailed discussion of the experimental implementation, under the $\mathbb{Z}_2$-symmetric assumption, is described in detail in \App{sec:algorithm}.

In addition to the full shape dependence of the celestial non-gaussianity, we also consider an azimuthally averaged version:
\begin{equation}
\langle Q_\cE \rangle_\phi (\xi) =\frac{
\int_{R_{12}<R_{23}<R_{13}} d\vec{n}_1 \, d\vec{n}_2 \, d\vec{n}_3 \, \delta(\xi - R_{12}/R_{23}) ~ \langle\mathcal{E}(\vec{n}_1)\, \mathcal{E}(\vec{n}_2)\, \mathcal{E}(\vec{n}_3)\rangle_\Phi ~ \langle\mathcal{E}^2(\vec{n}_1)\rangle_\Phi}
{\int_{R_{12}<R_{23}<R_{13}} d\vec{n}_1 \, d\vec{n}_2 \, d\vec{n}_3 \, \delta(\xi - R_{12}/R_{23}) ~ \langle\mathcal{E}(\vec{n}_1)\, \mathcal{E}(\vec{n}_2)\rangle_\Phi ~ \langle\mathcal{E}^2(\vec{n}_1)\, \mathcal{E}(\vec{n}_3)\rangle_\Phi}\,.
\label{eq:azimuthal_average}
\end{equation}
Being a one-dimensional distribution, this projection is easier to visualize than the full shape dependence, however, it does not allow us to study the shape of the non-gaussianities.

%%%%%%%%%%%%%%%%%%%%%%%%%%%%%%%%%%%%%%%%%%%%
\subsection{Theoretical Motivation} \label{sec:NG_motivation}
%%%%%%%%%%%%%%%%%%%%%%%%%%%%%%%%%%%%%%%%%%%%

At the first sight, the definition of the celestial non-gaussianity $Q_{\mathcal{E}}$ in \Eq{eq:NG_def} might not look straightforward, nor does it have the same form as the non-gaussianity $Q_\sigma$ of \Eq{eq:Ising_NG_def}.
We now describe in more detail the theoretical motivation for this definition in terms of the factorization of the three-point correlator in the squeezed limits.

\begin{figure}[t]
\begin{center}
\includegraphics[scale=0.7]{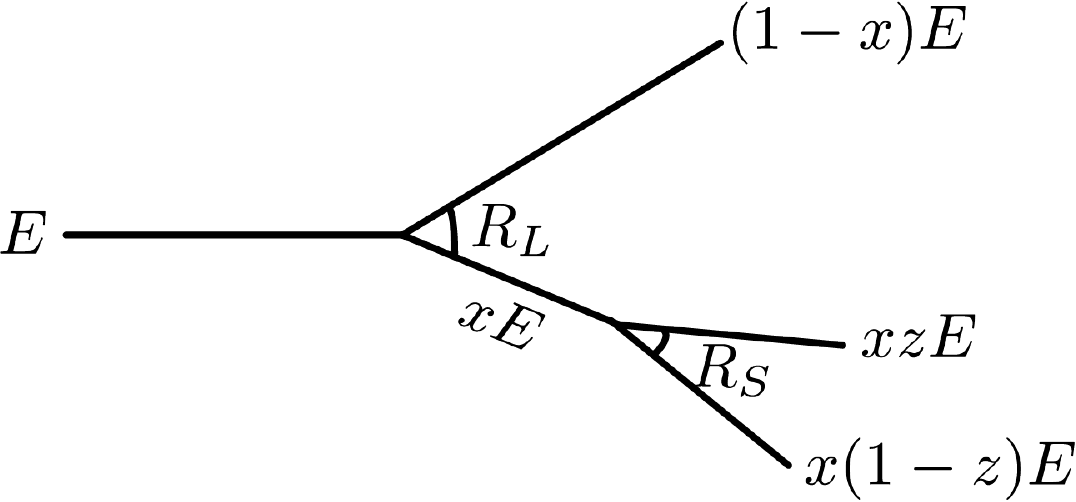}
\caption{An illustration of the squeezed limit of a $1 \to 3$ splitting from iterated $1 \to 2$ splittings. }
\label{fig:sequential}
\end{center}
\end{figure}

An intuitive motivation of \Eq{eq:Ising_NG_def} comes from momentum space particle splitting.
The EEEC in the squeezed limit is dominated by iterated $1 \to 2$ splittings.
At leading order in QCD, it requires two iterated $1 \to 2$ splittings, as shown schematically in \Fig{fig:sequential}, where the first branch has momentum fractions $x$ and $1-x$, while the second branch has momentum fractions $z$ and $1-z$.
The EEEC is then obtained by weighting the cross section with the energy of the final-state particles, $x^2 (1- x) z (1-z) E^3$.
This motivates us to put in the denominator of the ratio \Eq{eq:NG_def} a $\langle \mathcal{E} \mathcal{E} \rangle$ for the second splitting, which is weighted by $z (1 - z) E^2$, and a $\langle \mathcal{E}^2 \mathcal{E} \rangle$ for the first splitting, which is weighted by $x^2 (1-x) E^3$.

While this choice of denominator mimics the momentum fraction structure in the numerator, it has two issues: (a) the power of $E$ is not balanced between numerator and denominator;  and (b) $\langle \mathcal{E}^2 \mathcal{E} \rangle$ is not collinear safe since it is not linear in particle energies.
Interestingly, both issues can be overcome if we put a one-point $\langle \mathcal{E}^2 \rangle$ function in the numerator.
While $\langle \mathcal{E}^2 \rangle$ by itself is also collinear unsafe, it nicely cancels the collinear un-safety of $\langle \mathcal{E}^2 \mathcal{E} \rangle$ in the denominator, in the sense that the un-cancelled collinear divergences in perturbation theory for one-point function $\langle \mathcal{E}^2 \rangle$ and two-point correlator $\langle \mathcal{E}^2 \mathcal{E} \rangle$ are the same.
In the case that there are multiple flavors, this cancellation is not exact in the sense that the observable is strictly speaking collinear unsafe, nevertheless it has been observed in the study of track functions that the low moments of track/fragmentation functions, as appear in the calculation of the celestial non-gaussianity, are numerically very similar \cite{Jaarsma:2022kdd}. Therefore, while the celestial non-gaussianity is not technically collinear safe, we will see that it is numerically insensitive to hadronization. We therefore achieve an effectively collinear-safe ratio observable which is constructed from collinear-unsafe ingredients. Furthermore, it can be systematically computed in perturbation theory using the techniques of \Ref{Li:2021zcf}.
We believe that this construction can be generalized to more general energy weightings.

The above argument can be made rigorous using the leading-power perturbative light-ray OPE for the squeezed limit of the EEEC in QCD~\cite{Chen:2020adz,Chen:2021gdk}:
\beq
\langle\mathcal{E}(\vec{n}_1) \, \mathcal{E}(\vec{n}_2) \, \mathcal{E}(\vec{n}_3)\rangle_\Phi^{\mathrm{LP,\, LO}}
=\frac{1}{(2\pi)^2} \frac{2}{R_S^2}\frac{2}{R_L^2} {\mathcal{J}}\cdot {C}_{R_S}^{(1,1)}
\cdot {C}_{R_L}^{(1,2)}
\cdot \mathcal{S}_\Phi\,,
\eeq
where the details of the notation are explained in \App{sec:resum_formula}.
Roughly speaking, $\mathcal{J}$ and $\mathcal{S}_\Phi$ can be considered as projection vectors to select the correct matrix components.
The more interesting ingredients are the two OPE coefficients matrices $C_{R_S}^{(1,1)}$ and $C_{R_L}^{(1,2)}$ that respectively appear in the LO OPE for $\langle\mathcal{E}(\vec{n}_1)\mathcal{E}(\vec{n}_2)\rangle_\Phi^{\mathrm{LP,\, LO}}$ and $\langle\mathcal{E}^2(\vec{n}_1)\mathcal{E}(\vec{n}_3)\rangle_\Phi^{\mathrm{LP,\, LO}}$:
\begin{align}
\langle\mathcal{E}(\vec{n}_1)\mathcal{E}(\vec{n}_2)\rangle_\Phi^{\mathrm{LP,\, LO}} 
&= -\frac{1}{2\pi} \frac{2}{R_S^2} \mathcal{J} \cdot C_{R_S}^{(1,1)} \cdot \mathcal{S}_\Phi\,, \\
\langle\mathcal{E}^2(\vec{n}_1)\mathcal{E}(\vec{n}_3)\rangle_\Phi^{\mathrm{LP,\, LO}}
&= -\frac{1}{2\pi} \frac{2}{R_L^2} \mathcal{J} \cdot C_{R_L}^{(1,2)} \cdot \mathcal{S}_\Phi\,. 
\end{align}
 This formula implies that, neglecting matrix multiplication, the 3-point function factorizes into the product of two 2-point function with the energy weighting $(1,1)$ and $(1,2)$.

%%%%%%%%%%%%%%%%%%%%%%%%%%%%%%%%%%%%%%%%%%%%
\section{Theoretical Properties}
\label{sec:NG_analytic}
%%%%%%%%%%%%%%%%%%%%%%%%%%%%%%%%%%%%%%%%%%%%

While our ultimate goal is to study $Q_{\cE}$ in actual collider data, we begin by studying its properties using perturbative calculations as well as comparisons to parton showers.
In addition to gaining some intuition for its behavior, it is crucial to demonstrate that this observable is under perturbative control for jet energies accessible at the LHC.
We first consider the properties of the azimuthal-averaged $\langle Q_\cE \rangle_\phi$, highlighting its general behavior as well as how it can be understood in terms of celestial blocks.
We then consider the behavior of the full shape dependence of $Q_\cE$.

Throughout this section, we use the analytic results for the three-point correlator from \Ref{Chen:2019bpb}, as well as resummed results for the squeezed limits derived and discussed in detail in \App{sec:resum_formula}.
These factorized expressions can be embedded into jets at the LHC using the fragmenting jets formalism \cite{Procura:2009vm,Kang:2016ehg,Kang:2016mcy}, though we will not discuss this detail here.

%%%%%%%%%%%%%%%%%%%%%%%%%%%%%%%%%%%%%%%%%%%%
\subsection{Azimuthal Averaging}
\label{sec:blocks}
%%%%%%%%%%%%%%%%%%%%%%%%%%%%%%%%%%%%%%%%%%%%

\begin{figure}
\begin{center}
\subfloat[]{
\includegraphics[width=0.45\textwidth]{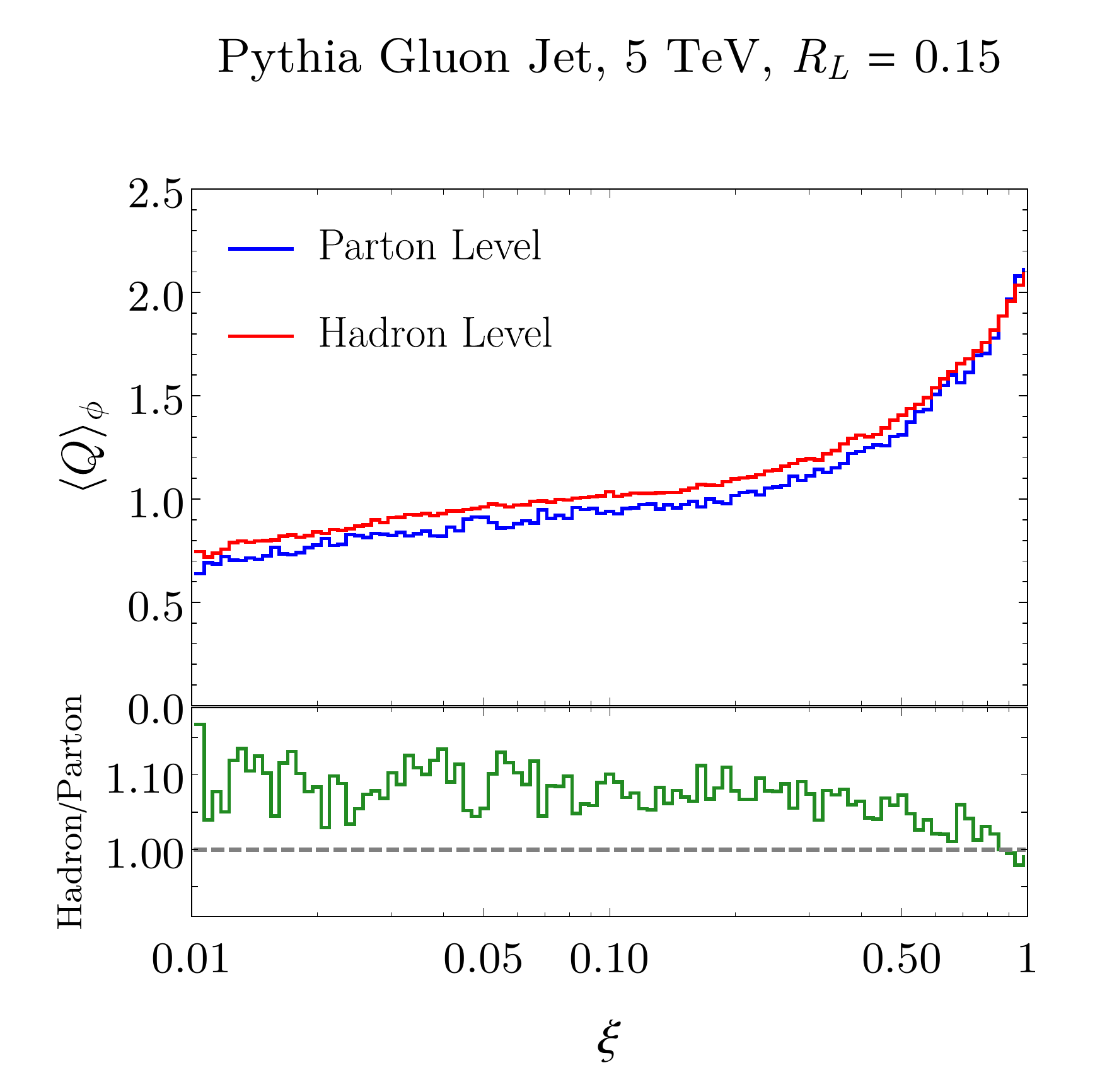}\label{fig:azi_5TeV_a}
}\qquad
\subfloat[]{
\includegraphics[width=0.45\textwidth]{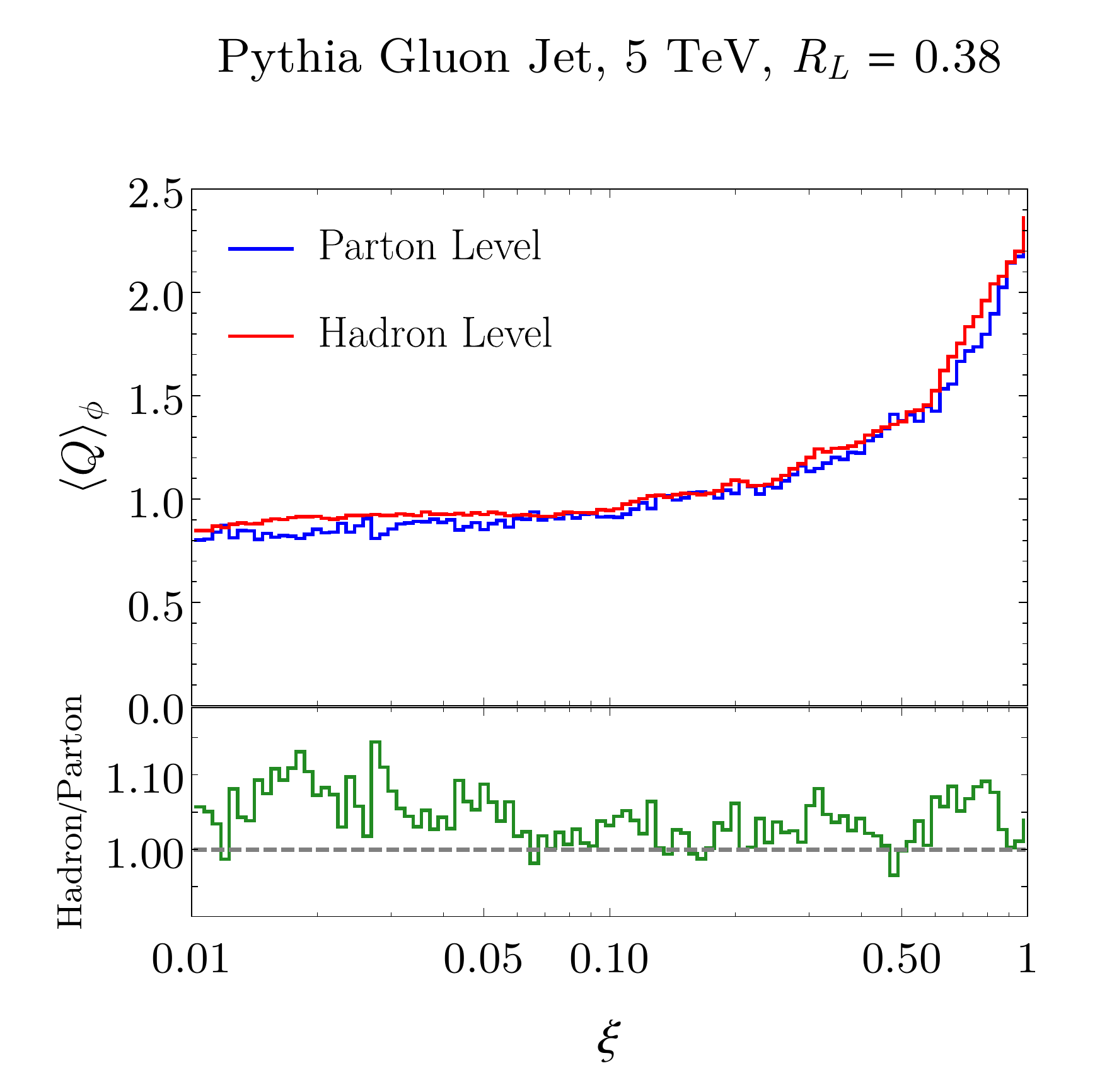}\label{fig:azi_5TeV_b}
}\qquad
\end{center}
\caption{
The azimuthally averaged celestial non-gaussianity for 5 TeV gluon jets for (a) $R_L = 0.15$ and (b) $R_L = 0.38$. 
We observe minimal corrections from hadronization and excellent perturbative control.
In the squeezed limit of $\xi \to 0$, $\langle Q \rangle_\phi (\xi)$ asymptotes to a constant, illustrating the correct construction of the celestial non-gaussianity ratio.
The celestial non-gaussianity increases monotonically away from the squeezed limit.}
\label{fig:azi_5TeV}
\end{figure}

\begin{figure}
\begin{center}
\subfloat[]{
\includegraphics[width=0.45\textwidth]{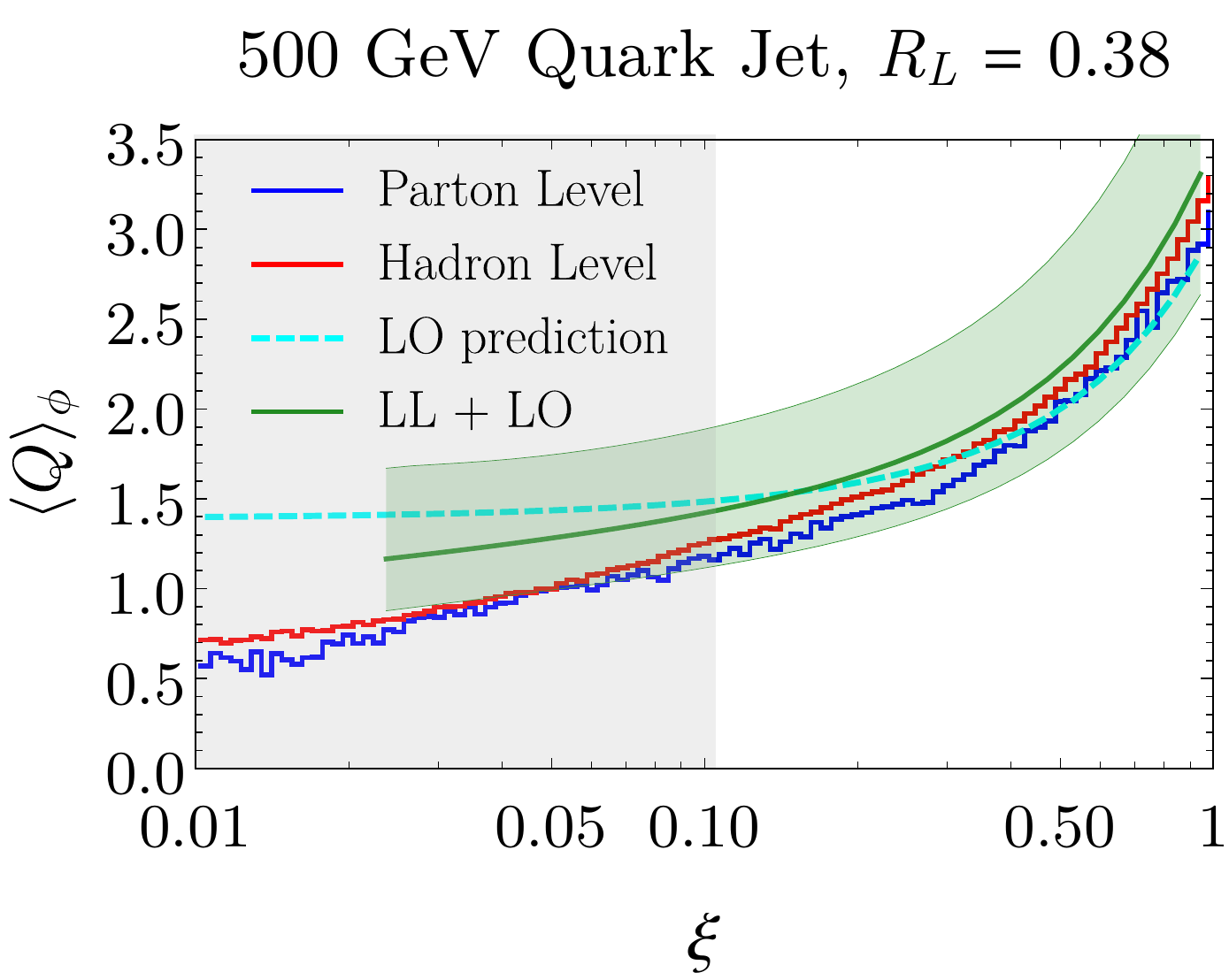}\label{fig:azi_500GeV_a}
}\qquad
\subfloat[]{
\includegraphics[width=0.45\textwidth]{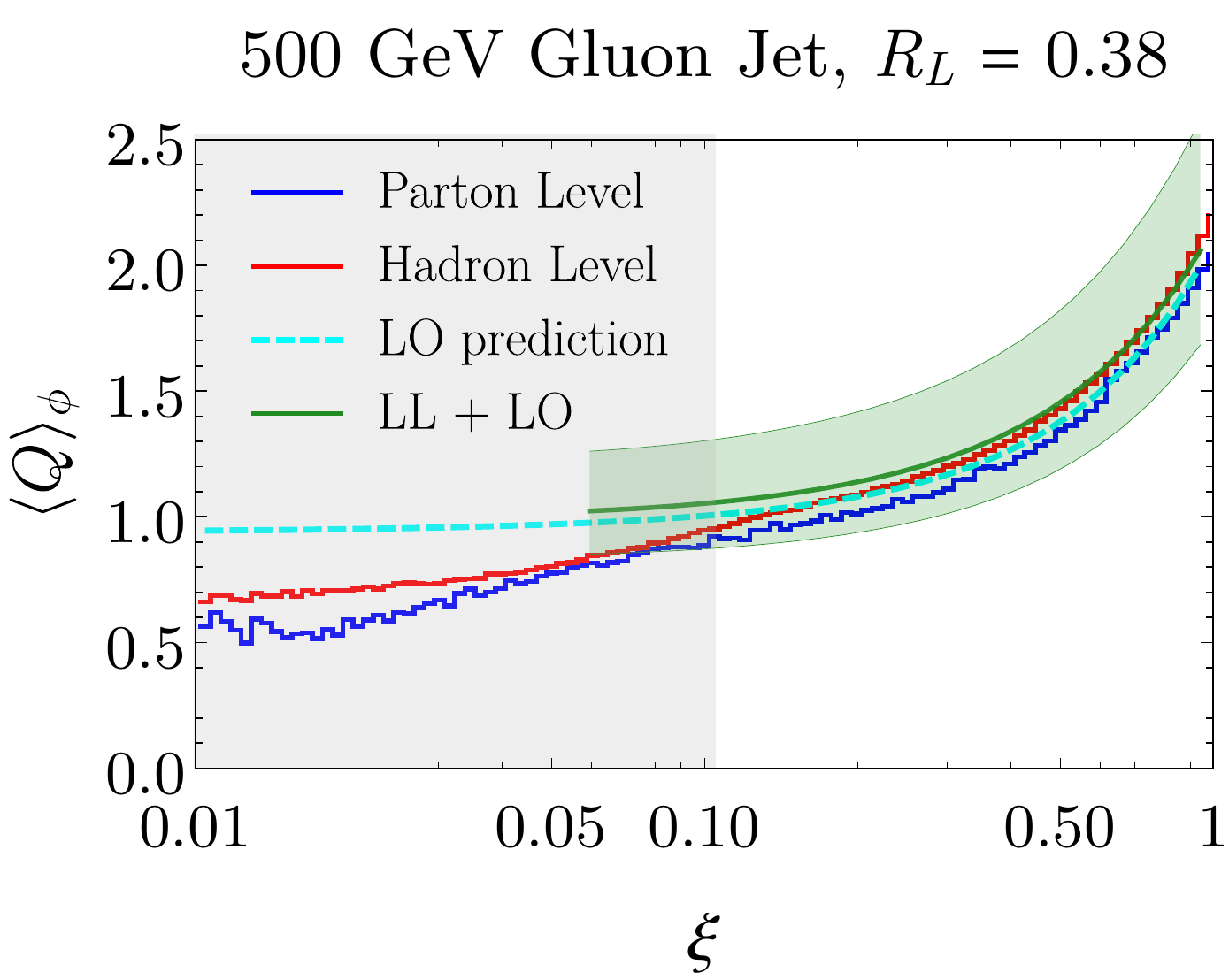}\label{fig:azi_500GeV_b}
}\\
\subfloat[]{
\includegraphics[width=0.45\textwidth]{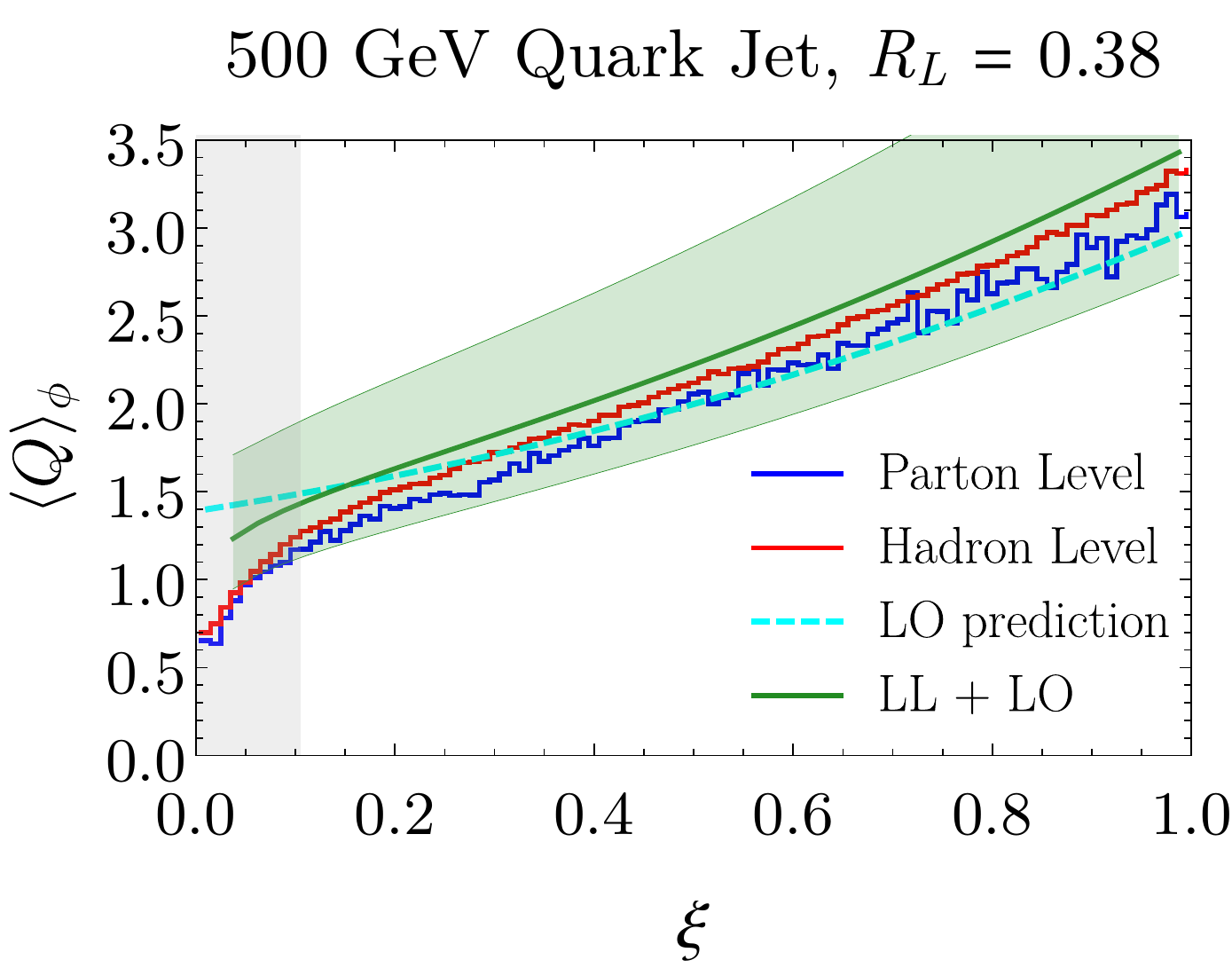}\label{fig:azi_500GeV_c}
}\qquad
\subfloat[]{
\includegraphics[width=0.45\textwidth]{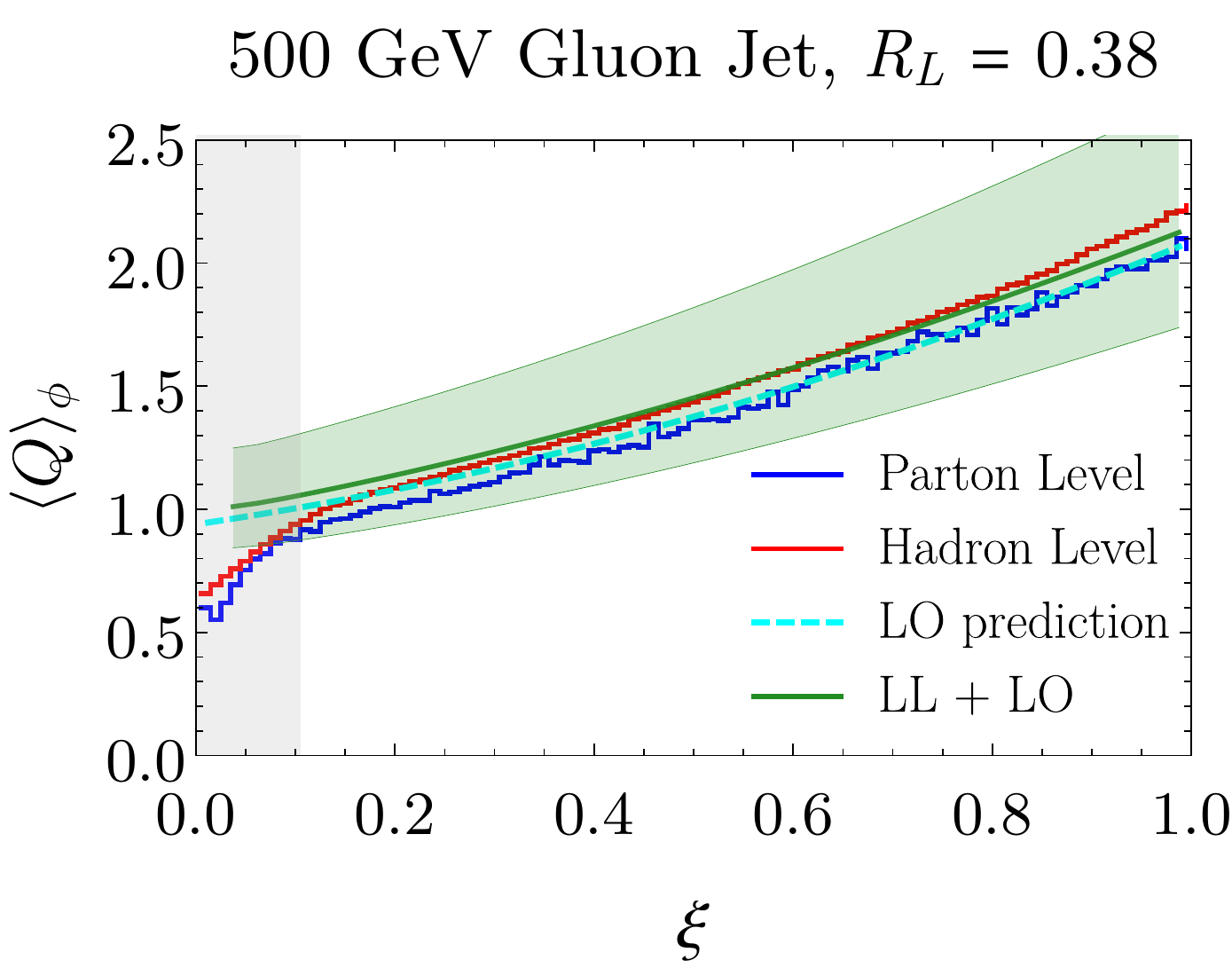}\label{fig:azi_500GeV_d}
}
\end{center}
\caption{
The azimuthally averaged celestial non-gaussianity for 500 GeV quark jets (left column) and gluon jets (right column).
The logarithmic (top row) and linear (bottom row) scalings respectively highlight the squeezed ($\xi\to 0$) and perturbative ($\xi\to 1$) limits.
The green error bands are from scale variation.
}
\label{fig:azi_500GeV}
\end{figure}

For simplicity, we start with the azimuthally-averaged non-gaussianity $\langle Q \rangle_\phi$, defined in \Eq{eq:azimuthal_average}.
We want to verify that the basic features of the celestial non-gaussianity are satisfied, namely that it flattens to a constant in the squeezed limit and that it is insensitive to hadronization.

We begin by considering jets at extremely high energies, namely 5 TeV anti-$k_T$ \cite{Cacciari:2008gp}, $R=0.5$ gluon jets simulated using \textsc{Pythia} 8.305 \cite{Sjostrand:2014zea,Sjostrand:2007gs} and clustered using \textsc{FastJet} 3.4.0 \cite{Cacciari:2011ma}, where these properties should be maximally apparent.
In \Fig{fig:azi_5TeV}, we show results for $\langle Q \rangle_\phi$ for two different values of $R_L$ at both parton level and hadron level.
We see clearly that the desired features of a non-gaussianity are indeed reproduced.
First, $\langle Q \rangle_\phi$ asymptotes to a constant (i.e.~becomes $\xi$ independent) in the squeezed limit of $\xi\to 0$, where the three-point correlator factorizes into a product of two-point correlators.
Second, despite being infrared and collinear unsafe, hadronization corrections are small, illustrating that the non-gaussianity is determined dominantly by perturbative physics.
Finally, we see a sharp rise in the non-gaussianity as $\xi\to 1$, showing that the three-point correlator deviates strongly from a product of two two-point correlators away from the squeezed limit. 
For any fixed jet energy, at smaller $R_L$, non-perturbative effects become larger, as illustrated by comparing the results for $R_L=0.15$ and $R_L=0.38$. When showing results in future sections, we will choose the largest value of $R_L$ possible for which we are unaffected by the jet boundary \cite{Komiske:2022enw}.

We now consider the more realistic energy of 500 GeV, which is the scale accessible with the CMS Open Data.
Results for both quark and gluon jets are shown in \Fig{fig:azi_500GeV}.
Since this is our ultimate energy of interest, we also show analytic results at leading order (LO) and with leading-logarithmic (LL) resummation.
In the squeezed limit ($\xi \to 0$, highlighted with the logarithmic axis), the non-gaussianity asymptotes to a constant, although not quite as clearly as at 5 TeV, and has minimal non-perturbative corrections.
In the perturbative regime ($\xi\to 1$, highlighted with the linear axis), the analytic results provide a good description of the parton shower behavior.
The error bands in green are the result of scale variation, and are quite large since we have only considered the LL result.

While any comparison between physically distinct systems comes with numerous caveats, it is amusing to compare to the level of non-gaussianity in high-energy jets to those found in the 2d and 3d Ising models.
In the Ising model, the non-gaussianity is minimized at $Q_{\text{min}}=1/\sqrt{6} \simeq 0.408$ in two-dimensions \cite{Belavin:1984vu} and $Q_{\text{min}}\simeq 0.683$ in three-dimensions \cite{Rychkov:2016mrc}.
Interestingly, the fractional variation in $Q$, by approximately a factor of $2$ (slightly larger in 2d, slightly smaller in 3d), is qualitatively similar to the QCD results shown in \Figs{fig:azi_5TeV}{fig:azi_500GeV}.
This provides some intuition for the amount of celestial non-gaussianity in the QCD energy flux, and also suggests that our measure is reasonable. 
We note also that Hofman and Maldacena found that at strong coupling, the three-point correlator of energy flow operators was also highly non-gaussian \cite{Hofman:2008ar}.

\subsection{Celestial Block Structure}

It is intuitive that the level of non-gaussianity increases as $\xi$ moves away from the squeezed $\xi \to 0$ region, since genuine $1 \to 3$ effects start to dominate over iterated $1 \to 2$ splittings.
It is interesting to understand if this intuition is a generic feature.
We now present an argument that  the monotonically increasing behavior of the non-gaussianity in $\xi$ on the line $\phi=0$, which corresponds to the flattened triangle configuration, is a consequence of Lorentz symmetry and the average null energy condition (ANEC) \cite{Tipler:1978zz,Klinkhammer:1991ki,Wald:1991xn,Hofman:2008ar,Faulkner:2016mzt,Hartman:2016lgu}. We will later show that the flattened triangle configuration dominates the shape of the non-gaussianity when there are poles associated to the massless propagators of the particles initiating the jets, and therefore the behavior in this region dominates the angular averaged non-gaussianity.  

\begin{figure}
\begin{center}
\subfloat[]{
\includegraphics[width=0.45\textwidth]{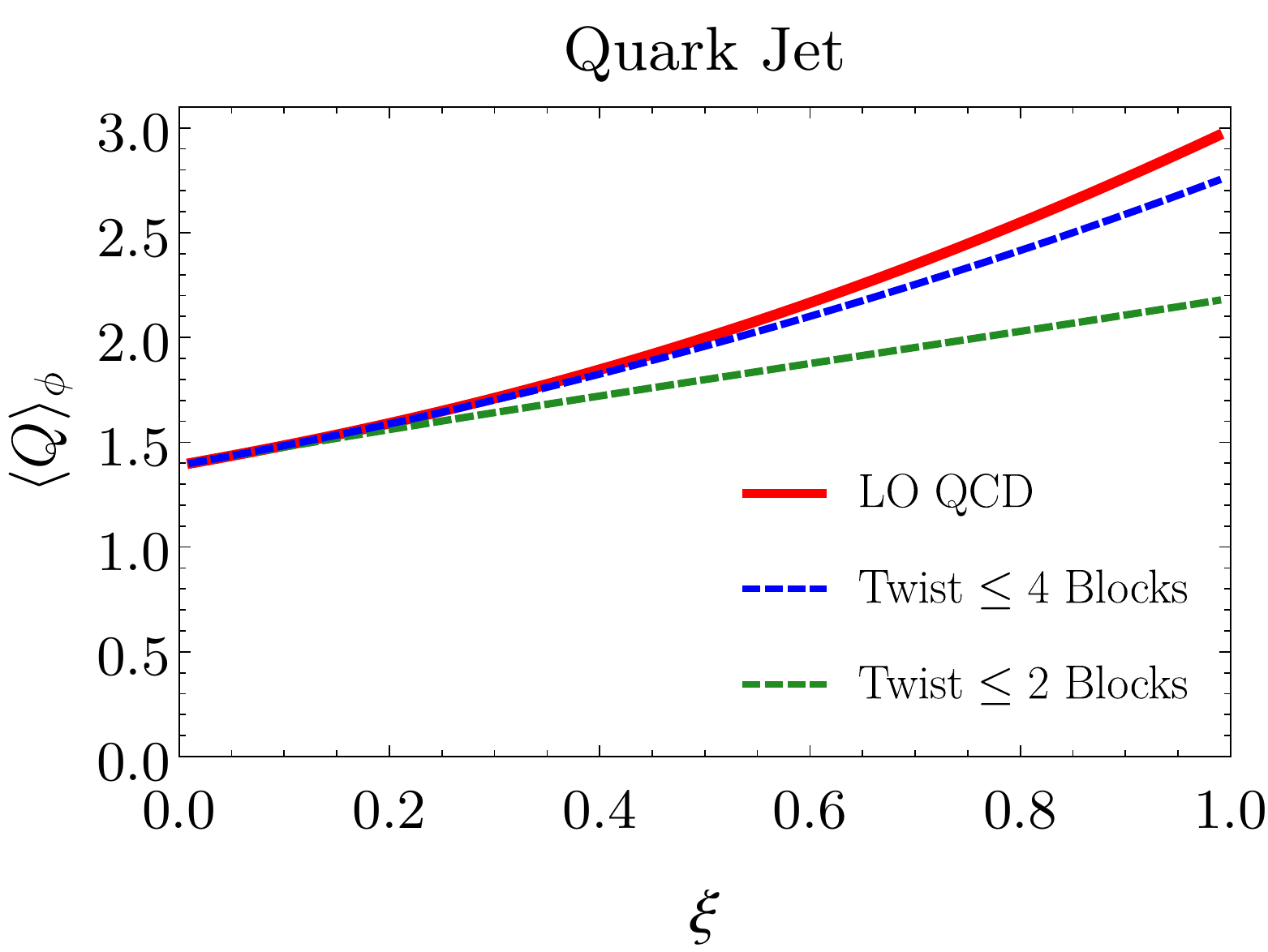}\label{fig:block_a}
}\qquad
\subfloat[]{
\includegraphics[width=0.45\textwidth]{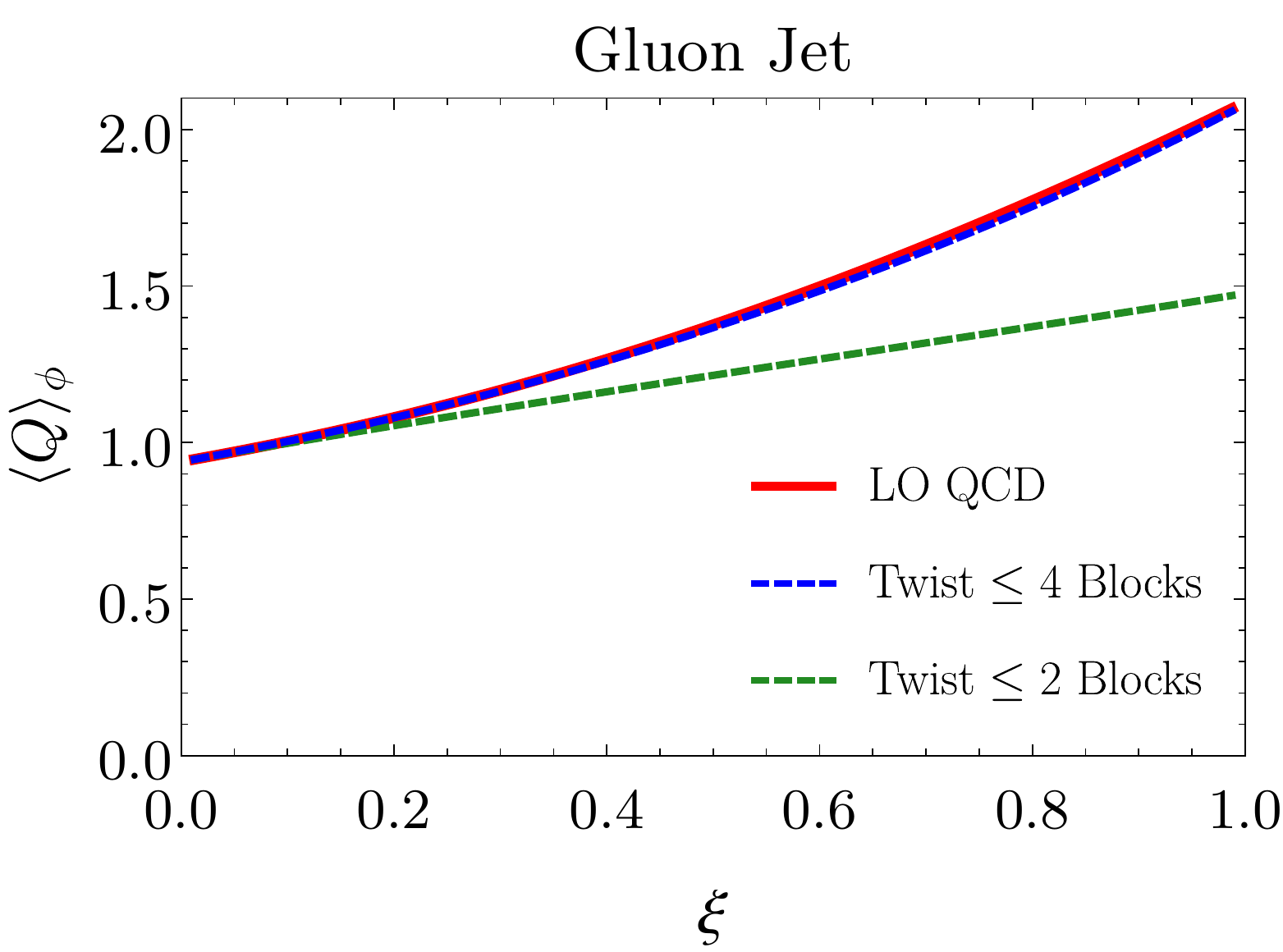}\label{fig:block_b}
}\qquad
%\captionsetup{font={footnotesize}}
\end{center}
\caption{
The celestial block expansion of the leading order result for the azimuthally averaged $\langle Q \rangle_\phi (\xi)$ for (a) quark jets and (b) gluon jets.
At LO, the result is independent of $R_L$.
The twist-4 expansion describes the dominant features of the non-gaussianity. The positivity of the non-gaussianity can the be derived from the properties of the celestial block expansion.}
\label{fig:block_expand}
\end{figure}

To understand the monotonicity of the non-gaussianity, we must first understand how the shape of the three-point correlator is constrained by the symmetries of the Lorentz group. It was shown in \Refs{Chang:2022ryc,Chen:2022jhb} that the three-point correlation function can be expanded in a basis of appropriate partial waves of definite quantum numbers under the Lorentz group, which are referred to as celestial blocks \cite{Chang:2022ryc,Chen:2022jhb}. This allows us to express the LO results for the quark and gluon three-point correlators as
\begin{equation}\label{eq:cb_expand}
G_{q/g} = \sum_{\delta,\,j} c_{\delta,j}^{(q/g)} G_{\delta,\,j}(\xi,\phi)\,,
\end{equation}
where the sum is over celestial quantum numbers $(\delta, j)$, which describe respectively the quantum number under boosts ($\delta$) and the transverse spin ($j$), $c_{\delta,j}^{(q/g)}$ are coefficients encoding the dynamics of the theory, and $G_{\delta,\,j}(\xi,\phi)$ are the celestial partial waves. More details can be found in \Refs{Chang:2022ryc,Chen:2022jhb}, and a short review is given in \App{sec:resum_formula}.

The celestial block expansion is an expansion about the squeezed limit.
In \Fig{fig:block_expand}, we show the celestial block expansion of $\langle Q_{\cE}\rangle_\phi$, including the leading twist-2 contribution and the twist-4 expansion.%
\footnote{
Intermediate states of light-ray OPEs are light transforms of twist operator $O_{\Delta, J}$ with dimension $\Delta$ and collinear spin $J$, where the twist is defined as $\Delta - J$. Similar to the twist expansion in deep ineleastic scattering, the light-ray OPE twist expansion provides a power expansion for timelike parton fragmentation. The expansion is systematic in a conformal field theory, while in perturbative QCD they receive corrections from running coupling. }
At LO, the result is independent of $R_L$. Here, we see that the block expansion converges well. In our coordinates, the leading block (twist-2, transverse spin $j=0$) has the form 
\begin{align}
G_{4,0}(\xi, \phi=0)=\left[_2F_1\left(1,2,4,\,\frac{\xi}{\xi+1}\right)\right]^2\,.
\end{align}
From the definition of the hypergeometric function $_2F_1(a,b,c,z) = \sum_{n=0}^{\infty} \frac{\Gamma(n+a) \Gamma(n+b)}{\Gamma(n+c)\Gamma(n)} \frac{z^n}{n!}$, we see that $_2F_1(a,b,c,z)$ should monotonically increase on its convergent domain $z\in (0,1)$ when $a,b,c>0$. This structure is fixed by Lorentz invariance. The coefficient of the leading block is positive by the ANEC, and hence this guarantees the growth of the non-gaussianity (at least in some small window), as one increases $\xi$ away from $0$. Due to the near cancellation between bosonic and fermionic contributions, the leading twist transverse spin-2 block gives a negligible contribution, and so we do not consider it \cite{Chen:2020adz,Chen:2021gdk}. The twist-4 transverse spin-0 and transverse spin-2 celestial blocks take the form of similar hypergeometric functions and are therefore also monotonic. Unlike for the OPE of local operators, the positivity of OPE coefficients in the light-ray OPE is less well understood. However, we again find that the OPE coefficients for the twist-4 operators are positive; it would be interesting to understand this better.\footnote{Starting at twist-6, the OPE coefficients in the EEEC are no longer found to be positive, which was identified as arising from soft (zero-mode) contributions \cite{Chen:2022jhb}. Again, it would be interesting to understand this issue better.} Therefore, while the leading twist-2 contribution guarantees the increase of $\langle Q_{\cE}\rangle_\phi$ in some small window of the squeezed limit, this extends to the whole range due to the positivity of the higher twist OPE coefficients.

%%%%%%%%%%%%%%%%%%%%%%%%%%%%%%%%%%%%%%%%%%%%
\subsection{Shapes of Non-Gaussianities}
\label{sec:shape_explain}
%%%%%%%%%%%%%%%%%%%%%%%%%%%%%%%%%%%%%%%%%%%%

\begin{figure}[t]
\begin{center}
\subfloat[]{
\includegraphics[width=0.45\textwidth]{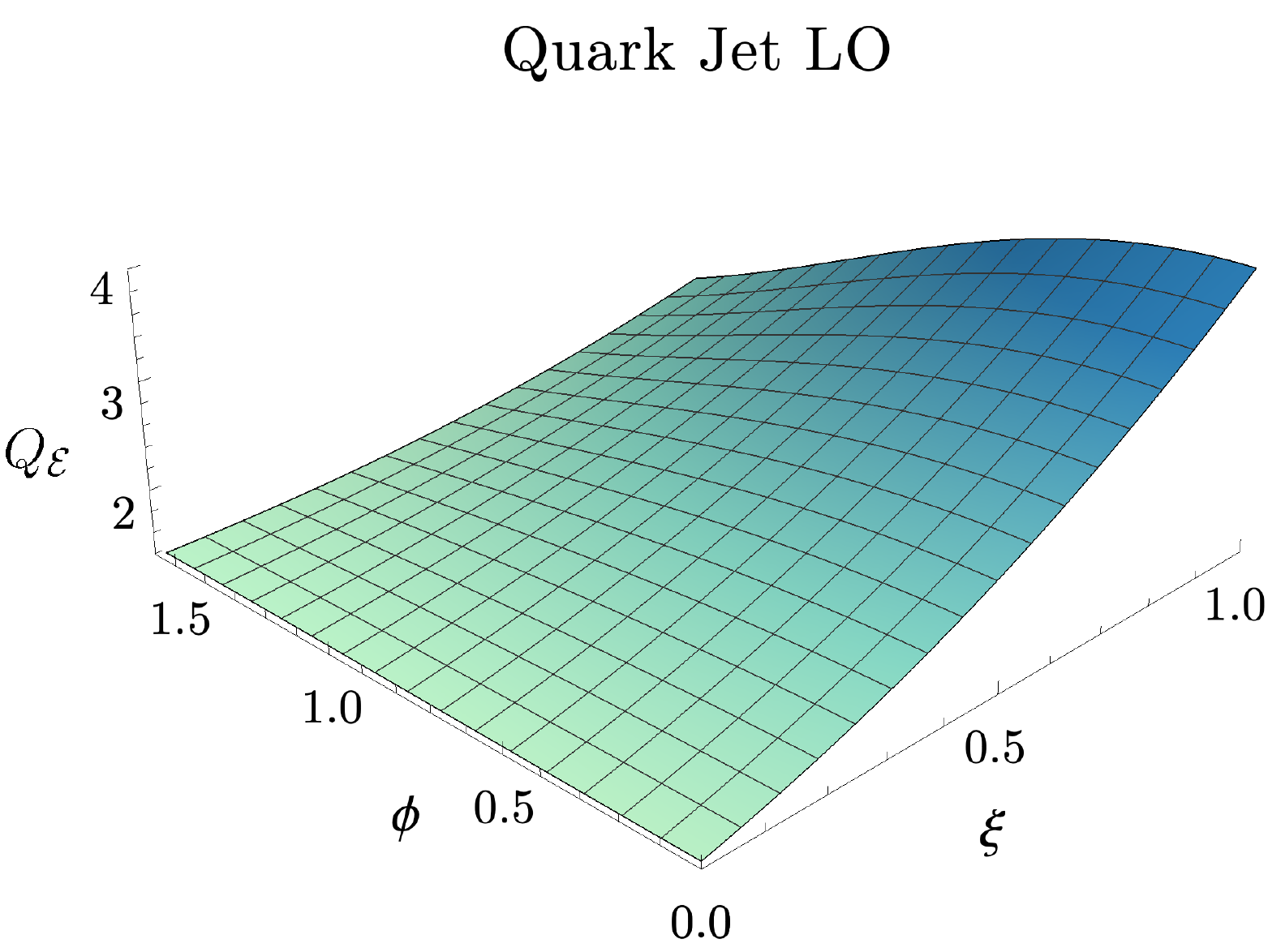}\label{fig:pert_shape_a}
}\qquad
\subfloat[]{
\includegraphics[width=0.45\textwidth]{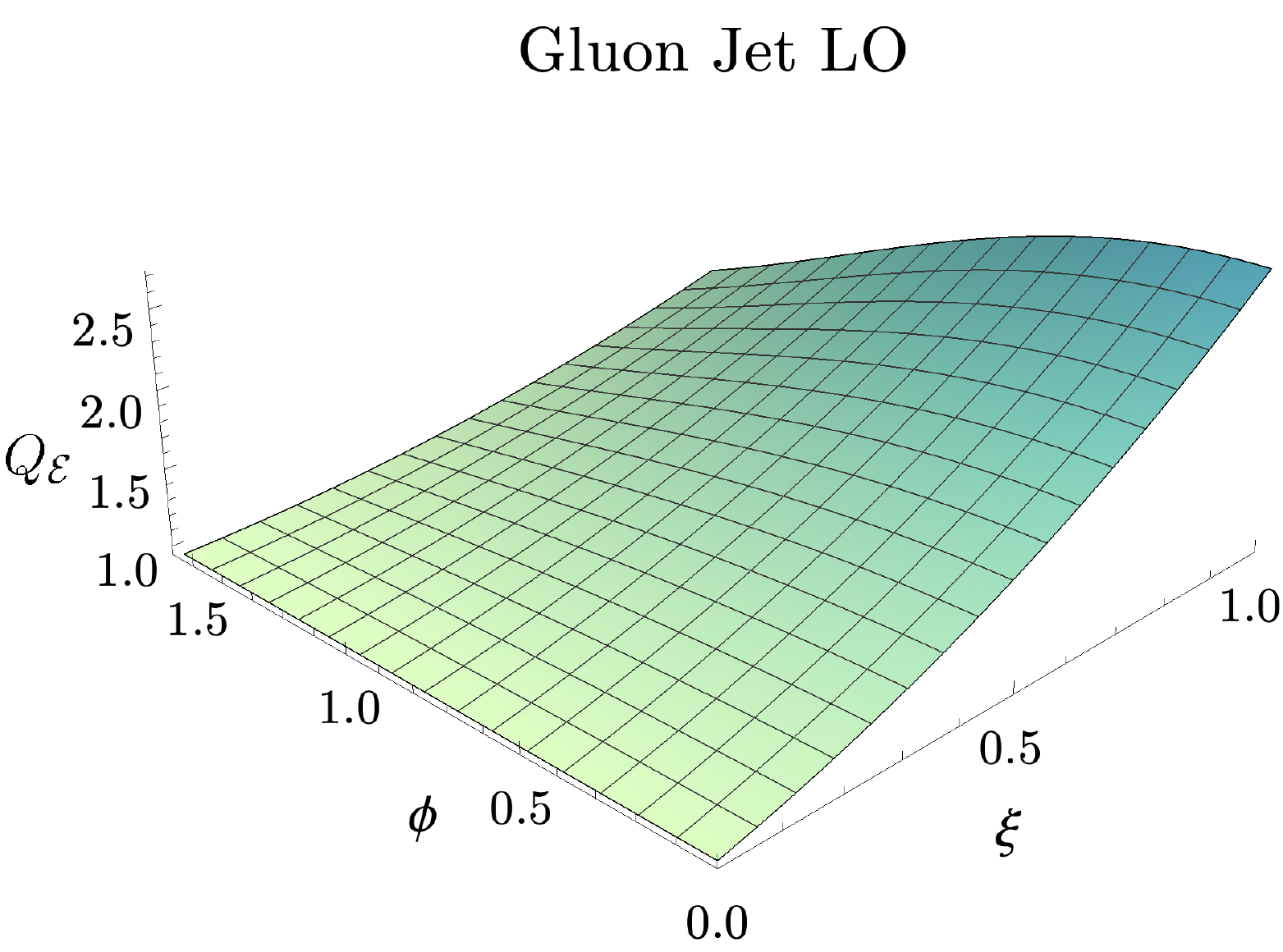}\label{fig:pert_shape_b}
}%
\end{center}
\caption{
The full shape dependence of the celestial non-gaussianity, $Q_\mathcal{E}$, in the leading order perturbative calculations for (a) quark jets and (b) gluon jets. The non-gaussianity is strongly peaked in the  ``flattened triangle" region, illustrating the presence of the propagator associated with the highly boosted quark or gluon state producing the jet.}
\label{fig:pert_shape}
\end{figure}

The most exciting aspect of the celestial non-gaussianity is that we can study its full shape dependence, much like what is done for cosmological correlators.
In \Fig{fig:pert_shape}, we show the shape dependence of the celestial non-gaussianity for both quark and gluon jets using the leading order calculation of \Ref{Chen:2019bpb}.
In both cases we find that it is peaked in the ``flattened triangle" region (illustrated in \Fig{fig:shapes_c}).
Furthermore, we see that the overall shape of the non-gaussianity is quite similar for both quark and gluon jets.
As we will see shortly, this is due to the fact that in both cases it is dominated by the leading twist celestial block, whose form is fixed by Lorentz symmetry.
While the shape is quite similar, the non-gaussianity is larger for quark jets than gluon jets. Unlike for many jet substructure observables that have a simple scaling with the Casimir factors, this is not true for the non-gaussianities. For example, for a three-point splitting function that is the exact iteration of the $1\to 2$ splitting functions, the celestial non-gaussianity is constructed to be unity independent of the color factors. The fact that the non-gaussianity is larger for quark jets illustrates a non-trivial feature of the $1\to 3$ splitting functions, namely that they are less close to iterated $1\to 2$ splittings for quark jets than gluon jets. It would be interesting to understand this more intuitively.

\begin{figure}[t]
\begin{center}
\subfloat[]{
\includegraphics[width=0.45\textwidth]{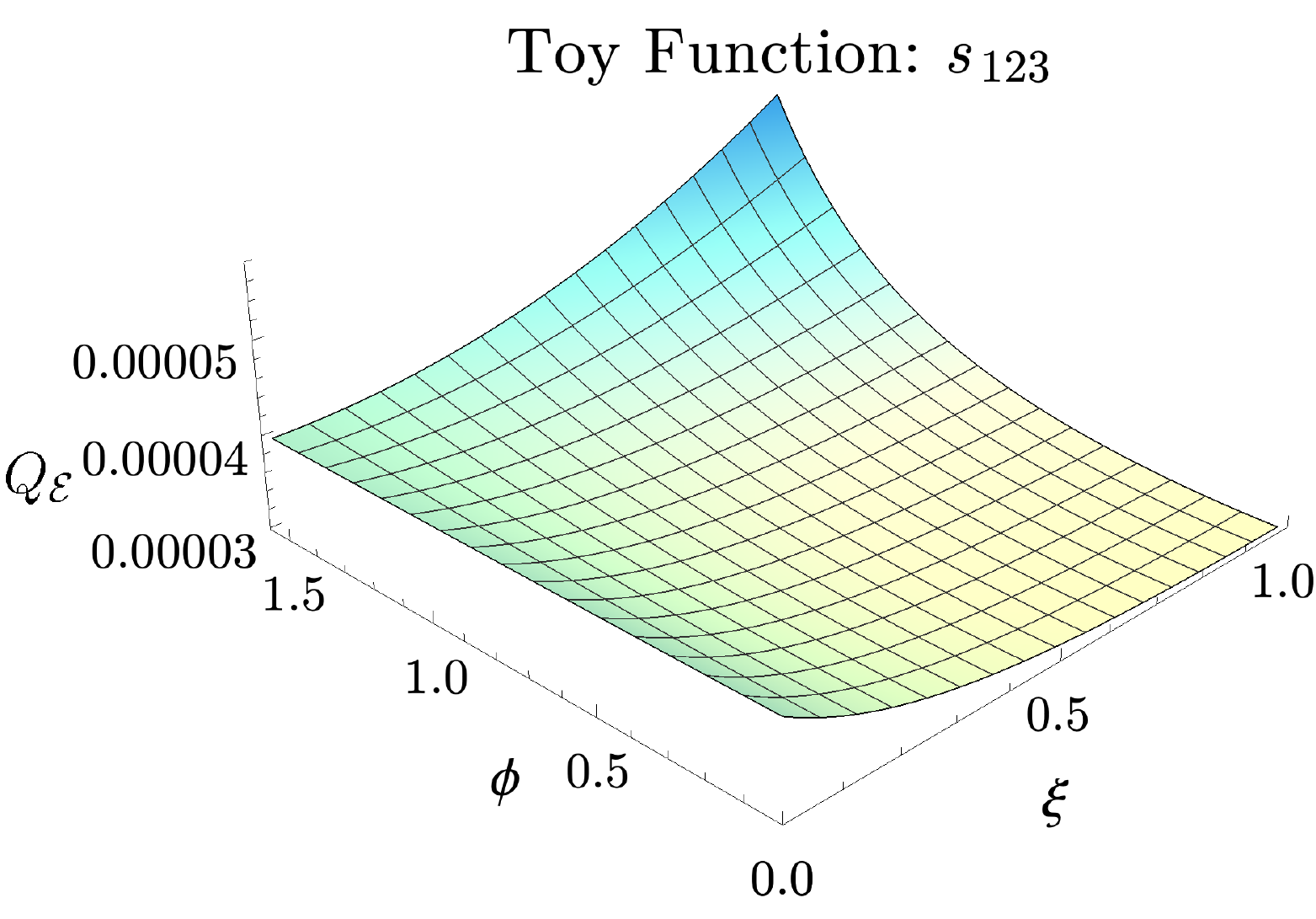}\label{fig:toy_a}
}\qquad
\subfloat[]{
\includegraphics[width=0.45\textwidth]{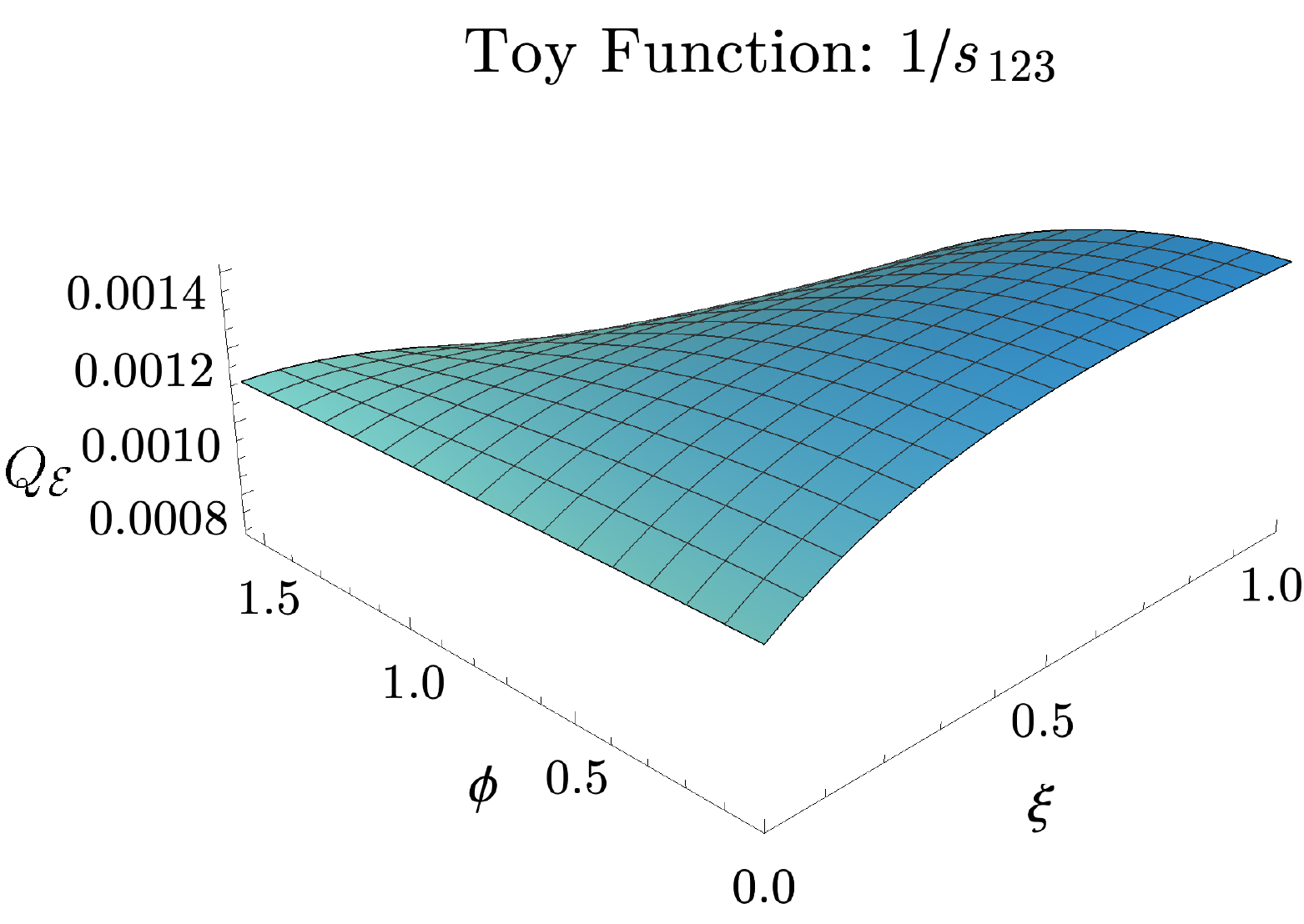}\label{fig:toy_b}
}
\end{center}
\caption{
A comparison of the shape dependence of the celestial non-gaussianity for toy functions:  (a) $s_{123}$ and (b) $1/s_{123}$.  We see that the function $s_{123}$ gives rise to a non-gaussianity that is peaked in the equilateral region, whereas the function $1/s_{123}$ gives rise to a non-gaussianity peaked in the flattened triangle region. }
\label{fig:toy}
\end{figure}

It is enlightening to compare the shape of the celestial non-gaussianity with non-gaussianities studied in other systems.
In the 2d and 3d Ising models, the non-gaussianity is peaked in equilateral triangle configurations \cite{Rychkov:2016mrc}, making it quite distinct from the case of celestial non-gaussianities.
The ``shapes of non-gaussianities" have been extensively studied for cosmological correlators, where one can find models of inflation that give rise to enhanced non-gaussianities for a wide variety of shapes; see e.g.~\cite{Babich:2004gb,Chen:2006nt} or more recently for the tensor case \cite{Cabass:2021fnw}.
For example, ghost inflation \cite{Arkani-Hamed:2003juy}, DBI \cite{Alishahiha:2004eh,Silverstein:2003hf}, and general higher-derivative operators have non-gaussianities peaked in the equilateral region \cite{Babich:2004gb,Chen:2006nt,Baumann:2011su}; solid inflation is peaked in the squeezed limit \cite{Endlich:2012pz}; and excited initial states lead to enhanced non-gaussianities in the flattened limit \cite{Chen:2006nt,Holman:2007na,Meerburg:2009ys,LopezNacir:2011kk,Flauger:2013hra,Green:2020whw,Green:2022fwg}.
This last example has been studied in detail recently, trying to use the absence of enhancement of non-gaussianity in the flattened limit to prove the quantum nature of cosmological fluctuations \cite{Green:2020whw,Green:2022fwg}.

\begin{figure}
\begin{center}
\subfloat[]{
\includegraphics[width=0.45\textwidth]{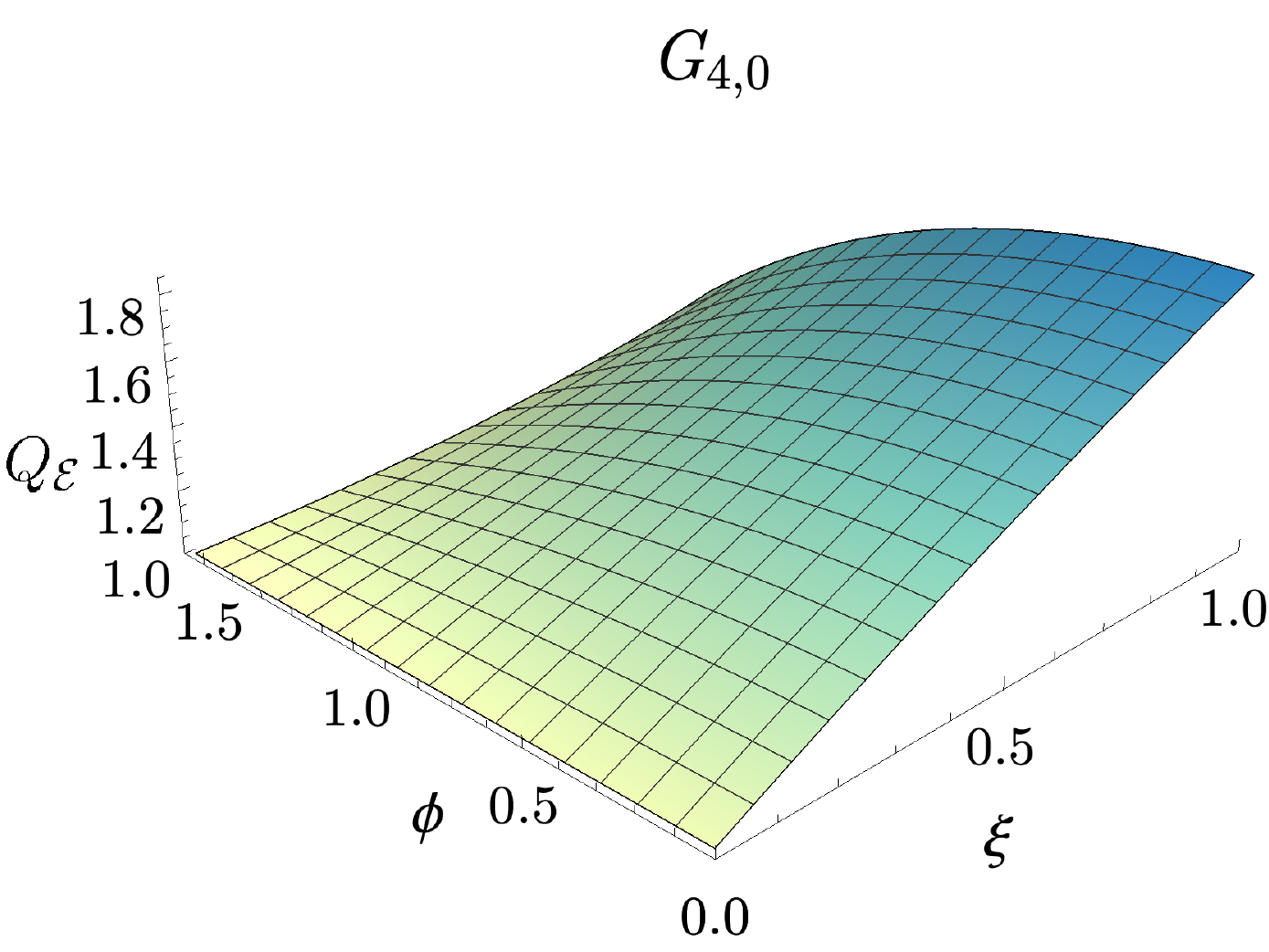}\label{fig:G40_3d}
}\qquad 
\subfloat[]{
\includegraphics[width=0.45\textwidth]{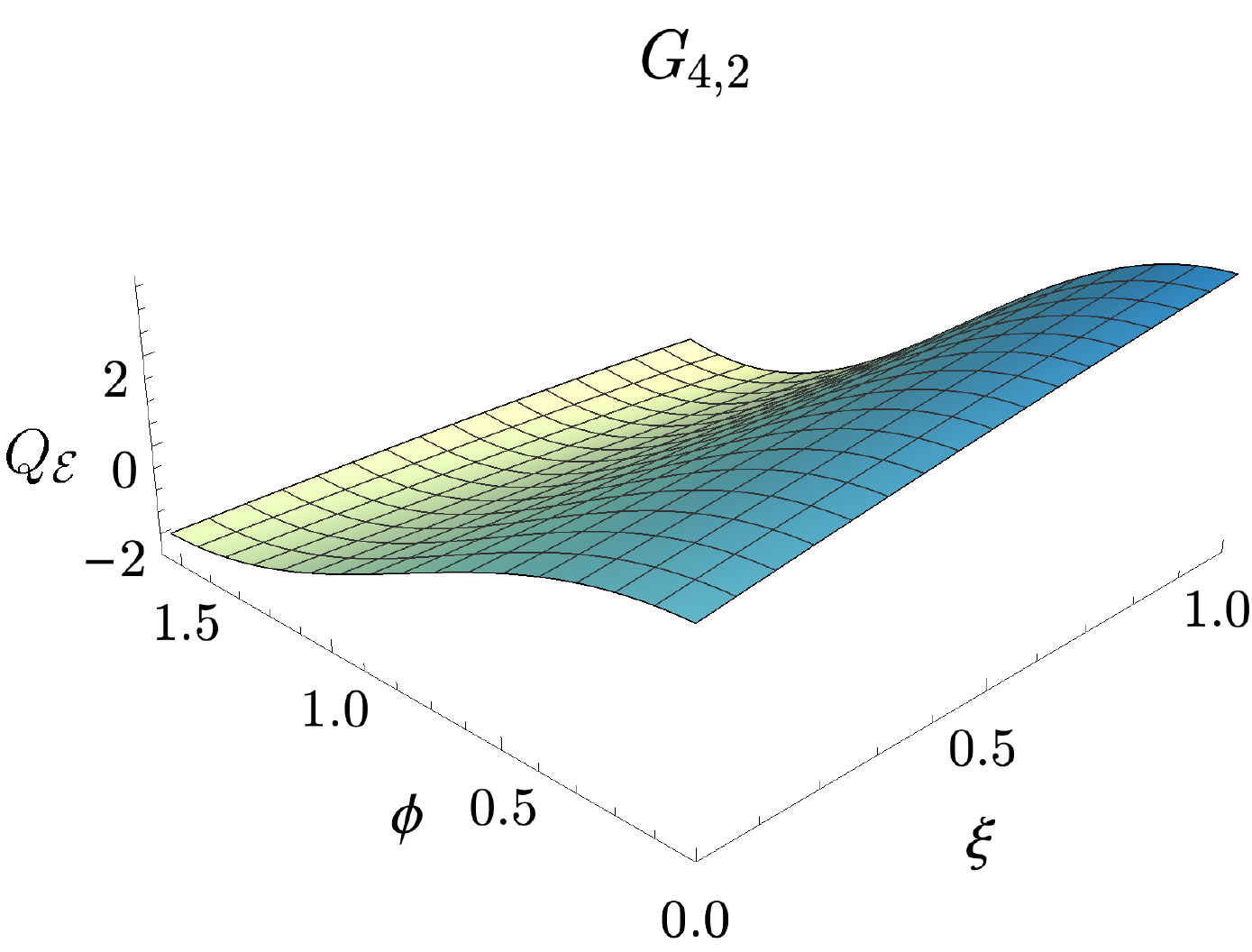}\label{fig:G42_3d}
}\\
\subfloat[]{
\includegraphics[width=0.45\textwidth]{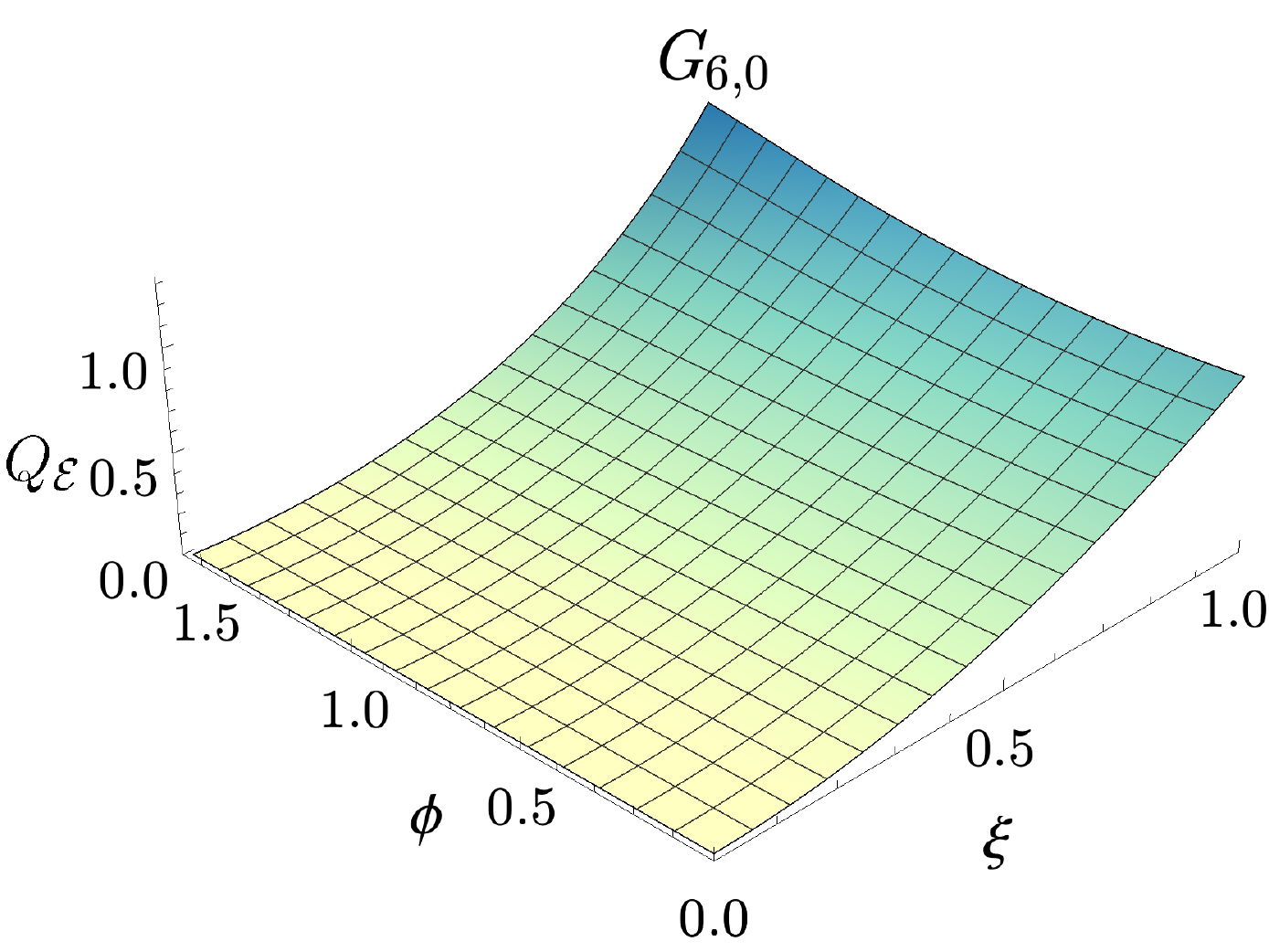}\label{fig:G60_3d}
}\qquad 
\subfloat[]{
\includegraphics[width=0.45\textwidth]{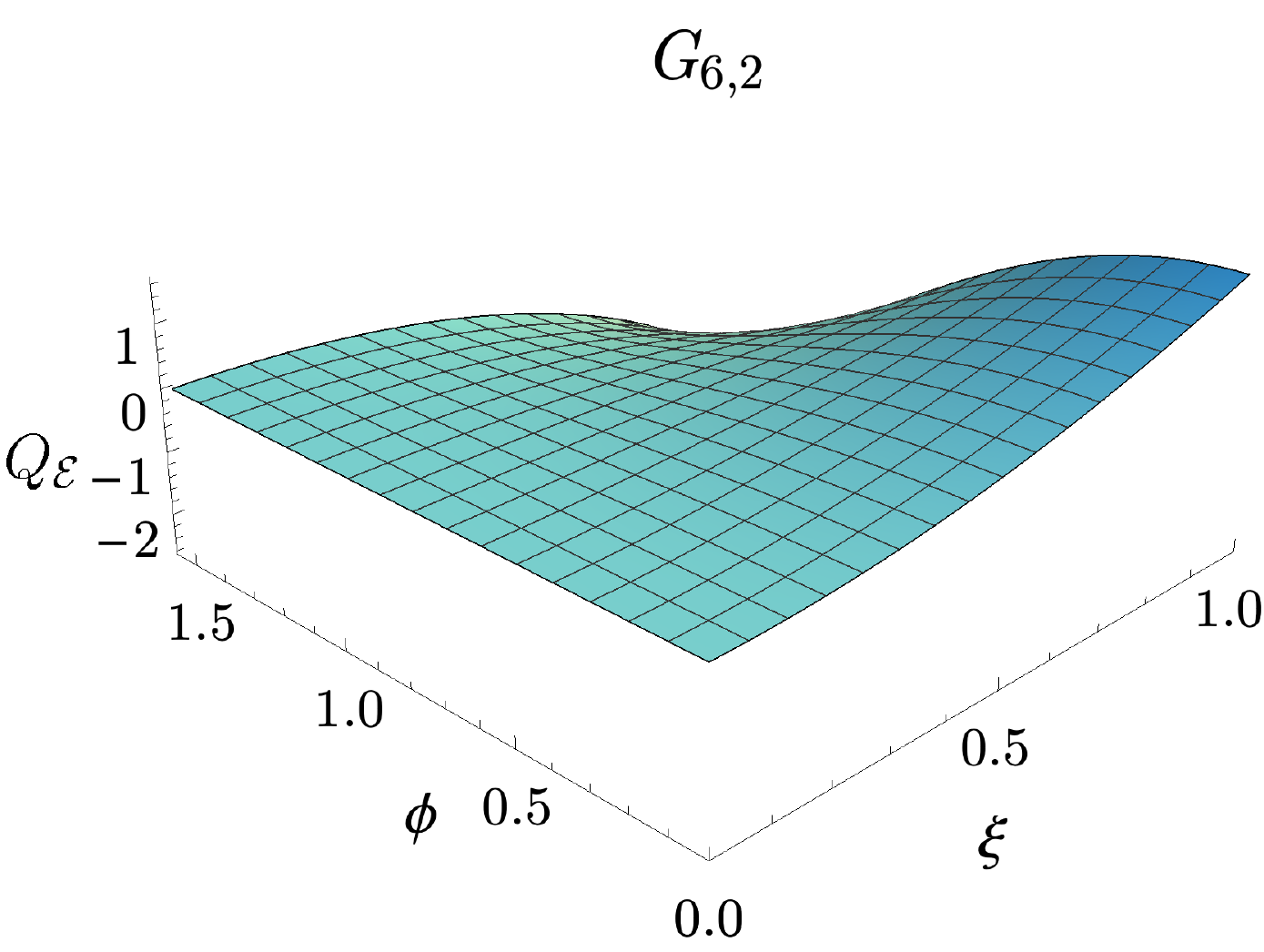}\label{fig:G62_3d}
}
\end{center}
\caption{
The celestial non-gaussianity for twist-2 blocks $G_{4, j}$ (top row) and twist-4 blocks $G_{6, j}$ (bottom row), with $j = 0$ (left column) and $j = 2$ (right column).  The leading twist block with zero transverse spin---$G_{4, 0}$ in (a)---provides the dominant contribution to the shape for high-energy jets, since it appears with the largest pre-factor in the celestial block expansion.}
\label{fig:twist2_blocks}
\end{figure}

Much like in the cosmological case, we can also gain some intuition for the shape of the celestial non-gaussianity by studying the shape of $Q_\cE$ for certain toy functions, instead of the full QCD splitting functions.
In \Fig{fig:toy_a}, we show the result for the toy function $s_{123} = s_{12} + s_{23} + s_{13}$.
This is peaked in the equilateral triangle configuration, and is thus completely different in shape from the QCD result.
This can be viewed analogously to the case of a higher-derivative operator for cosmological correlators.
By contrast, in \Fig{fig:toy_b} we show the result for the toy function $1/s_{123}$, which provides a simplified description of the pole structure for the initiating parton in QCD.
Here, we see that the result is maximized for a flattened triangle configuration.
This should be viewed in analogy with the case of an excited initial state in cosmology~\cite{Green:2020whw,Green:2022fwg}.

While the full case of QCD is of course more complicated than a single pole in $s_{123}$, the enhancement in the flattened limit arises for the same general reason, namely due to the presence of the propagator associated with the highly boosted quark/gluon state.
This is reflected in the monotonically decreasing behavior as a function of $\phi$ at large $\xi$.
We find it satisfying that for the first time we are starting to be able to understand the flux of energy within jets at this level of detail.
This shape is robust, and we will see that it is well borne out in CMS Open Data.

As discussed in \Sec{sec:blocks}, another way of analyzing the three-point correlation function is to expand it in celestial blocks.
Since the celestial block expansion converges rapidly, it is interesting to plot the shape dependence of the contributions to the non-gaussianities for the low twist blocks.
This is analogous to the visualization of the spherical harmonics for the shape of atomic orbits.
We show the contributions to the non-gaussianity for the twist-2 and twist-4 blocks in \Fig{fig:twist2_blocks}.
The blocks with a non-zero transverse spin $j$ exhibit a clear modulation in $\phi$.
We see that the contribution from the leading twist, $j=0$ block, $G_{4,0}$, has a similar shape to the full result, showing that it captures most of the shape dependence.

We emphasize that blocks themselves, like spherical harmonics, are purely kinematic and their contributions should be weighted by the numerical coefficients given in \Eqs{eq:quark_block_approx}{eq:gluon_block_approx} for quark and gluon jets, respectively.
For example, while $G_{4,2}$ has quite a different shape, it is negligible in both quark and gluon jets because the corresponding coefficients are small due to the cancellation between bosonic and fermionic statistics~\cite{Chen:2020adz,Chen:2021gdk}.

\begin{figure}
\begin{center}
\subfloat[]{
\includegraphics[width=0.45\textwidth]{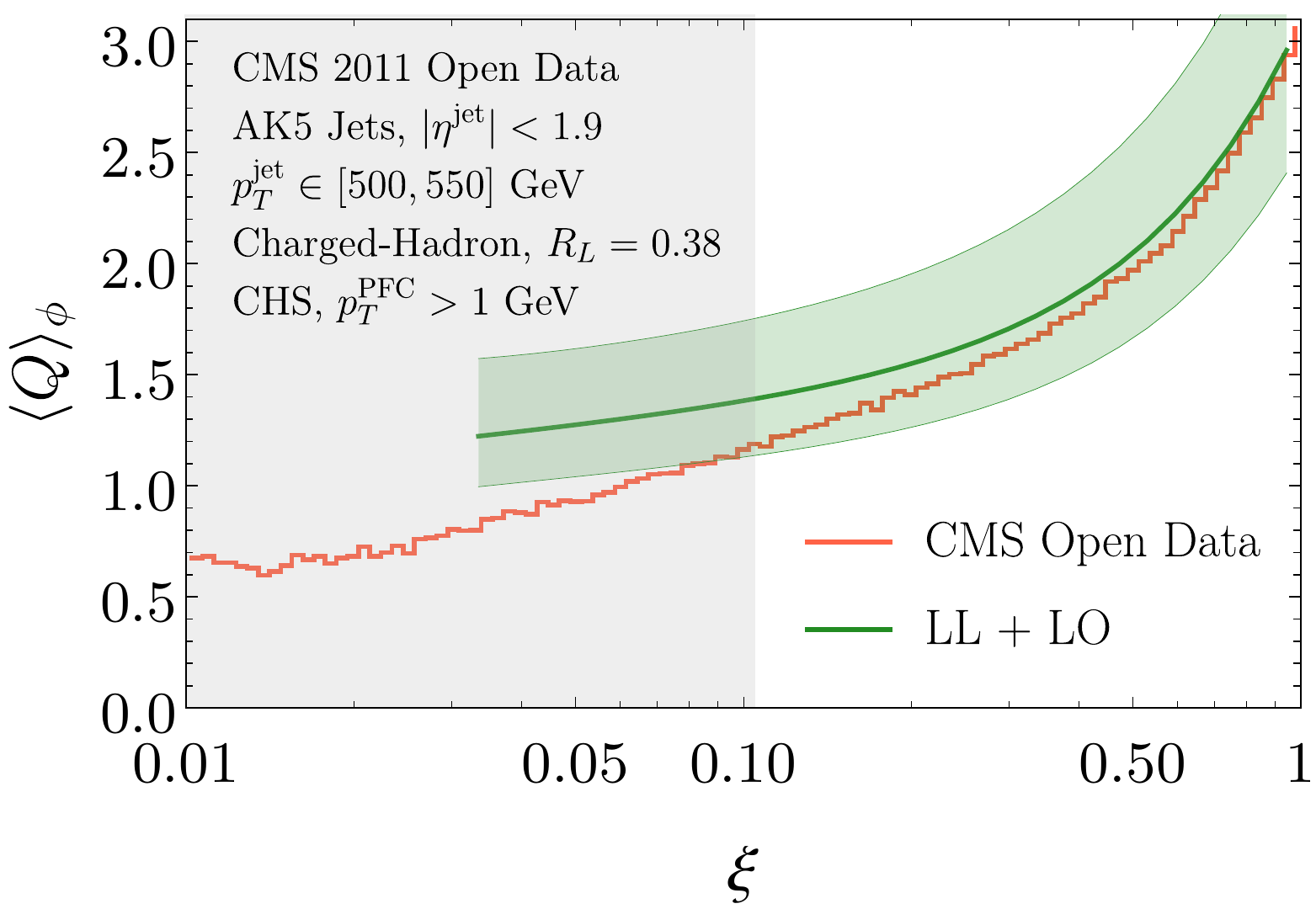}\label{fig:OD_average_a}
}
\subfloat[]{
\includegraphics[width=0.45\textwidth]{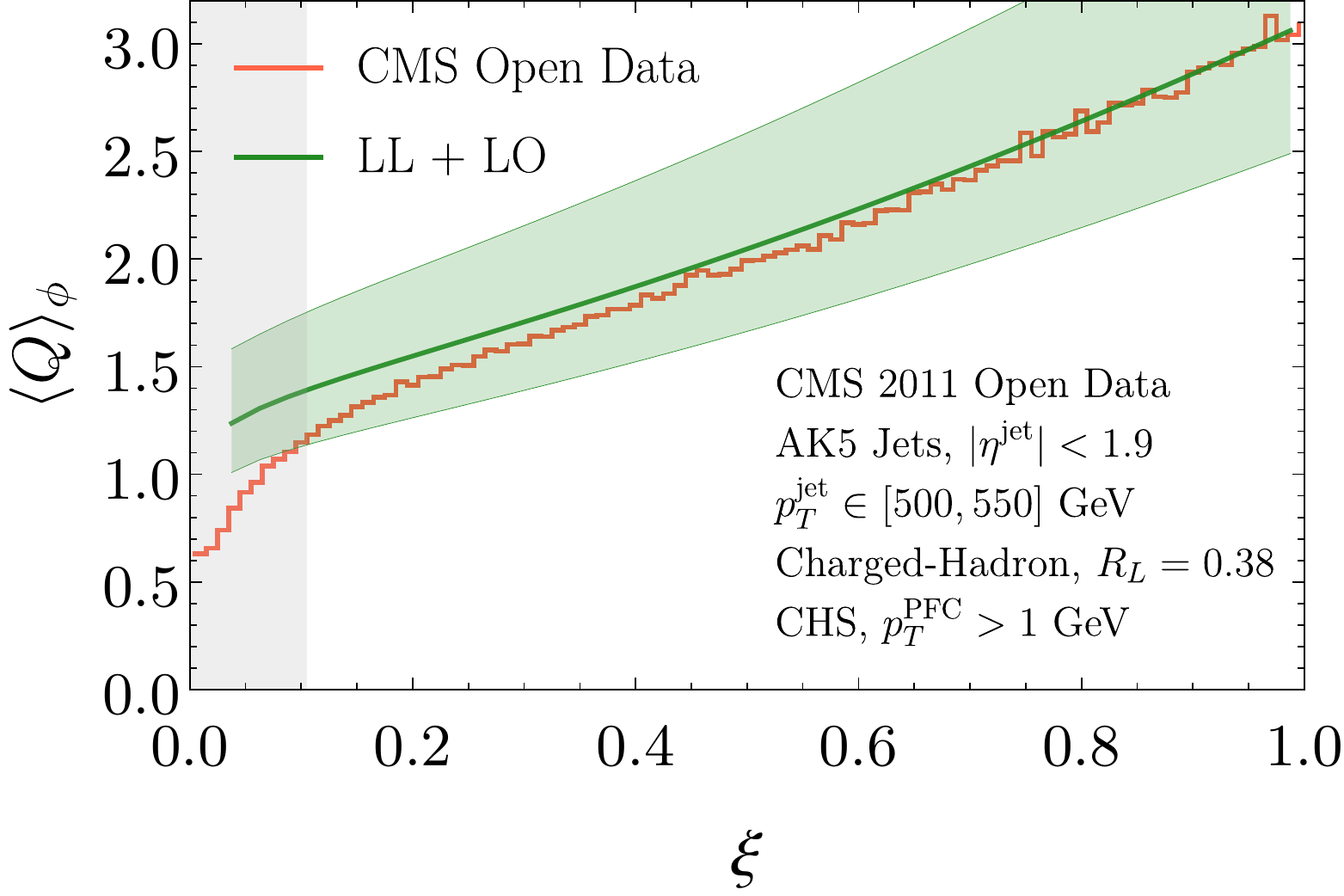}\label{fig:OD_average_b}
}\qquad
\end{center}
\caption{
The angular averaged non-gaussianity $\langle Q \rangle_\phi$, plotted using CMS Open Data, and compared with our LL+LO calculation with (a) logarithmic and (b) linear scales. Excellent agreement is seen between data and our calculation in the perturbative regime. As expected, the result is strongly peaked in the $\xi\to 1$ limit. }
\label{fig:OD_average}
\end{figure}

%%%%%%%%%%%%%%%%%%%%%%%%%%%%%%%%%%%%%%%%%%%%
\section{Celestial Non-Gaussianities in CMS Open Data}\label{sec:open_data}
%%%%%%%%%%%%%%%%%%%%%%%%%%%%%%%%%%%%%%%%%%%%

We now investigate the behavior of celestial non-gaussianities inside jets on actual LHC data.
This is possible due to the remarkable release~\cite{CERNOpenDataPortal} of research-grade data by the CMS collaboration~\cite{Chatrchyan:2008aa,CMS:OpenAccessPolicy}.
For examples of past analyses using these datasets, see e.g.~\cite{Larkoski:2017bvj,Tripathee:2017ybi,PaktinatMehdiabadi:2019ujl,Cesarotti:2019nax,Komiske:2019fks,Lester:2019bso,Apyan:2019ybx,Komiske:2019jim,Bhaduri:2019zkd,refId0,An:2021yqd,Elgammal:2021rne}.

In \Ref{Komiske:2022enw}, the three-point correlator was first studied using a jet dataset from the CMS 2011A Open Data~\cite{CMS:JetPrimary2011A}, which has been made public in the ``MIT Open Data'' (MOD) format by \Refs{Komiske:2019jim,komiske_patrick_2019_3340205}.
Here, we follow precisely this analysis, but extend it to the study of non-gaussianities.
In particular, we select $R = 0.5$ anti-$k_t$~\cite{Cacciari:2008gp} jets with $p_T \in [500,550]$ GeV and pseudo-rapidity $|\eta| < 1.9$.
Following \Refs{Komiske:2019jim}, we use charged hadron subtraction (CHS)~\cite{CMS:2014ata} and restrict to particle flow candidates (PFCs) with $p_T > 1$ GeV; this is done to mitigate pileup and minimize acceptance effects.

We compute the celestial non-gaussianity using only charged particles, capitalizing on the fantastic track reconstruction performance of CMS~\cite{CMS:2014pgm}.
The use of charged particles allows us to study three-point correlators with exceptional angular resolution, and it also minimizes the impact of detector distortions.
We do not perform any unfolding in this paper, so for this reason we cannot include systematic error bars.
(For ease of display, we omit statistical error bars, though they can be inferred from the size of bin-to-bin fluctuations.)
As with the scaling features of the energy correlators studied in \Ref{Komiske:2022enw}, we find that the non-gaussianity $Q_{\mathcal{E}}$ is robust, and the expected features are clearly reproduced in the CMS Open Data.
Nevertheless, it would be highly interesting to properly unfold the data to perform quantitative studies.

We start in \Fig{fig:OD_average} by ploting the azimuthally-averaged celestial non-gaussianity $\langle Q \rangle_\phi$, as extracted from CMS Open Data, along with our theoretical predictions.
At LL+LO, our perturbative predictions require knowing the quark/gluon fraction, which we have extracted from \textsc{Pythia} to be $43\%$ quarks and $57\%$ gluons.
As before, we plot the results with both logarithmic and linear scales to emphasize the squeezed ($\xi\to 0$) and perturbative ($\xi\to 1$) limits, respectively.
We see that our calculations describe the Open Data results remarkably well, emphasizing that we have identified a robust perturbative quantity within the three-point correlator.
Since we have only performed a LL calculation, the uncertainty band from scale variation is quite large, and it would be interesting to improve this.

\begin{figure}
\begin{center}
\subfloat[]{
\includegraphics[scale=0.45]{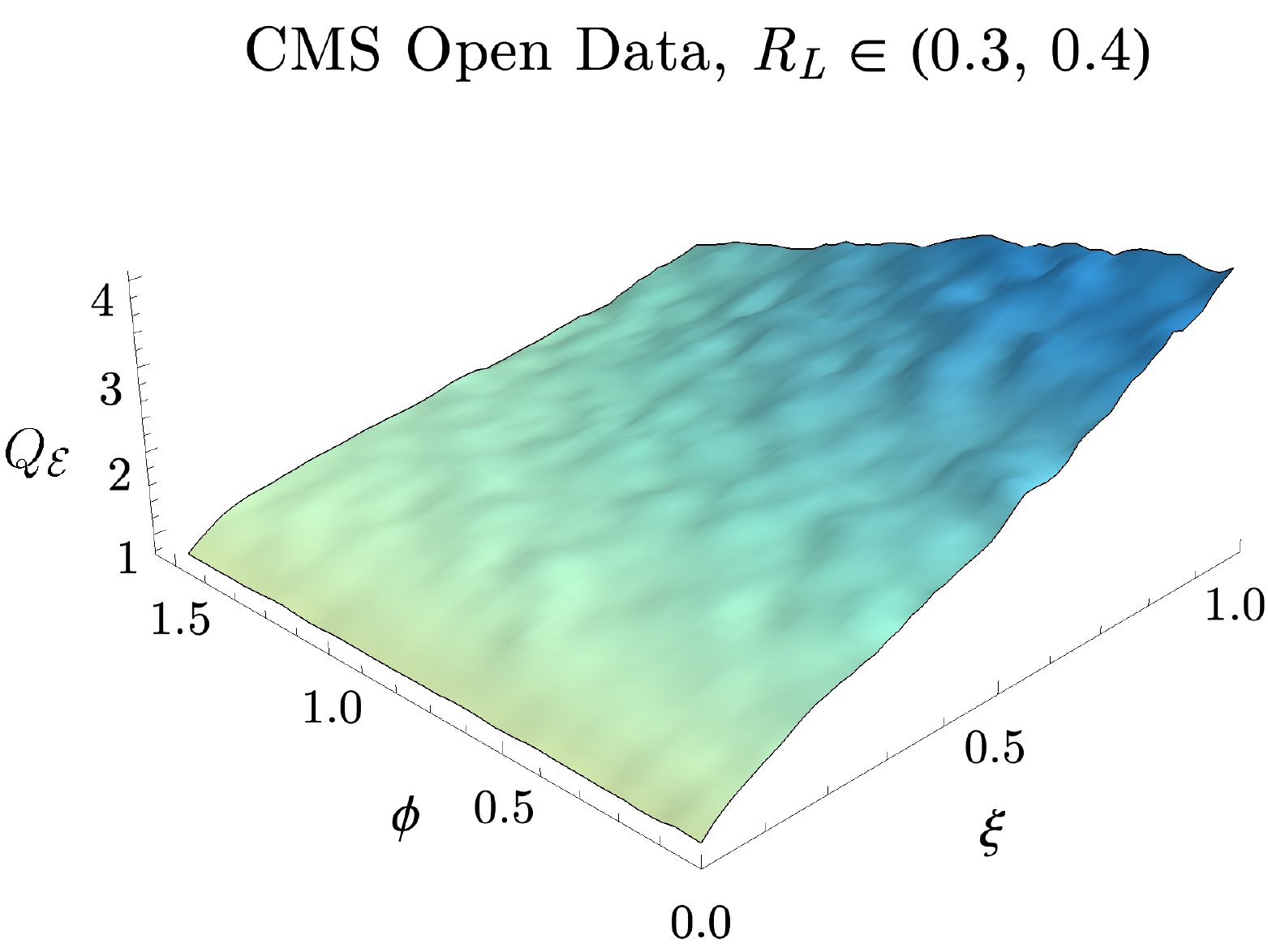}\label{fig:OD_shape_a}
}\qquad
\subfloat[]{
\includegraphics[scale=0.45]{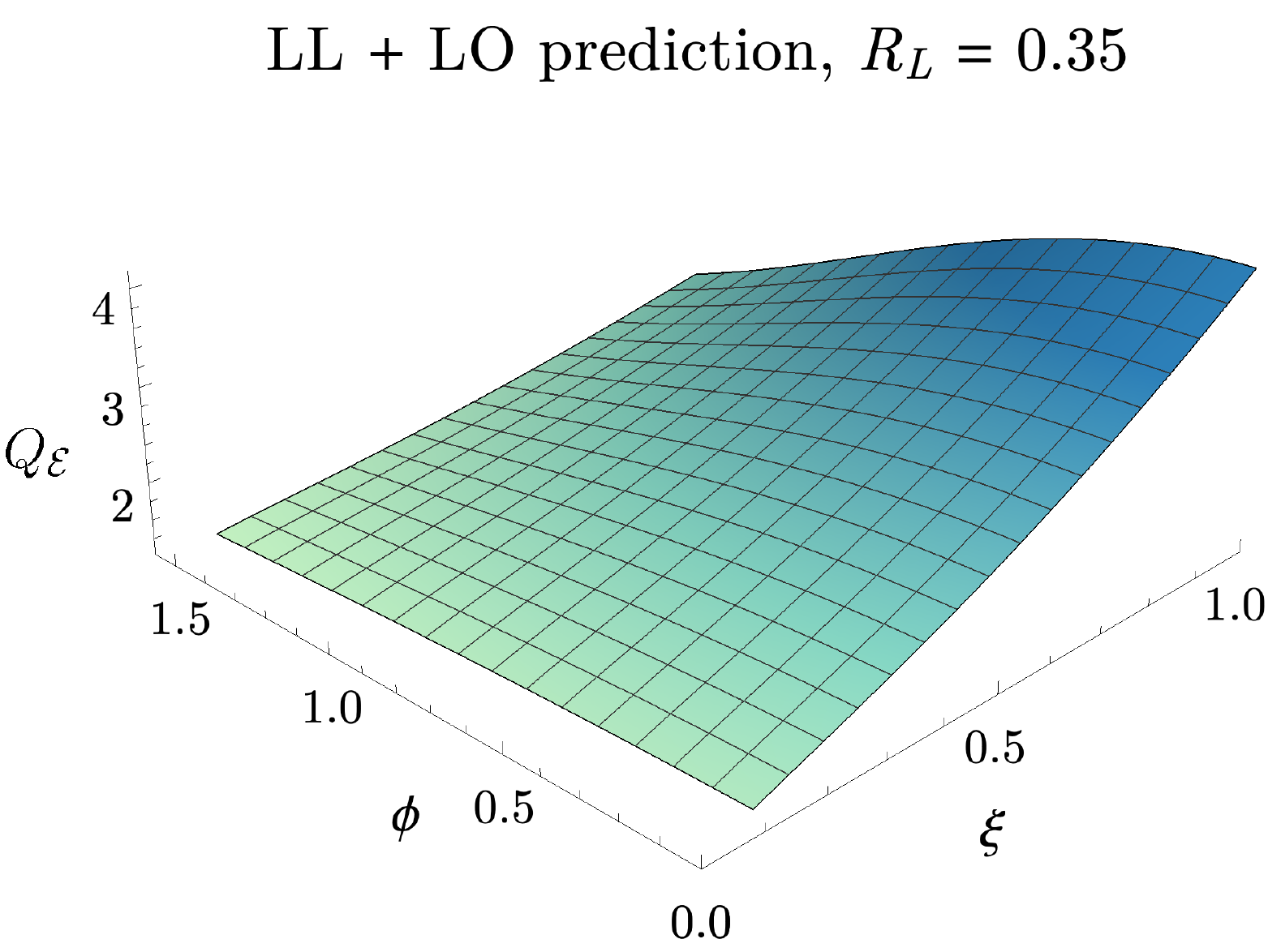}\label{fig:OD_shape_b}
}\qquad
\end{center}
\caption{The full shape dependence of the celestial non-gaussianity $Q_\mathcal{E}$  in (a) CMS Open Data and (b) our theoretical prediction at LL+LO. The non-gaussianity is strongly peaked in the  ``flattened triangle" region. To our knowledge, this is the first study of non-gaussianities in QCD energy flux.}
\label{fig:OD_shape}
\end{figure}

In \Fig{fig:OD_shape_a}, we show the shape dependence of the celestial non-gaussianity on CMS Open Data, and in \Fig{fig:OD_shape_b} we show our analytic prediction at LL+LO.
As expected from the analysis in \Sec{sec:shape_explain}, the data exhibits an enhanced non-gaussianity in the ``flattened triangle" region.
We find it quite remarkable that we can study the energy flux inside QCD jets at this level of detail, with control over the shape dependence of the three-point correlations.
To our knowledge, this is the first study of non-gaussianities of QCD energy flux, so we are finally able to match the beautiful studies of the shapes of non-gaussianities for cosmological correlators, but in the collider context!
We believe that these observables will be extremely useful for increasing the perturbative accuracy of parton shower Monte Carlo programs used at the LHC.

\begin{figure}[t]
\begin{center}
\includegraphics[scale=0.6]{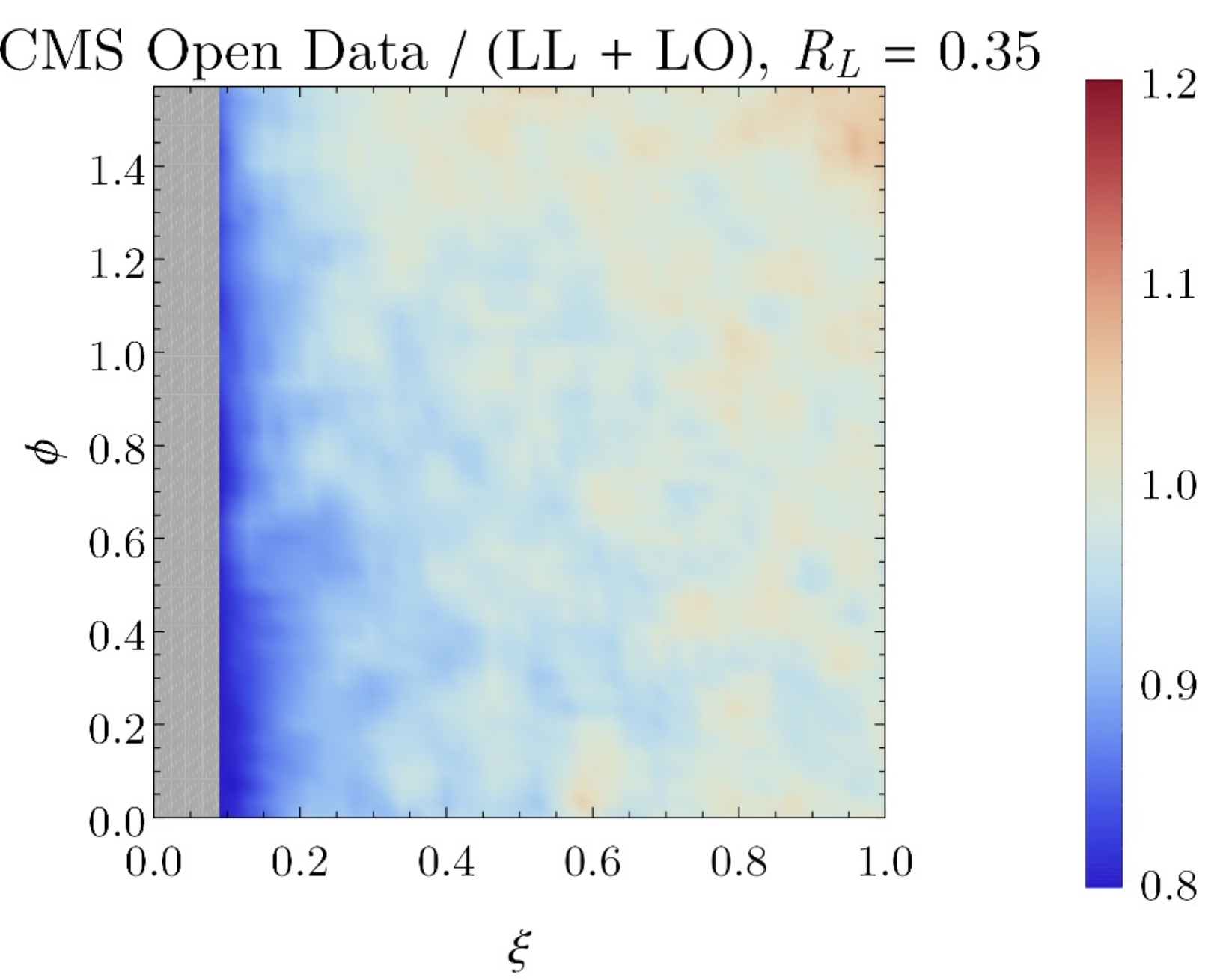}
\caption{
The ratio of the CMS Open Data to the LL+LO perturbative prediction. Perturbative control is lost in the squeezed limit ($\xi \to 0$), which we have illustrated in grey. Excellent agreement is observed in the perturbative regime, particularly for a LL+LO prediction.}
\label{fig:OD_shape_error}
\end{center}
\end{figure}

In \Fig{fig:OD_shape_error} we show the ratio between the CMS Open Data and our LO+LL calculation.
Overall, we find remarkably good agreement in the perturbative ($\xi\to 1$) limit, even though we only have a LO+LL calculation and have not incorporated any non-perturbative corrections.
This illustrates the robustness of this observable, as well as the fact that the celestial non-gaussianity isolates the perturbative component of the three-point correlator.
There are deviations in the squeezed ($\xi\to 0$) limit, which are of course expected, since non-perturbative corrections should dominate in the squeezed limit which probes lower scales.
All in all, we find it quite remarkable to see these theoretical predictions borne out in real LHC data.

%%%%%%%%%%%%%%%%%%%%%%
\section{Conclusions}\label{sec:conc}
%%%%%%%%%%%%%%%%%%%%%%

In the last few years, there has been remarkable progress in our theoretical understanding of jet substructure, allowing for the first time the study of multi-point correlations of energy flow within high-energy jets at the LHC.
This progress has been driven both by advances in perturbative calculations in quantum field theory, as well as by new techniques in the study of light-ray operators from the conformal bootstrap program.

In this paper, we took the next step by defining a notion of ``celestial non-gaussianity", allowing us to isolate the robust features of three-point correlations of energy flow in jets.
The celestial non-gaussianity is designed to capture features that go beyond iterated $1\to 2$ limits.
We studied the properties of this celestial non-gaussianity using both analytic results and parton shower simulations, and we found that the shape of the celestial non-gaussianity is highly peaked in the ``flattened triangle" configuration due to the initial-state propagator.

Using CMS Open Data, we were able to directly study the celestial non-gaussianity inside high-energy jets and compare with our analytic results.
To our knowledge, this is the first study of non-gaussianities in QCD energy flux, and it shows that we have theoretical control over multi-point correlation functions.
We find this extremely exciting, particularly since the remarkable LHC dataset allows these higher-point correlation functions to be directly measured with exceptional resolution.

Although we have emphasized the robustness of the celestial non-gaussianities, to perform true unfolded measurements of these observables will require several developments.
First, our analysis relied crucially on the use of charged-particle tracks for angular resolution.
One of the key features of energy correlators is that they can be easily calculated on tracks \cite{Chen:2020vvp,Li:2021zcf,Jaarsma:2022kdd} using the track function formalism \cite{Chang:2013rca,Chang:2013iba}.
This motivates further understanding of track functions theoretically, as well as their experimental measurement.
Second, the ability to calculate increasingly sophisticated features of the statistical properties of energy flow within jets also motivates the development of unfolding techniques to handle the complete phase space.
Recent examples of this using deep learning are \textsc{OmniFold} \cite{Andreassen:2019cjw} and conditional invertible neural networks~\cite{Bellagente:2020piv}, and it will be important to further develop these approaches on LHC data.

Achieving theoretical control over multi-point correlators has a number of applications for improving our understanding of jet substructure.
There has been significant recent work on improving the theoretical accuracy of parton showers~\cite{Li:2016yez,Hoche:2017hno,Hoche:2017iem,Dulat:2018vuy,Gellersen:2021eci,Hamilton:2020rcu,Dasgupta:2020fwr,Hamilton:2021dyz,Karlberg:2021kwr}, with the goal of better describing energy flow within jets at the LHC.
The ability to compute and measure higher-point correlators will provide crucial theoretical data for the development of higher-order parton showers.
These parton showers can then be used more broadly in increasingly sophisticated searches for new physics at the LHC.

Another application of the celestial non-gaussianities is to understanding the interactions of jets with the medium in heavy-ion collisions, where the use of jet substructure techniques has seen significant recent interest~\cite{Andrews:2018jcm,Cunqueiro:2021wls}.
One difficulty with current measurements is that it is often hard to disentangle multiple possible sources of modification due to the medium.
Much like the study of the shapes of non-gaussianity in inflation, the shapes of the celestial non-gaussianity contain significantly more information about energy flow within jets, and may therefore prove useful in unravelling different sources of medium modification.
Another interesting feature of the celestial non-gaussianities is that their shape is quite similar for both quark and gluon jets, even though their normalizations are different. 
While for many jet substructure observables, the observed modification in the medium can be explained by a modification of the quark/gluon fractions for the jets, a modification of the non-gaussianity would illustrate a genuine modification from interaction with the medium.
The triple collinear splitting functions in the medium are known \cite{Fickinger:2013xwa}, but to our knowledge have not yet been used in applications.
We believe that this is an interesting avenue for future study.

The results in this paper are a significant step in our understanding of the energy flow within jets at the LHC, demonstrating that we have robust control over three-point correlations.
Going forward, it will be important to continue to push towards higher-point correlators, with the next obvious step being four-point correlators.
Shapes of four-point correlators have been studied in the cosmological context~\cite{Arroja:2009pd,Chen:2009bc,Hindmarsh:2009es,Senatore:2010jy,Bartolo:2010di,Lewis:2011au}.
Our success in isolating the perturbative physics of three-point correlation functions of energy flow suggests that this may also be possible for four-point correlators, allowing them to robustly measured within high-energy jets at the LHC.
An initial step in this direction will be figuring out how to normalize the four-point correlators to isolate physics beyond the iterated $1 \to 2$ limit while also suppressing non-perturbative contributions.

An efficient understanding of multi-point correlators will also require the development of improved theoretical tools.
Motivated by the progress in the understanding of cosmological correlators, we are optimistic that insights from scattering amplitudes and the conformal bootstrap can have an impact, and many of the required theoretical ingredients are already available~\cite{DelDuca:2019ggv,DelDuca:2020vst}.
We believe that higher-point non-gaussianities will be interesting both for improving our understanding of quantum field theory and precision studies of QCD, but also for better exploiting jet substructure to hunt for ever more subtle features of potential new physics encoded in the radiation pattern within jets.

%%%%%%%%%%%%%%%%%%%%%%%%%%%%%%%%%%%%%%%%%%
\begin{acknowledgments}
%%%%%%%%%%%%%%%%%%%%%%%%%%%%%%%%%%%%%%%%%%
	
We thank Cyuan-Han Chang, David Simmons-Duffin, Patrick Komiske, Jingjing Pan for many useful discussions.
We thank the referee for the suggestion of a symmetric definition, which is now explored in \App{sec:new_ratio}.
H.C. is supported by the Natural Science Foundation of
China under contract No. 11975200.
I.M. is supported by start-up funds from Yale University.	
J.T. is supported by the U.S. Department of Energy (DOE) Office of High Energy Physics under contract DE-SC0012567.
H.X.Z. is supported by the Natural Science Foundation of
China under contract No.~11975200.

%%%%%%%%%%%%%%%%%%%%%%%%%%%%%%%%%%%%%%%%%%
\end{acknowledgments}
%%%%%%%%%%%%%%%%%%%%%%%%%%%%%%%%%%%%%%%%%%

\appendix

%%%%%%%%%%%%%%%%%%%%%%%%%%%%%%%%%%%%%%%%%%
\section{Observable Definitions}
\label{sec:def_app}
%%%%%%%%%%%%%%%%%%%%%%%%%%%%%%%%%%%%%%%%%%

In this appendix, we provide precise definitions of the energy correlator observables appearing in our definition of the celestial non-gaussianity.
Code for the individual energy correlators is available at \Ref{EEC_github}. 
A detailed description of our analysis algorithms for computing the celestial non-gaussianities is provided in \App{sec:algorithm}.

As in \Sec{sec:NG}, we consider an unpolarized source, so that the correlation functions depend only on the angles between the energy flow operators.
We order the angles as $R_S<R_M<R_L$. The definitions of the correlation functions appearing in the celestial non-gaussianity are 
\begin{align}
\langle \mathcal{E}^2(\vec{n}_1)\rangle_\Phi 
&= \frac{1}{4\pi}\sum_n \int d\sigma^{(\Phi)}_n \sum_{1\leq a\leq n} E_a^2\, ,\\
\langle \mathcal{E}(\vec{n}_1) \mathcal{E}(\vec{n}_2)\rangle_\Phi 
&= \frac{1}{8 \pi^2 R_S} \sum_n \int d\sigma^{(\Phi)}_n \sum_{1\leq a, b\leq n} E_a E_b\, \delta(R_S - R_{ab})\,,\\
\langle \mathcal{E}^2(\vec{n}_1) \mathcal{E}(\vec{n}_3)\rangle_\Phi 
&= \frac{1}{8 \pi^2 R_L} \sum_n \int d\sigma^{(\Phi)}_n \sum_{1\leq a, b\leq n} E_a^2 E_b\, \delta(R_L - R_{ab})\,,\\
\langle \mathcal{E}(\vec{n}_1) \mathcal{E}(\vec{n}_2) \mathcal{E}(\vec{n}_3)\rangle_\Phi
&=\frac{A}{4\pi^2}  \sum_n \! \int \! d\sigma^{(\Phi)}_n \!\!\!\!\!\!\!
\sum_{1\leq a, b, c\leq n} \!\!\!\!\!\!
 E_a E_b E_c \, \delta(R_S \! - \! R_{ab}) \, \delta(R_M \!- \! R_{bc}) \, \delta(R_L \!- \! R_{ac}) .
\end{align}
We will always consider energy flow within jets within a narrow central rapidity bin, and therefore we have $\langle \mathcal{E}^2(\vec{n}_1)\rangle_\Phi 
=\langle \mathcal{E}^2\rangle_\Phi $.
In these expressions, the $n$-particle differential cross section $d\sigma_n^{(\Phi)}$ generated by the initial state $| \Phi \rangle$ is:
\begin{equation}
d\sigma_n^{(\Phi)}=\prod_{k=1}^{n} \frac{d^3\vec{p}_k}{(2\pi)^3 2E_k} \left| \langle p_1,p_2,\cdots, p_n | \Phi\rangle \right|^2\,.
\end{equation}
The prefactor $A$ comes from the Jacobian of changing $\vec{n}_i$ to $R_{ij}$, which in the collinear limit is
\begin{equation}
A = \frac{1}{4 R_S R_M R_L} \sqrt{2 R_S^2 R_M^2 + 2R_M^2 R_L^2 + 2 R_S^2 R_L^2 
-R_S^4 -R_M^4 -R_L^4}\, .
\end{equation}
%

%%%%%%%%%%%%%%%%%%%%%%%%%%%%%%%%%%%%%%%%%%
\section{Analysis Algorithms}
\label{sec:algorithm}
%%%%%%%%%%%%%%%%%%%%%%%%%%%%%%%%%%%%%%%%%%

We now describe how we compute the celestial non-gaussianity on collections of hadrons.
We consider only the $\mathbb{Z}_2$-symmetric case though the extension to the asymmetric case is relatively easy by including sign information for $\phi$. 
We first collect the complete set of coordinates used for convenience:
\begin{itemize}
\item $R_{ij}$: the angle between detectors/particles $i$ and $j$.
For $e^+ e^-$ colliders, $R_{ij}=\arccos(\vec{n}_i\cdot \vec{n}_j)$, while for hadron colliders,  $R_{ij}=\sqrt{\Delta y_{ij}^2 +\Delta \phi_{ij}^2}$.
\item $(R_S, R_M, R_L)$: angles in ascending order.
\item Cross-ratios $(u,\,v)$ and $(z,\,\bar{z})$: $u=(R_S/R_L)^2=z\bar{z},\; v=(R_M/R_L)^2=(1-z)(1-\bar{z})$.
The range of $\arg z$ depends on $|z|$:  
\beq
\arg z \in \left[\arctan\sqrt{4|z|^2-1} \, \Theta(1/2<|z|\leq 1), \, \arctan \frac{\sqrt{1-|z|^2}}{|z|} \right]\,,
\eeq
where $\Theta(1/2<|z|\leq 1)$ is the product of Heaviside functions $\Theta(|z|-1/2) \times \Theta(1-|z|)$.
\item Deformed cross-ratios from \Eq{eq:transf}: $\xi = R_S/R_M = \frac{|z|}{|1-z|}$, $\sin\phi = \sqrt{1-\frac{(R_L-R_M)^2}{R_S^2}}$.
\end{itemize}

To plot the ratio appearing in the definition of the celestial non-gaussianity, we need to lift the denominator in \Eq{eq:NG_def} from a two-variable function to a three-variable function by regarding it as a constant function with respect to $R_M$.
For simplicity, we do not consider contact terms in the description, which amounts to including an additional symmetry factor in the weighting.
The algorithm we use for plotting the ratio in $(\xi, \phi, R_L)$ space is the following:  
\begin{enumerate}
\def\labelenumi{\arabic{enumi}.}
\item
  Choose bins for each variable by specifying ticks:

  \begin{itemize}
  \item
    \((\xi_0,\, \xi_1,\, \dots ,\, \xi_{N_1})\)
  \item
    \((\phi_0,\, \phi_1,\, \dots ,\, \phi_{N_2})\)
  \item
    \((R^L_0,\, R^L_1,\, \dots ,\, R^L_{N_3})\), where we denote the center value of \(i\)-th bin of \(R_L\) by
    \(\widetilde{R}^L_i = (R^L_{i-1} + R^L_i)/2\).
  \end{itemize}
\item
  Prepare one variable, one 1D histogram, and two 3D histograms to fill
  in the data:
  \begin{itemize}
  \item
    a real number \(H_{\mathrm{E}^2}\) \(\leftrightarrow\)
    \(\langle\mathcal{E}^2(\vec{n}_1)\rangle\)
  \item
    a 1D histogram \(H_{\mathrm{E^2E}}\) with \(N_3\) bins
    \(\leftrightarrow\)
    \(\langle\mathcal{E}^2(\vec{n}_1)\,\mathcal{E}(\vec{n}_3)\rangle\)
  \item
    first 3D histogram \(H_{\mathrm{EEE}}\) with
    \(N_1\times N_2\times N_3\) bins \(\leftrightarrow\)
    \(\langle\mathcal{E}(\vec{n}_1)\, \mathcal{E}(\vec{n}_2)\,\mathcal{E}(\vec{n}_3)\rangle\)
  \item
    second 3D histogram \(H_{\mathrm{EE}}\) with
    \(N_1\times N_2\times N_3\) bins \(\leftrightarrow\)
    \(\langle\mathcal{E}(\vec{n}_1)\,\mathcal{E}(\vec{n}_2)\rangle\)
  \end{itemize}
\item
  For each given event, fill in the data in the following way:

  \begin{itemize}
  \item  (Optional) Normalize the energy.
  \item
    {Perform a one-particle loop for \(\langle\mathcal{E}^2(\vec{n}_1)\rangle\)}:
    \begin{enumerate}
    	\item For the \(a\)-th particle, add \(E_a^2\) to \(H_{\mathrm{E^2}}\).
	 \end{enumerate}
  \item Perform a two-particle loop for \(\langle\mathcal{E}^2(\vec{n}_1)\mathcal{E}(\vec{n}_3)\rangle\):
	\begin{enumerate}
   	 \item Select each pair of (un-ordered) particles \((a,b)\);
    \item  Calculate their angle \(R_{ab}\);
   \item Add \(E_a^2 E_b + E_a E_b^2\) to \(i\)-th bin of    \(H_{\mathrm{E^2E}}\) if \(R^L_{i-1}<R_{ab}<R^L_{i}\) .
    \end{enumerate}
  \item Perform a three-particle loop for    \(\langle\mathcal{E}(\vec{n}_1)\mathcal{E}(\vec{n}_2)\mathcal{E}(\vec{n}_3)\rangle\):
    	\begin{enumerate}

    \item Select each unordered three-particle combination
    \((a,\,b,\,c)\);

   \item Calculate angles \((R_{ab},\,R_{bc},\,R_{ac})\) and
    sort them in an ascending way \((R_S,\,R_M,\,R_L)\);

    \item Calculate \((\xi,\,\phi,\, R_L)\) from the
    definitions
    \(\xi = R_S/R_M,\;\phi = \arcsin \sqrt{1-\frac{(R_L-R_M)^2}{R_S^2}}\);

   \item Add \(6 E_a E_b E_c\) to \((i,\,j,\,k)\)-th bin of
    \(H_{\mathrm{EEE}}\) if
    \(\xi_{i-1} < \xi < \xi_{i},\;\phi_{j-1} < \phi < \phi_{j},\; R^L_{k-1}<R_L < R^L_{k}\).
    
        \end{enumerate}

  \item Perform a two-particle loop for \(\langle\mathcal{E}(\vec{n}_1)\mathcal{E}(\vec{n}_2)\rangle\): \\
      {\emph{N.b.}: we need to fill multiple bins for a single data point in this part.}

	\begin{enumerate}

    \item Select each pair of (un-ordered) particles \((a,b)\);

    \item Calculate their angle \(R_{ab}\) and energy weight
    \(w_{ab} = 2 E_a E_b\);

    \item Loop all central values
    \(\widetilde{R}^L_{k=1,2,\dots,N_3}\). If
    \(R_{ab} > \widetilde{R}^L_k\), do nothing; else, continue following these steps for each value of $k$.
  
     \item Calculate all values of \(\{\arg z_*^{(\alpha)}\}\)
    at intersections of constant \(|z|=R_{ab}/R_L\) curve with constant
    \(\xi=(\xi_0,\, \xi_1,\, \dots ,\, \xi_{N_1})\) lines and constant
    \(\phi=(\phi_0,\, \phi_1,\, \dots ,\, \phi_{N_2})\) lines. \emph{N.b.}: the range of \(\arg z\) depends on \(|z|\), so
    intersections do not happen for every \(\xi_i,\,\phi_j\);

    \item For each
    \(\xi_i\in \left[\,|z|,\,\min\{1,\,\frac{|z|}{1-|z|}\}\,\right]\),
    calculate
    \[\arg z^{(i,1)}_* = \arccos\left((|z|^{-1} +|z| -|z|\xi_i^{-2})/2\right);\]

    \item For each
    \(\phi_j\in \left[\varphi, \pi/2\right]\), calculate
    \(\arg z^{(j,2)}_* = \arccos(\cos \phi_j + \frac{1}{2}|z| \sin^2 \phi_{j})\),
    where we define \(\varphi=0\) when \(0\leq|z|\leq 1/2\) and
    \(\varphi=\arcsin \frac{\sqrt{2|z|-1}}{|z|}\) when
    \(1/2 <|z|\leq 1\);

    \item Put the above solutions together and sort them in
    a non-descending way \(\{\arg z_*^{(\alpha)}\}\).
    \emph{N.b.}: if we want to give \(\xi\text{-}R_L\) plots by
    integrating out \(\phi\), we need to add end points
    \(\Theta(1/2<|z|<1) \arctan\sqrt{4|z|^2-1}\) and
    \(\arctan \frac{\sqrt{1-|z|^2}}{|z|} \), which are the boundaries of
    \(\arg z\), to the list \(\{\arg z_*^{(\alpha)}\}\) before sorting.

    \item Calculate the interval lengths and central values
    from all the neighboring pairs in \(\{\arg z_*^{(\alpha)}\}\):
    \(\{\Delta^{(\alpha)}=\arg z_*^{(\alpha+1)}-\arg z_*^{(\alpha)}\}\)
    and
    \(\{\arg \widetilde{z}_*^{(\alpha)}=(\arg z_*^{(\alpha)}+\arg z_*^{(\alpha+1)})/2 \}\).

    \item For each element in the center-length pair set
    \(\{(\arg \widetilde{z}_*^{(\alpha)},\Delta^{(\alpha)})\}\),
    calculate the corresponding point in \((\xi,\phi,R_L)\) space from
    \((\arg \widetilde{z}_*^{(\alpha)}, |z|, \widetilde{R}^L_k)\) using
    \(\xi^{(\alpha)} = |z|/\sqrt {v^{(\alpha)}}\), \(\phi^{(\alpha)} = \arcsin \sqrt{1-(1-\sqrt{v^{(\alpha)}})^2/|z|^2}\),
    where
    \(v^{(\alpha)}=1+|z|^2-2|z|\cos (\arg \widetilde{z}^{(\alpha)}_*)\).

    \item For each point
    \((\xi^{(\alpha)},\phi^{(\alpha)},\widetilde{R}^L_k)\) above, we
    fill the corresponding bin with weight
    $w_{ab}=2E_aE_b$ to
    \(H_{\mathrm{EE}}\).
  \end{enumerate}
  
  	\end{itemize}

\item Finally, construct the ratio plot \(\widetilde{H}\) from
  \(H_{\mathrm{E^2}},\, H_{\mathrm{E^2E}}\, H_{\mathrm{EEE}},\, H_{\mathrm{EE}}\)
  obtained above:
  
  \[\widetilde{H}(i,j,k) = \frac{\pi}{6}\frac{H_{\mathrm{E^2}} \times H_{\mathrm{EEE}}(i,j,k)}{H_{\mathrm{E^2E}}(k) \times H_{\mathrm{EE}}(i,j,k)}\]

  where \(H_{\mathrm{E^2E}}(k)\) is value of the \(k\)-th bin in the
  \(H_{\mathrm{E^2E}}\), and \(H_{\mathrm{EEE}}(i,j,k)\) and
  \(H_{\mathrm{EE}}(i,j,k)\) are the values of \((i,j,k)\)-th bin in
  \(H_{\mathrm{EEE}}\) and \(H_{\mathrm{EE}}\), respectively.
\end{enumerate}

%%%%%%%%%%%%%%%%%%%%%%%%%%%%%%%%%%%%%%%%%%
\section{Resummed Predictions for Celestial Non-Gaussianities \label{sec:resum_formula}}
%%%%%%%%%%%%%%%%%%%%%%%%%%%%%%%%%%%%%%%%%%

In this appendix, we present the expressions used for the analytic predictions of the resummed azimuthally averaged non-gaussianities.

In the collinear limit, using leading-power results for $\langle \mathcal{E}(\vec{n}_1) \, \mathcal{E}(\vec{n}_2) \rangle_\Phi $ and $\langle  \mathcal{E}^2(\vec{n}_1) \, \mathcal{E}(\vec{n}_3)\rangle_\Phi$ is a good approximation.
The NNLL resummation for collinear $\langle \mathcal{E}(\vec{n}_1) \, \mathcal{E}(\vec{n}_2) \rangle_\Phi $ is available from \Ref{Dixon:2019uzg}, but we only use LL approximation in this work:
\begin{equation}
\langle \mathcal{E}(\vec{n}_1) \mathcal{E}(\vec{n}_2) \rangle_\Phi^{\mathrm{LP,\, LL}} = -\frac{1}{\pi R_S^2}\mathcal{J}\cdot C_{R_S}^{(1,1)}\cdot
\left[\frac{\alpha_S(Q)}{\alpha_S(R_S Q)}\right]^{\frac{\gamma^{(0)}(3)}{\beta_0}}
\cdot \mathcal{S}_\Phi \,.
\end{equation}
Note that $\langle  \mathcal{E}^2(\vec{n}_1) \, \mathcal{E}(\vec{n}_3)\rangle_\Phi$ and $\langle \mathcal{E}^2 \rangle$ are not collinear-safe quantities.
We therefore need to introduce an IR regulator scale $\Lambda_{\mathrm{IR}}$ and a non-perturbative function $\mathcal{T}$ (related to the track function for energy correlators \cite{Chang:2013rca,Chang:2013iba,Elder:2017bkd,Chen:2020vvp,Li:2021zcf,Jaarsma:2022kdd}) for the double energy weighting.
Then, we can write down corresponding LL approximation for these quantities in the similar form as $\langle \mathcal{E}(\vec{n}_1) \mathcal{E}(\vec{n}_2) \rangle_\Phi $:
\begin{eqnarray}
\langle \mathcal{E}^2\rangle_\Phi^{\mathrm{LL}} 
&=&\mathcal{T}\cdot
\left[\frac{\alpha_S(Q)}{\alpha_S(\Lambda_{\mathrm{IR}})}\right]^{\frac{\gamma^{(0)}(3)}{\beta_0}}
\cdot \mathcal{S}_\Phi\,, \\
\langle\mathcal{E}^2(\vec{n}_1)\mathcal{E}(\vec{n}_3)\rangle_\Phi^{\mathrm{LP,\, LL}}
&=& -\frac{1}{\pi R_L^2}\mathcal{T}\cdot
\left[\frac{\alpha_S(R_L Q)}{\alpha_S(\Lambda_{\mathrm{IR}})}\right]^{\frac{\gamma^{(0)}(3)}{\beta_0}}
\cdot
C_{R_L}^{(1,2)}
\cdot
\left[\frac{\alpha_S(Q)}{\alpha_S(R_L Q)}\right]^{\frac{\gamma^{(0)}(4)}{\beta_0}}
\cdot \mathcal{S}_\Phi\,.
\end{eqnarray}
It has been observed in the study of track functions that the values of the low moments of quark and gluon track functions are quite similar \cite{Jaarsma:2022kdd}, so we will make the assumption $\mathcal{T} \propto \mathcal{J}$ in the calculations of this paper.
Because $\mathcal{T}$ is a vector and the OPE coefficients and anomalous dimensions are matrices, this non-perturbative function does not cancel when taking ratios.

 The LO results for collinear EEEC are given as the functions $G_q(z),\, G_g(z)$ in \Ref{Chen:2019bpb}, where the $z$ and its complex conjugate $\bar{z}$ is defined by the relation
\begin{equation}
z\bar{z} = (R_S/R_L)^2,\quad (1-z)(1-\bar{z}) = (R_M/R_L)^2\,.
\end{equation}
In the following, the source $\Phi$ denotes the mixing of collinear quark and gluon source with quark percentage $x\in [0,1]$, for example, $G_\Phi(z) \equiv x\, G_q(z) + (1-x)\, G_g(z)$. The relation between the LO fully differential distribution (normalized to jet energy) $\langle \mathcal{E}(\vec{n}_1)\mathcal{E}(\vec{n}_2)\mathcal{E}(\vec{n}_3)\rangle_\Phi^{\mathrm{LO}}$ and $G_\Phi(z)$ is 
\begin{equation}
\langle \mathcal{E}(\vec{n}_1)\mathcal{E}(\vec{n}_2)\mathcal{E}(\vec{n}_3)\rangle_\Phi^{\mathrm{LO}} = \frac{64}{\pi^2 R_L^4} G_\Phi(z)\,.
\end{equation}

In the squeezed limit, $\xi\ll 1$, we need to resum large logarithms $\ln (R_S/R_L)$.
Here it is interesting to keep subleading powers (higher twist) since we are also interested in the region of finite $\xi$.
Therefore, we divide the collinear LO EEEC $\langle \mathcal{E}(\vec{n}_1) \, \mathcal{E}(\vec{n}_2) \, \mathcal{E}(\vec{n}_3) \rangle_{\Phi}^{\mathrm{LO}}$ result~\cite{Chen:2019bpb} into leading power and sub-leading powers (which refers to the squeezed limit):
\begin{equation}
\langle \mathcal{E}(\vec{n}_1)\, \mathcal{E}(\vec{n}_2)\, \mathcal{E}(\vec{n}_3) \rangle_{\Phi}^{\mathrm{LO}} 
= \langle \mathcal{E}(\vec{n}_1)\, \mathcal{E}(\vec{n}_2)\, \mathcal{E}(\vec{n}_3) \rangle_{\Phi}^{\mathrm{LP,\,LO}} 
+ \langle \mathcal{E}(\vec{n}_1)\, \mathcal{E}(\vec{n}_2)\, \mathcal{E}(\vec{n}_3) \rangle_{\Phi}^{\mathrm{sub\text{-}LPs,\, LO}}.
\end{equation}
Our LL approximation for EEEC is to resum both $\ln R_S/R_L$ and  $\ln R_L$ for the LP contribution~\cite{Chen:2020adz,Chen:2021gdk} and to resum only $\ln R_L$ for sub-leading powers contributions:

In the squeezed limit, the leading power contribution of $\langle \mathcal{E}(\vec{n}_1)\mathcal{E}(\vec{n}_2)\mathcal{E}(\vec{n}_3)\rangle_\Phi^{\mathrm{LO}}$ can be obtained from iterative light-ray OPE $\mathcal{E}\mathcal{E} \mathcal{E} \to \vec{\mathbb{O}}^{[3]}\mathcal{E} \to \vec{\mathbb{O}}^{[4]}$, where we restrict ourselves in the twist-2 light-ray operators $\vec{\mathbb{O}}^{[J]}$ defined as the integral of local twist-2 operators $\vec{\mathcal{O}}^{[J]}$:
\begin{align}
\vec{\mathbb{O}}^{[J]}=\left( \mathbb{O}_q^{[J]}, \mathbb{O}_g^{[J]}, \mathbb{O}_{\tilde g,+}^{[J]}, \mathbb{O}_{\tilde g,-}^{[J]}  \right)^T &=\lim_{r\to \infty} r^2 \int\limits_0^{\infty} dt\, \vec{\mathcal{O}}^{[J]}(t,r\vec{n})\,,\\
\mathcal{O}_q^{[J]}&=\frac{1}{2^J}\bar \psi \gamma^+ ( iD^+)^{J-1}\psi\,,  \\
\mathcal{O}_g^{[J]}&=-\frac{1}{2^J} F_a^{\mu +}(i D^+)^{J-2}F_a^{\mu +}\,,  \\
\mathcal{O}_{\tilde g,\lambda}^{[J]}&=-\frac{1}{2^J} F_a^{\mu +}(i D^+)^{J-2}F_a^{\nu +}\epsilon_{\lambda, \mu}\epsilon_{\lambda, \nu}\,.
\end{align}
The plus component is defined as the scalar product with the null vector $\bar{n}^\mu=(1, -\vec{n})$. The transverse spin-$2$ operators have been projected onto polarization vector basis $\epsilon^\mu$ labelled by helicities $\lambda$.

The relevant light-ray OPEs in this paper are
\begin{align}\label{eq:EE_OPE}
\mathcal{E}(\vec n_1) \mathcal{E}(\vec n_2) = -\frac{1}{2\pi} \frac{2}{R_S^2}   {\cal J} \cdot C_{R_S}^{(1,1)} \cdot\vec {\mathbb{O}}^{[3]} (\vec n_1)  + \text{higher twist}\,, \\
\vec {\mathbb{O}}^{[3]} (\vec n_1) \mathcal{E}(\vec n_3)  =-\frac{1}{2\pi}\frac{2}{R_L^2} \;\, C_{R_L}^{(1,2)} \cdot\vec {\mathbb{O}}^{[4]} (\vec n_1) +\text{higher twist}\,. \label{eq:OE_OPE}
\end{align}
$\mathcal{J} = (1, 1, 0, 0)$ is the projection vector that does not distinguish the measurement of quark or gluon. On the free parton level, $\mathcal{E}$ and $\mathcal{E}$ can relate to $\vec{\mathbb{O}}^{[J]}$ using $\mathcal{J}$:
\begin{equation}
\mathcal{E} = \mathcal{J}\cdot \vec{\mathbb{O}}^{[2]},\quad \mathcal{E}^2 =\mathcal{J}\cdot \vec{\mathbb{O}}^{[3]}\, . 
\end{equation} 
The OPE coefficient matrices takes the form
\begin{equation}
C_{R_S}^{(1,1)} = \widehat C_{\phi_S}(2)- \widehat C_{\phi_S}(3)  \,,\quad
C_{R_L}^{(1,2)} =  \widehat C_{\phi_L}(3)- \widehat C_{\phi_L}(4)  \,,
\end{equation}
where the matrix $\widehat{C}_{\phi}(J)$ is 
\begin{align}\label{eq:structure_constants}
\widehat C_\phi(J)&=
\begin{pmatrix}
\gamma_{qq}(J)&&2n_f \gamma_{qg}(J)&&2n_f \gamma_{q\tilde g}(J) e^{-2i\phi}/2 &&2n_f \gamma_{q\tilde g}(J) e^{2i\phi}/2 \\
\gamma_{gq}(J)&& \gamma_{gg}(J)&&\gamma_{g\tilde g}(J) e^{-2i\phi}/2 &&\gamma_{g\tilde g}(J) e^{2i\phi}/2\\
\gamma_{\tilde gq}(J) e^{2i\phi} &&\gamma_{\tilde gg}(J) e^{2i\phi} && \gamma_{\tilde g \tilde g}(J)&&\gamma_{\tilde g \tilde g,\pm}(J) e^{4i\phi}\\
\gamma_{\tilde gq}(J) e^{-2i\phi} &&\gamma_{\tilde gg}(J) e^{-2i\phi} &&\gamma_{\tilde g \tilde g,\pm}(J)e^{-4i\phi}&& \gamma_{\tilde g \tilde g}(J)\\
\end{pmatrix}\,.
\end{align}
The convention of the perturbative expansion of $\gamma_{ab}(J)$ in this paper is $\displaystyle \frac{\alpha_s}{4\pi} \gamma_{ab}^{(0)}(J)+\mathcal{O}\left(\alpha_s^2\right)$ and values of $\gamma_{ab}^{(0)}(J)$ are
\beq
\begin{split} \label{eq: gamma_values}
&\gamma_{qq}^{(0)}(J)=C_F\left( 4\left(\psi^{(0)}(J+1)+\gamma_E\right)-\frac{2}{J(J+1)}-3\right),\quad 
\gamma_{qg}^{(0)}(J)=-T_F \frac{2(J^2+J+2)}{J(J+1)(J+2)},\\
&\gamma_{gq}^{(0)}(J)=-C_F \frac{2(J^2+J+2)}{(J-1)J(J+1)}, \\
&\gamma_{gg}^{(0)}(J)= 4 C_A \left( \psi^{(0)}(J+1)+\gamma_E -\frac{1}{(J-1)J}-\frac{1}{(J+1)(J+2)} \right)-\beta_0,\\
&\gamma^{(0)}_{\tilde g \tilde g} (J)=4C_A (\psi ^{(0)}(J+1)+\gamma_E) -\beta_0\,, \quad
\gamma^{(0)}_{\tilde gq}(J)=C_F\frac{2}{(J-1)J}\,,\qquad  \gamma^{(0)}_{\tilde gg}(J)=C_A\frac{2}{(J-1)J}\,, \\
&\gamma^{(0)}_{q\tilde g}(J)= - T_F\frac{8}{(J+1)(J+2)}\,, \qquad \gamma^{(0)}_{\tilde g \tilde g,\pm} (J)=0\,,  \quad
\gamma^{(0)}_{g\tilde g}(J)= C_A \left(\frac{8}{(J+1)(J+2)}+3\right) -\beta_0\,,
\end{split}
\eeq
where $\psi^{(0)}(z) = \Gamma'(z)/\Gamma(z)$ is the digamma function, and $\beta_0 = 11/3 C_A - 4/3 n_f T_F$ is the one-loop beta function in QCD. 

Combining \Eqs{eq:EE_OPE}{eq:OE_OPE}, we obtain the LP contribution:
\beq
\langle\mathcal{E}(\vec{n}_1)\mathcal{E}(\vec{n}_2)\mathcal{E}(\vec{n}_3)\rangle_\Phi^{\mathrm{LP,\, LO}}
=\frac{1}{(2\pi)^2} \frac{2}{R_S^2}\frac{2}{R_L^2} {\mathcal{J}}\cdot {C}_{R_S}^{(1,1)}
\cdot {C}_{R_L}^{(1,2)}
\cdot \mathcal{S}_\Phi\,.
\eeq
Here, $\mathcal{S}_\Phi$ stands for the one point function $\langle \vec{\mathbb{O}}^{[4]} \rangle_\Phi$, representing the hard function for the source $\Phi$.
The explicit expressions are:
\begin{align}
&\langle\mathcal{E}(\vec{n}_1)\mathcal{E}(\vec{n}_2)\mathcal{E}(\vec{n}_3)\rangle_q^{\mathrm{LP,\, LO}} \\
&= \frac{1}{\pi^2 R_S^2 R_L^2}\left[\frac{1}{75} C_F(91 C_A+240 C_F+13 n_f T_F) + \frac{1}{45} C_F  (C_A-2 n_f T_F)\left(\frac{z}{\bar{z}}+\frac{\bar{z}}{z}\right) \right]\,,\nonumber\\
\nonumber\\
&\langle\mathcal{E}(\vec{n}_1)\mathcal{E}(\vec{n}_2)\mathcal{E}(\vec{n}_3)\rangle_g^{\mathrm{LP,\, LO}} \\
&= \frac{1}{\pi^2 R_S^2 R_L^2}\left[\frac{1}{25} \left(98 C_A^2+14 C_A n_f T_F+15 C_F n_f T_F\right)+\frac{1}{45} C_A \left(\frac{z}{\bar{z}}+\frac{\bar{z}}{z}\right) (C_A-2 n_f T_F)\right]\,.\nonumber
\end{align}
With these results, we can obtain the LO predictions (see \Fig{fig:pert_shape}) for the boundary values of $Q_{\mathcal{E}}$ at $\xi=0$:
\begin{eqnarray}
\lim_{\xi\to0} Q_{\mathcal{E}}^q = \frac{\langle \mathcal{E}(\vec{n}_1)\mathcal{E}(\vec{n}_2)\mathcal{E}(\vec{n}_3) \rangle_q^{\mathrm{LP,\,LO}}}{\frac{3C_F}{\pi R_S^2}\frac{3C_F}{2\pi R_L^2}}= \frac{139}{100}-\frac{2}{135}\cos 2\phi,\\
\lim_{\xi\to0} Q_{\mathcal{E}}^g = \frac{\langle \mathcal{E}(\vec{n}_1)\mathcal{E}(\vec{n}_2)\mathcal{E}(\vec{n}_3) \rangle_g^{\mathrm{LP,\,LO}}}{\frac{14C_A+10T_F}{5\pi R_S^2}\frac{7C_A +5 T_F}{5\pi R_L^2}}= \frac{2074}{2209}-\frac{40}{6627}\cos 2\phi.
\end{eqnarray}

The subleading powers contribution is obtained by subtracting $\langle\mathcal{E}(\vec{n}_1)\mathcal{E}(\vec{n}_2)\mathcal{E}(\vec{n}_3)\rangle_\Phi^{\mathrm{LP,\, LO}}$ from the fixed order result $\langle\mathcal{E}(\vec{n}_1)\mathcal{E}(\vec{n}_2)\mathcal{E}(\vec{n}_3)\rangle_\Phi^{\mathrm{LO}}$:
\beq
\langle\mathcal{E}(\vec{n}_1)\mathcal{E}(\vec{n}_2)\mathcal{E}(\vec{n}_3)\rangle_\Phi^{\mathrm{sub\text{-}LPs,\, LO}} 
=\langle\mathcal{E}(\vec{n}_1)\mathcal{E}(\vec{n}_2)\mathcal{E}(\vec{n}_3)\rangle_\Phi^{\mathrm{LO}}
 - \langle\mathcal{E}(\vec{n}_1)\mathcal{E}(\vec{n}_2)\mathcal{E}(\vec{n}_3)\rangle_\Phi^{\mathrm{LP,\, LO}}\,. 
\eeq

To get LL result, we make use of the renormalization group equation for the twist-2 light-ray operators:
\begin{align}\label{eq:resum}
\frac{d}{d\ln \mu^2} \vec {\mathbb{O}}^{[J]} = - \hat \gamma(J) \cdot  \vec {\mathbb{O}}^{[J]}\,, 
\end{align}
where
\begin{align}\label{eq:anom_dim}
\hat \gamma(J)=
\begin{pmatrix}
\gamma_{qq}(J)&&2n_f \gamma_{qg}(J)&&0\\
\gamma_{gq}(J)&& \gamma_{gg}(J)&&0\\
0&&0&& \gamma_{\tilde g \tilde g}(J)\mathbf{1}
\end{pmatrix}\,,
\end{align}
where $\mathbf{1}$ is the $2\times 2$ identity matrix.
At LL accuracy, this results in inserting the RG factor $\displaystyle \left[\frac{\alpha_S(Q_2)}{\alpha_S(Q_1)}\right]^{\frac{\hat\gamma^{(0)}(J)}{\beta_0}}$ when there is a large hierarchy between scale $Q_1$ and $Q_2$. Therefore, we can obtain the LL resummation for each piece, 
\begin{eqnarray}
\langle \mathcal{E}^2\rangle_\Phi^{\mathrm{LL}} 
&=&\mathcal{T}\cdot
\left[\frac{\alpha_S(Q)}{\alpha_S(\Lambda_{\mathrm{IR}})}\right]^{\frac{\gamma^{(0)}(3)}{\beta_0}}
\cdot \mathcal{S}_\Phi\,, \\
\langle \mathcal{E}(\vec{n}_1) \mathcal{E}(\vec{n}_2) \rangle_\Phi^{\mathrm{LP,\, LL}} 
&=& -\frac{1}{\pi R_S^2}\mathcal{J}\cdot C_{R_S}^{(1,1)}\cdot
\left[\frac{\alpha_S(Q)}{\alpha_S(R_S Q)}\right]^{\frac{\gamma^{(0)}(3)}{\beta_0}}
\cdot \mathcal{S}_\Phi \,,\\
\nonumber \\
\langle\mathcal{E}^2(\vec{n}_1)\mathcal{E}(\vec{n}_3)\rangle_\Phi^{\mathrm{LP,\, LL}}
&=& -\frac{1}{\pi R_L^2}\mathcal{T}\cdot
\left[\frac{\alpha_S(R_L Q)}{\alpha_S(\Lambda_{\mathrm{IR}})}\right]^{\frac{\gamma^{(0)}(3)}{\beta_0}}
\cdot C_{R_L}^{(1,2)} \cdot
\left[\frac{\alpha_S(Q)}{\alpha_S(R_L Q)}\right]^{\frac{\gamma^{(0)}(4)}{\beta_0}}
\cdot \mathcal{S}_\Phi\,,\\
\nonumber \\
\langle \mathcal{E}(\vec{n}_1) \mathcal{E}(\vec{n}_2) \mathcal{E}(\vec{n}_3) \rangle_{\Phi}^{\mathrm{LL}}
&=& \frac{1}{\pi^2 R_S^2 R_L^2} {\mathcal{J}}\cdot {C}_{R_S}^{(1,1)}\cdot
\left[\frac{\alpha_S(R_L Q)}{\alpha_S(R_S Q)}\right]^{\frac{\gamma^{(0)}(3)}{\beta_0}}
\cdot {C}_{R_L}^{(1,2)}\cdot
\left[\frac{\alpha_S(Q)}{\alpha_S(R_L Q)}\right]^{\frac{\gamma^{(0)}(4)}{\beta_0}}
\cdot \mathcal{S}_\Phi
\nonumber\\
&&\!\!\! +\; \langle \mathcal{E}(\vec{n}_1) \mathcal{E}(\vec{n}_2) \mathcal{E}(\vec{n}_3) \rangle^{\mathrm{sub\text{-}LPs,\, LO}} \cdot 
\left[\frac{\alpha_S(Q)}{\alpha_S(R_L Q)}\right]^{\frac{\gamma^{(0)}(4)}{\beta_0}}\cdot \mathcal{S}_\Phi\,.
\end{eqnarray}
These results allow us to compute the resummed predictions for the celestial non-gaussianity shown in the main text.

\section{Celestial Block Expansion \label{sec:block_expansion}}
%%%%%%%%%%%%%%%%%%%%%%%%%%%%%%%%%%%%%%%%%%

In this appendix, we give some details of the celestial block expansion for collinear EEEC.
From a practical perspective, the blocks serve as a good approximation for the more complicated full results.
The LO results $G_q(z),\, G_g(z)$ in \Ref{Chen:2019bpb} used as the ingredients for the resummation in \App{sec:resum_formula}, are not very convenient to use.
In the following, we provide the approximations based on blocks for $G_q(z),\, G_g(z)$ in the region $\Omega=\{z| \mathrm{Re}\, z < 1/2,\; \mathrm{Im}\,z>0,\, |1-z|<1\}$ for the interested reader to use.
The block structure can as well help us understand the shape of the non-gaussianity in this paper. 

The celestial blocks are the partial wave expansion basis for observables like energy correlators.
The reason for favoring such a basis is that the Lorentz symmetry in the Minkowski space is isomorphic to the conformal symmetry on the celestial sphere.
This is much like the familiar example that the spherical harmonics $Y_{\ell m}$ are the eigenfunctions of total angular momentum $\mathbf{L}^2$.
While the components $L_i=-i\epsilon_{ijk}x_j \partial_k$ have a simple form in Cartesian coordinates, $\mathbf{L}^2$ acts on a function with only angular dependence as:
\begin{equation}
    \mathbf{L}^2=-\left[ \frac{1}{\sin\theta}\frac{\partial}{\partial \theta}\left(\sin\theta \frac{\partial}{\partial\theta}\right)
    +\frac{1}{\sin^2\theta}\frac{\partial^2}{\partial\phi^2}\right].
\end{equation}
The spherical harmonics are the solutions to the equation:
\begin{equation}
    \mathbf{L}^2 Y_{\ell m}(\theta,\phi)=\ell(\ell+1)Y_{\ell m}(\theta,\phi).
\end{equation}
The counterpart of angular momentum in Lorentz group is $\frac{1}{2}M^{\mu\nu}M_{\mu\nu}$.
Therefore, one natural choice of basis for energy correlator $\langle\mathcal{E}(n_1) \mathcal{E}(n_2) \dots \rangle$ is to find the eigenfunctions of the Casimir differential operator:
\begin{equation}
    C_2=-\frac{1}{2}(M_{1}^{\mu \nu}+M_{2}^{\mu\nu})(M_{1\,\mu \nu}+M_{2\,\mu\nu})
    =-\frac{1}{2}\left[\left( n_1^\mu \frac{\partial}{\partial n_1^\nu}-n_1^\nu \frac{\partial}{\partial n_1^\mu}\right) 
    +\left( n_2^\mu \frac{\partial}{\partial n_2^\nu}-n_2^\nu \frac{\partial}{\partial n_2^\mu}\right)\right]^2\,.
\end{equation}
with the eigenvalue $\lambda_{\delta,j}=\delta(\delta-2)+j^2$:
\begin{equation}
C_2 \langle\mathcal{E}(n_1) \mathcal{E}(n_2) \dots \rangle = \lambda_{\delta,j} \langle\mathcal{E}(n_1) \mathcal{E}(n_2) \dots \rangle\,.
\end{equation}
The null vector $n_i^\mu=(1,\vec{n}_i)$ determines the direction of the calorimeter $\mathcal{E}(\vec{n}_i)$.
The parameters $\delta$ and $j$, called celestial dimension and transverse spin respectively, are the eigenvalues to the boost generator $\vec{n}\cdot\vec{K}$ and rotation generator $\vec{n}\cdot\vec{J}$.
The solutions to this differential equation are the celestial blocks.
The 2-point celestial blocks for EEC are discussed in \Ref{Kologlu:2019mfz,Chang:2020qpj}, while the 3-point blocks for EEEC are shown in \Ref{Chang:2022ryc,Chen:2022jhb}.

The 3-point celestial blocks coincide with 2D conformal block~\cite{Dolan:2000ut} in the collinear limit.
The convention for the block $G_{\delta,j}(z,\bar{z})$ is:
\begin{align}
\nonumber
G_{\delta,j}(z,\bar{z}) =\frac{1}{1+\delta_{0,j}} \Bigg[ &z^{\frac{\delta-j}{2}} \,_2F_1\left(\frac{\delta-j}{2}, \frac{\delta-j-2}{2}, \delta-j, z\right)
\bar{z}^{\frac{\delta+j}{2}} \,_2F_1\left(\frac{\delta+j}{2}, \frac{\delta+j-2}{2}, \delta+j, \bar{z}\right)
 \\
&
+ (z\leftrightarrow \bar{z})\Bigg] \,.
\end{align}
We will expand the LO results $G_q,\, G_g$  onto such a basis:
\begin{equation}
G_{q/g} = \sum_{\delta,\,j} c_{\delta,j}^{(q/g)} G_{\delta,\,j}(z,\bar{z}).
\end{equation}
The physical meaning behind this partial wave expansion is the light-ray OPE described in \App{sec:resum_formula}.
More details can be found in \Refs{Chang:2022ryc,Chen:2022jhb}.
Since the blocks, which describe the contribution of a whole family of operators, are independent of the coordinate we choose, $G_{\delta,j}(z,\bar{z})$ here is essentially the same function as $G_{\delta,j}(\xi,\phi)$ in \Eq{eq:cb_expand} with the relation:
\beq
G_{\delta,j}(z,\bar{z}) = G_{\delta,j}(\xi,\phi)\Big|_{\xi\to|z|/|1-z|,\,\phi\to \arcsin \sqrt{1-\left(1-|1-z|\right)^2/|z|^2}}.
\eeq

\begin{figure}[t]
\begin{center}
\subfloat[]{
\includegraphics[width=6cm]{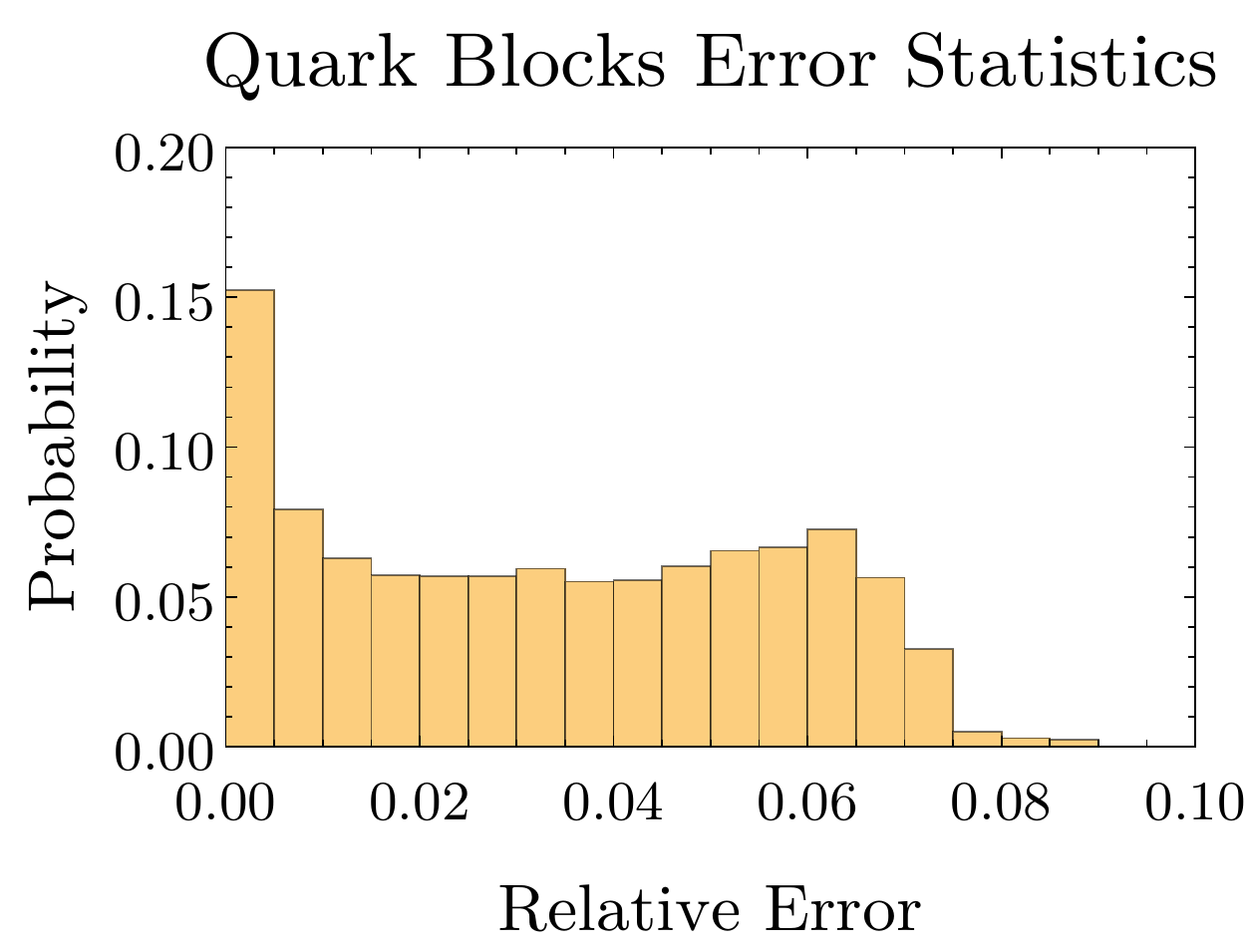}
}\qquad\quad
\subfloat[]{
\includegraphics[width=6cm]{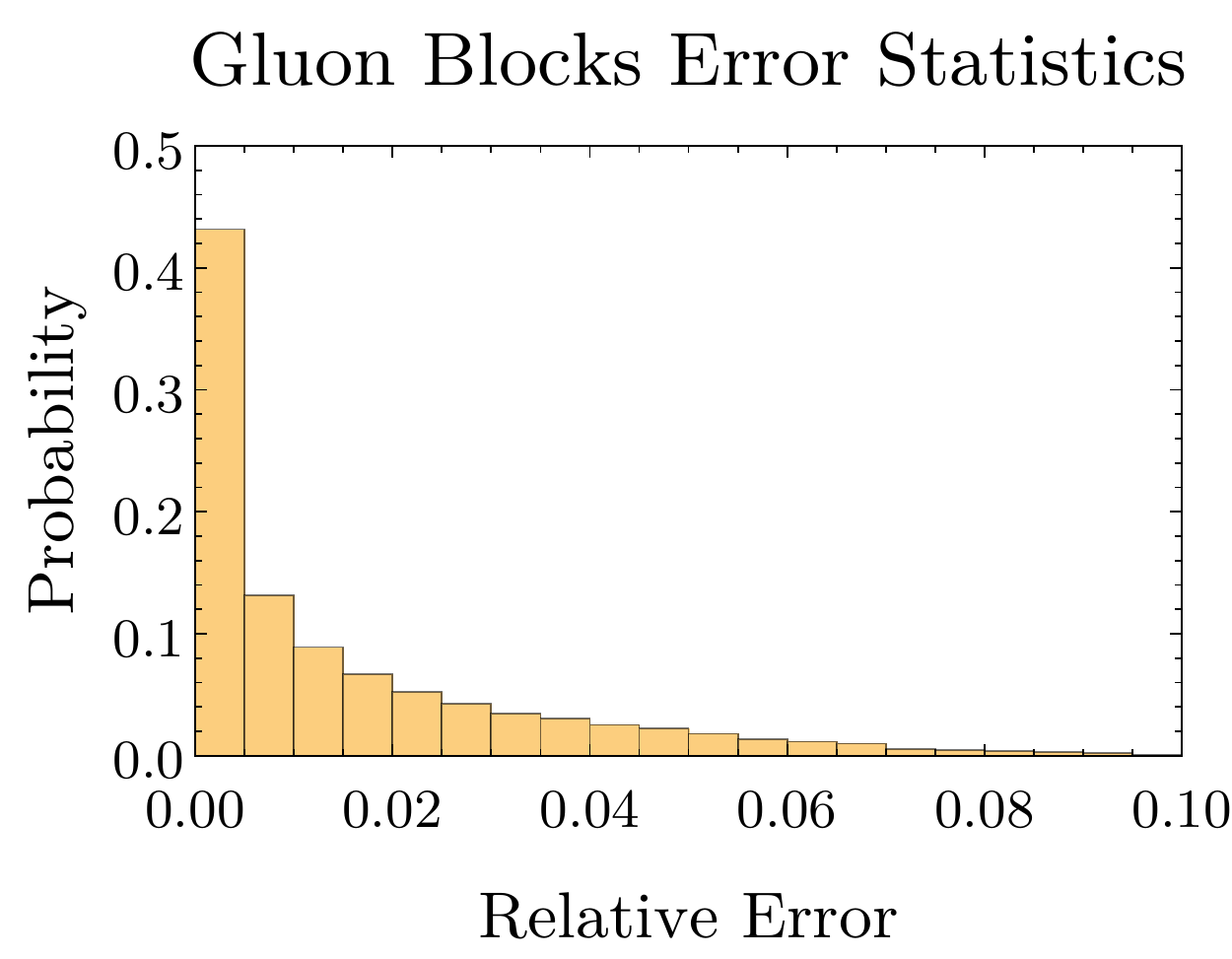}
}
\caption{
The probabilities of relative error of twist $\leq 4$ blocks approximations $G_{q/g}^{\text{block approx}} $ for (a) quark jet and (b) gluon jet. The histograms are obtained by using about 14,000 uniform sampling points in region $\Omega$. We see that the truncation to twist-4 blocks only induce few percent relative error for every sampling point, indicating that such an approximation captures the most feature of the shape dependence. This reflects the fact that the celestial blocks are good basis for energy correlators.}
\label{fig:block_approx_error}
\end{center}
\end{figure}

Numerically, truncation to twist-4 blocks offers quite good approximations for $G_q,\, G_g$:
\begin{align}
&G_q^{\text{block approx}} = \frac{1}{(z\bar{z})^3}\Bigg[  
\frac{C_F (91 C_A+240 C_F+13 n_f T_F)}{4800} G_{4,0}  + \frac{C_F (C_A-2 n_f T_F)}{2880} G_{4,2}\nonumber\\
& +\!\left(
\frac{13}{48} \pi ^2 C_A C_F \!-\! \frac{1077733 C_A C_F}{403200} \!-\! \frac{17}{48} \pi ^2 C_F^2\!+\! \frac{50929 C_F^2}{14400}\!+\!\frac{111199 C_F n_f T_F}{134400}\!-\!\frac{1}{12} \pi ^2 C_F n_f T_F
\!\right)G_{6,0} \nonumber\\
&\hspace{2cm}
+\left(
-\frac{1548703 C_A C_F}{1008000}+\frac{5}{32} \pi ^2 C_A C_F-\frac{13}{48} \pi ^2 C_F^2+\frac{12859 C_F^2}{4800}+\frac{163 C_F n_f T_F}{504000}
\right) G_{6,2}
\Bigg]\nonumber\\
&\quad
=\frac{1}{(z\bar{z})^3}\Bigg[ 
\frac{139}{800} G_{4,0}-\frac{1}{1080}G_{4,2}+\left(\frac{19 \pi ^2}{108}-\frac{2987143}{1814400}\right) G_{6,0}+\left(\frac{31 \pi ^2}{216}-\frac{58043}{42000}\right) G_{6,2}
\Bigg]\nonumber\\
&\quad
=\frac{1}{(z\bar{z})^3}\Bigg[ 
0.17375\, G_{4,0}-0.000925926\, G_{4,2}+0.0899662\, G_{6,0}+0.0344948\, G_{6,2}
\Bigg]\,, \label{eq:quark_block_approx}
\end{align}
\begin{align}
&G_g^{\text{block approx}} = \frac{1}{(z\bar{z})^3}\Bigg[  
\frac{98 C_A^2+14 C_A n_f T_F+15 C_F n_f T_F}{1600} G_{4,0}
+ \frac{C_A (C_A-2 n_f T_F)}{2880} G_{4,2} \nonumber\\
&\hspace{2cm} + 
\left(
-\frac{5}{32} \pi ^2 C_A^2+\frac{318193 C_A^2}{201600}-\frac{120899 C_A n_f T_F}{201600}+\frac{1}{16} \pi ^2 C_A n_f T_F-\frac{13 C_F n_f T_F}{6400}
\right)G_{6,0} \nonumber\\
&\hspace{2cm}+
\left(
-\frac{1}{6} \pi ^2 C_A^2+\frac{834469 C_A^2}{504000}-\frac{129587 C_A n_f T_F}{126000}+\frac{5}{48} \pi ^2 C_A n_f T_F+\frac{C_F n_f T_F}{400}
\right)G_{6,2} \Bigg] \nonumber\\
& \quad
= \frac{1}{(z\bar{z})^3}
\Bigg[
\frac{1037 }{1600}G_{4,0}-\frac{1}{480} G_{4,2}+\left(\frac{1303753}{134400}-\frac{15 \pi ^2}{16}\right) G_{6,0}+\left(\frac{402979}{56000}-\frac{23 \pi ^2}{32}\right) G_{6,2}
\Bigg]\nonumber\\
& \quad
= \frac{1}{(z\bar{z})^3}
\Bigg[
0.648125\, G_{4,0}-0.00208333\, G_{4,2}+0.447789\, G_{6,0}+0.102275\, G_{6,2}
\Bigg]\,, \label{eq:gluon_block_approx}
\end{align}
where we have set $C_F=4/3,\; C_A=3,\; T_F=1/2,\;n_f=5$ for the numerical coefficients.
The probabilities of the relative error using block approximations $G_q^{\text{block approx}},\, G_g^{\text{block approx}} $ are shown in \Fig{fig:block_approx_error}. 

\begin{figure}[t]
\begin{center}
\subfloat[]{
\includegraphics[width=6cm]{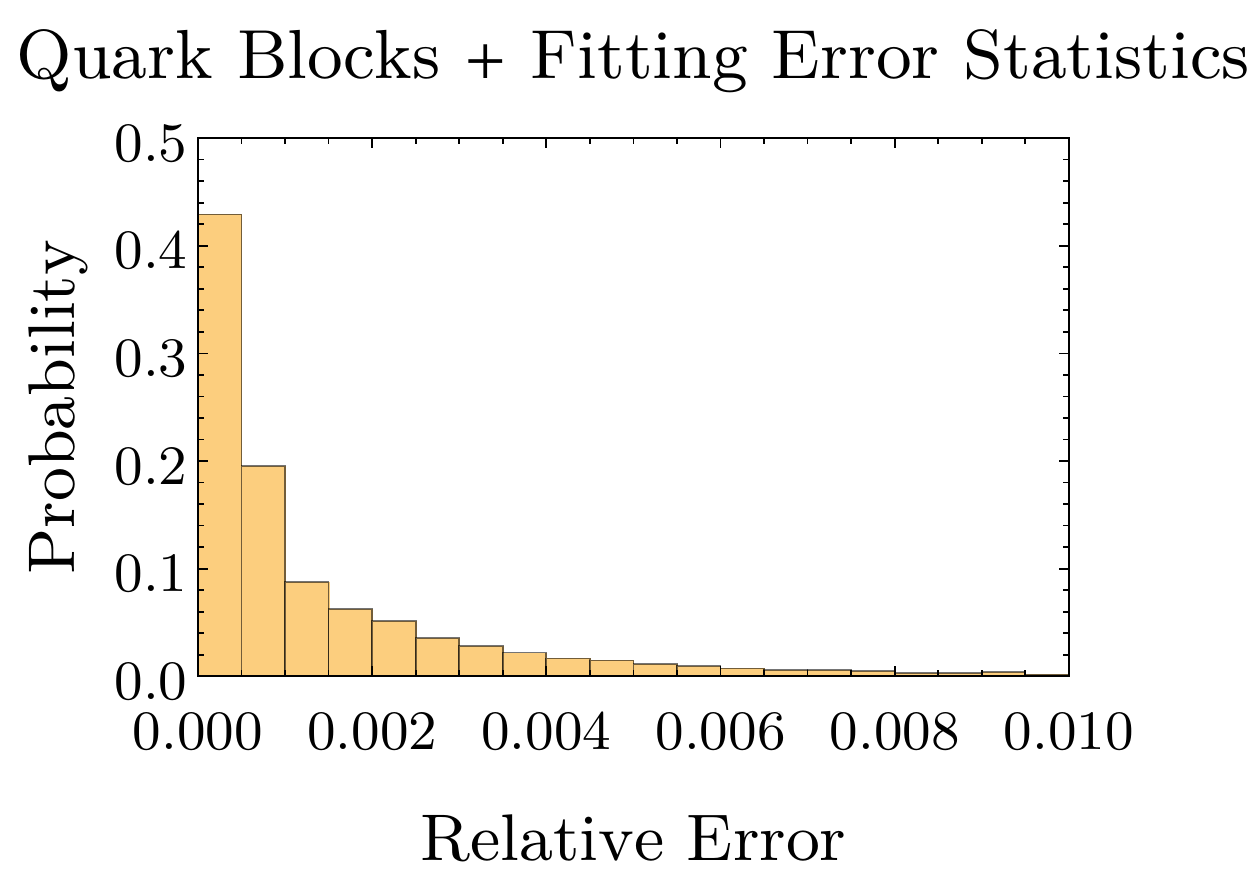}
}\qquad \quad
\subfloat[]{
\includegraphics[width=6cm]{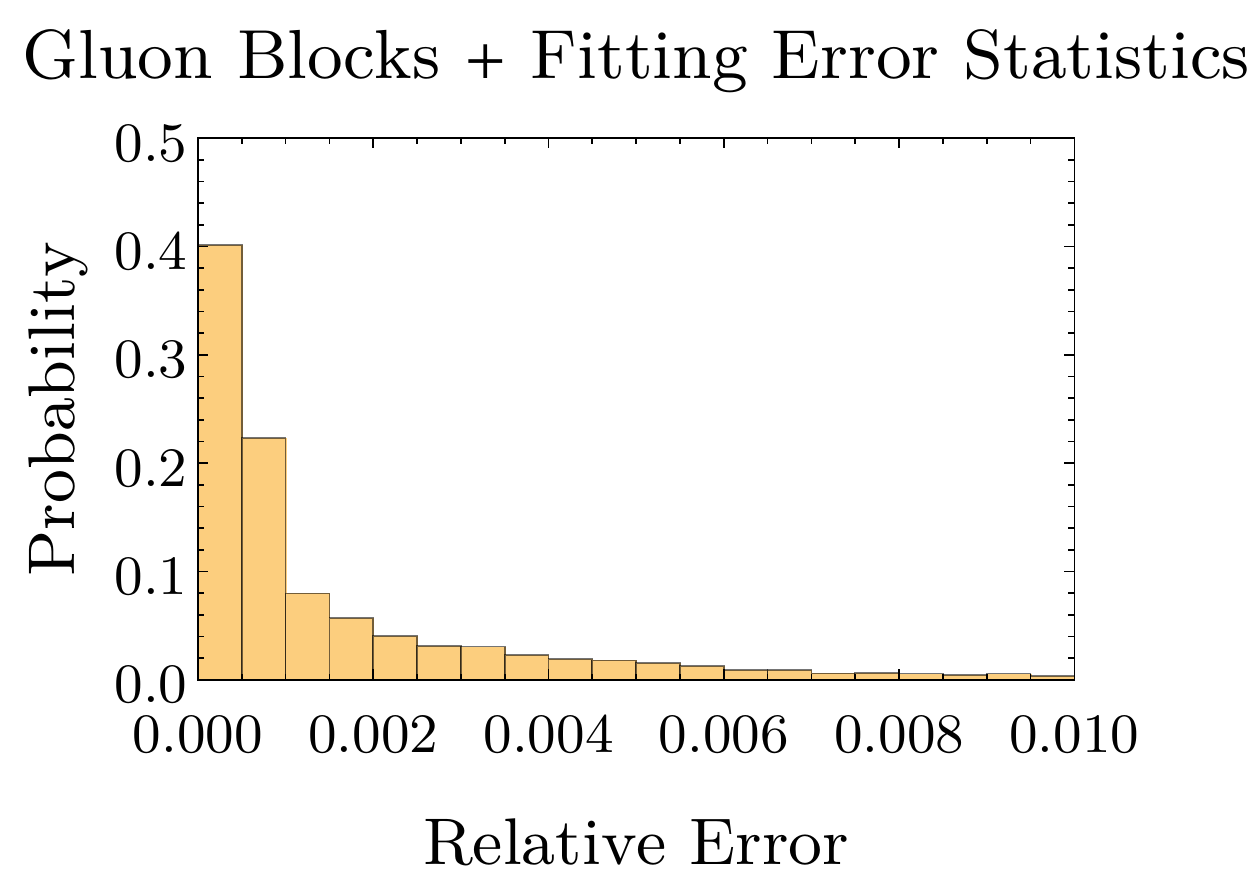}
}
\caption{
The probabilities of relative error of blocks approximations accompanied with remainders fitting $G_{q/g}^{\text{approx}}$ for (a) quark jet and (b) gluon jet. The histograms are obtained by using about 14,000 uniform sampling points in region $\Omega$. We see that with polynomial fitting for the remainders, the maximum relative error can be controlled to about 1\% while the mean relative error is about 0.1\%.}
\label{fig:fit_approx_error}
\end{center}
\end{figure}

For more accurate uses, we can further fit the remainders of block truncations with polynomials in the region $\Omega$:
\begin{equation}
G_{q/g}^{\text{approx}} (z) =  G_{q/g}^{\text{block approx}}(z) + \Delta_{q/g}(z)\,.
\end{equation}
We parametrize the complex number $z$ with its real part $x=\mathrm{Re}\, z$ and its imaginary part $y = \mathrm{Im}\, z$. If we only use the total degree $\leq 3$ monomials, the following polynomials give pretty good approximation:
\begin{align}
\Delta_q(z)\equiv \Delta_q(x,y) = 1.23344 x^3-2.30287 x^2 y+0.090218 x^2+0.952898 x y^2+0.169187 x y\nonumber\\
+0.027774 x-0.021798 y^3-0.229819 y^2+0.098435 y-0.004479 \,,\\
\Delta_g(z)\equiv \Delta_g(x,y) = 4.10647 x^3-8.35197 x^2 y+0.127413 x^2+3.90207 x y^2+0.154011 x y \;\; \nonumber \\
+0.099061 x+0.383566 y^3-1.25266 y^2+0.339035 y-0.018823 \,.
\end{align}
The relative error probabilities of using $G_{q/g}^{\text{approx}} (z)$ are shown in \Fig{fig:fit_approx_error}.

%%%%%%%%%%%%%%%%%%%%%%%%%%%%%%%%%%%%%%%%%%
\section{Symmetric Generalizations \label{sec:new_ratio}}
%%%%%%%%%%%%%%%%%%%%%%%%%%%%%%%%%%%%%%%%%%

In the main text, we used a definition for the non-gaussianity with an asymmetric denominator; see $Q_{\mathcal{E}}$ in \Eq{eq:NG_def}.
While we illustrated that this definition had a number of desirable features, it is also interesting to explore alternative definitions.
In particular, we can introduce a definition of a celestial non-gaussianity with a symmetric denominator, more similar to the definition of the non-gaussianity for the Ising model, given in \Eq{eq:Ising_NG_def}.

We define the symmetric celestial non-gaussianity, $\widetilde{Q}_{\mathcal{E}}$, as:
\beq
\widetilde{Q}_{\mathcal{E}} = \frac{\langle\mathcal{E}(\vec{n}_1)\, \mathcal{E}(\vec{n}_2)\, \mathcal{E}(\vec{n}_3)\rangle  \langle\mathcal{E}^2(\vec{n}_1)\rangle }
{\langle\mathcal{E}(\vec{n}_1)\, \mathcal{E}(\vec{n}_2)\rangle  \langle\mathcal{E}^2(\vec{n}_1)\, \mathcal{E}(\vec{n}_3)\rangle
+\langle\mathcal{E}(\vec{n}_1)\, \mathcal{E}(\vec{n}_2)\rangle   \langle\mathcal{E}^2(\vec{n}_2)\, \mathcal{E}(\vec{n}_3)\rangle
+\langle\mathcal{E}(\vec{n}_2)\, \mathcal{E}(\vec{n}_3)\rangle  \langle\mathcal{E}^2(\vec{n}_1)\, \mathcal{E}(\vec{n}_3)\rangle
 }\,. 
\eeq
In the squeezed limit, this symmetric definition reduces (up to normalization) to $Q_{\mathcal{E}}$, but it can behave quite differently away from the squeezed limit.
In this appendix, we provide a brief numerical study of this modified definition using parton shower Monte Carlo and CMS Open Data, leaving a more detailed theoretical treatment and comparison between theory and data for future work.

\begin{figure}
\begin{center}
\subfloat[]{
\includegraphics[width=0.35\textwidth]{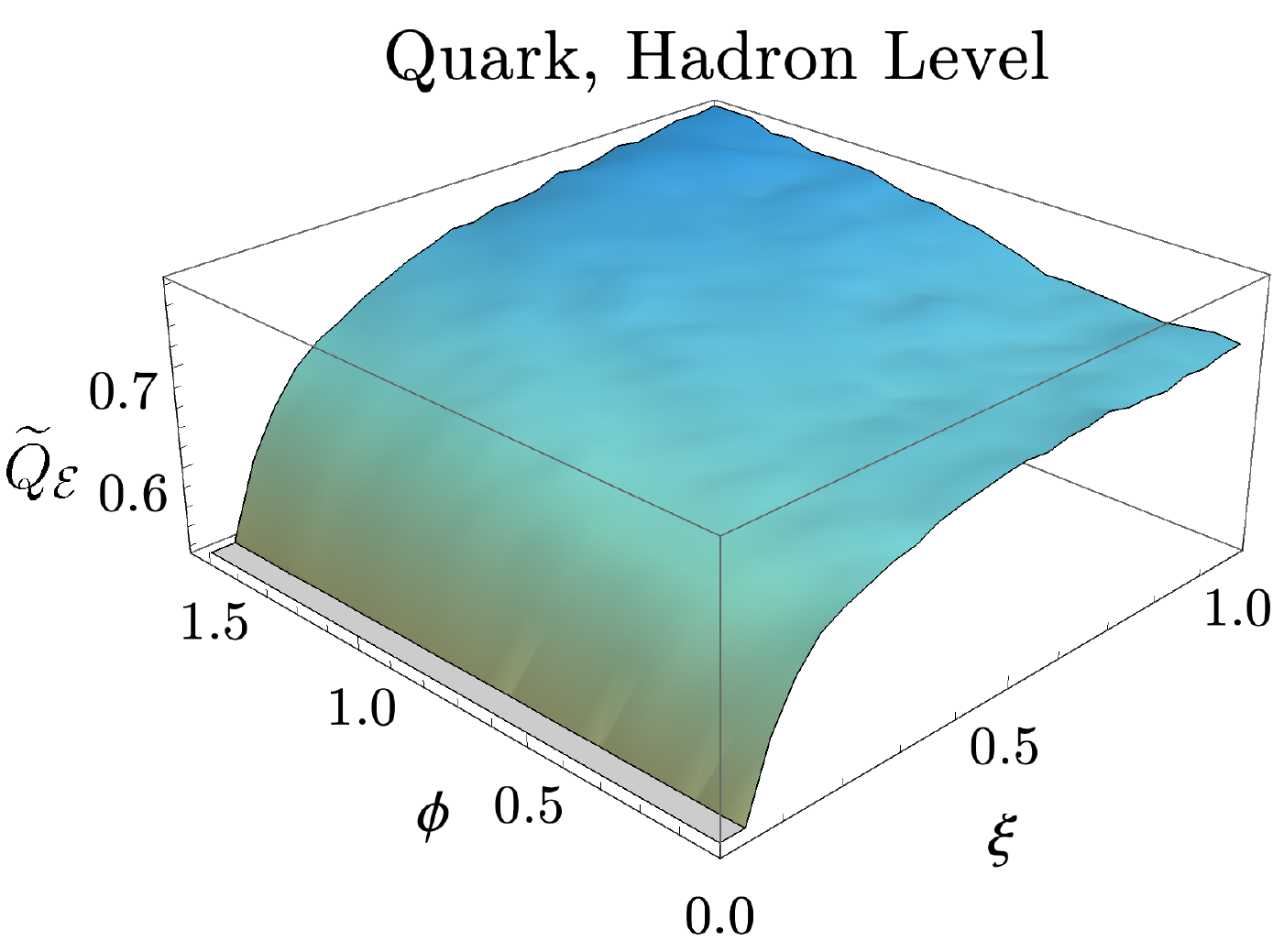}\label{fig:new_hadron_quark}
}\qquad 
\subfloat[]{
\includegraphics[width=0.35\textwidth]{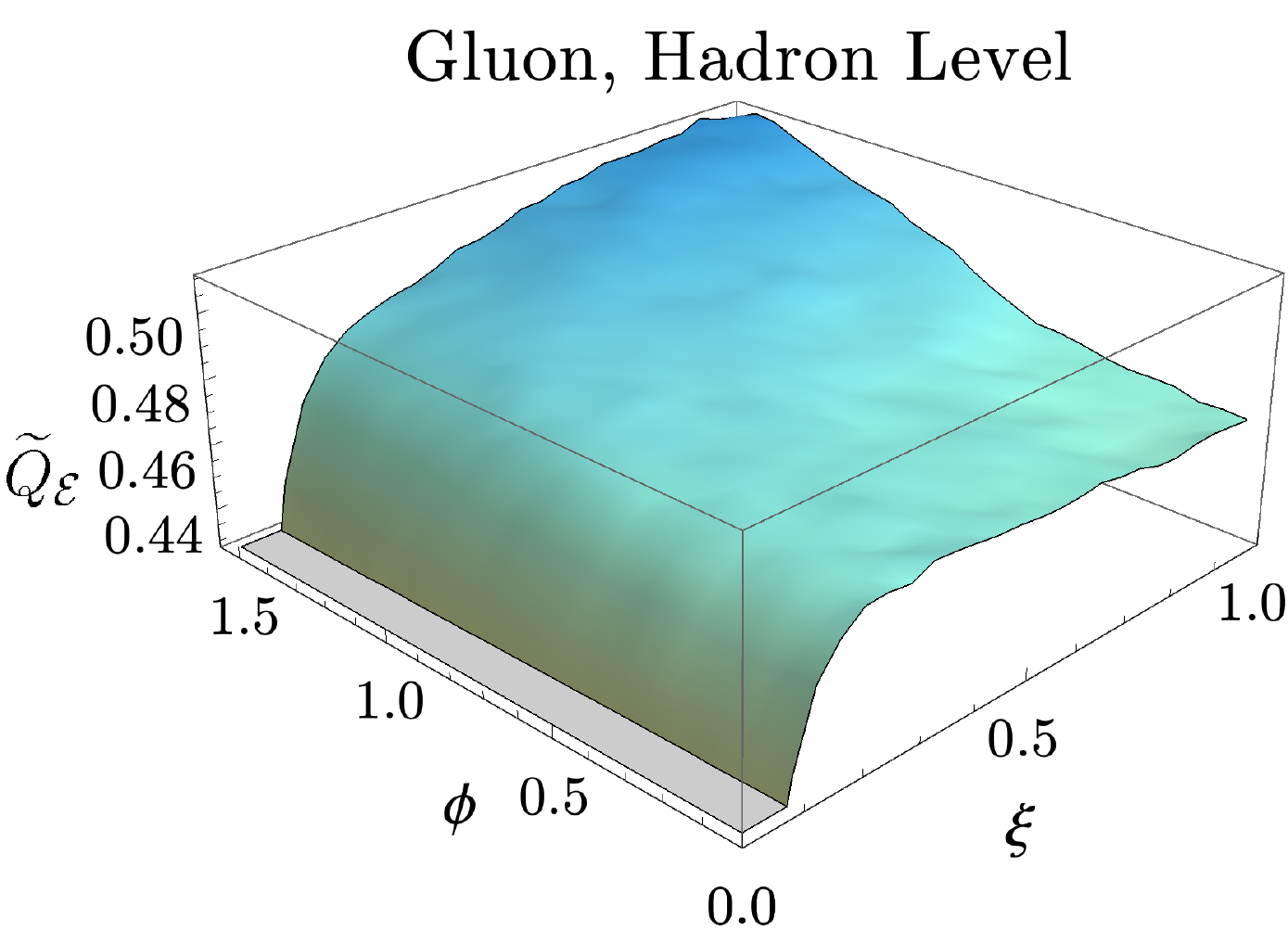}\label{fig:new_hadron_gluon}
}\\
\subfloat[]{
\includegraphics[width=0.35\textwidth]{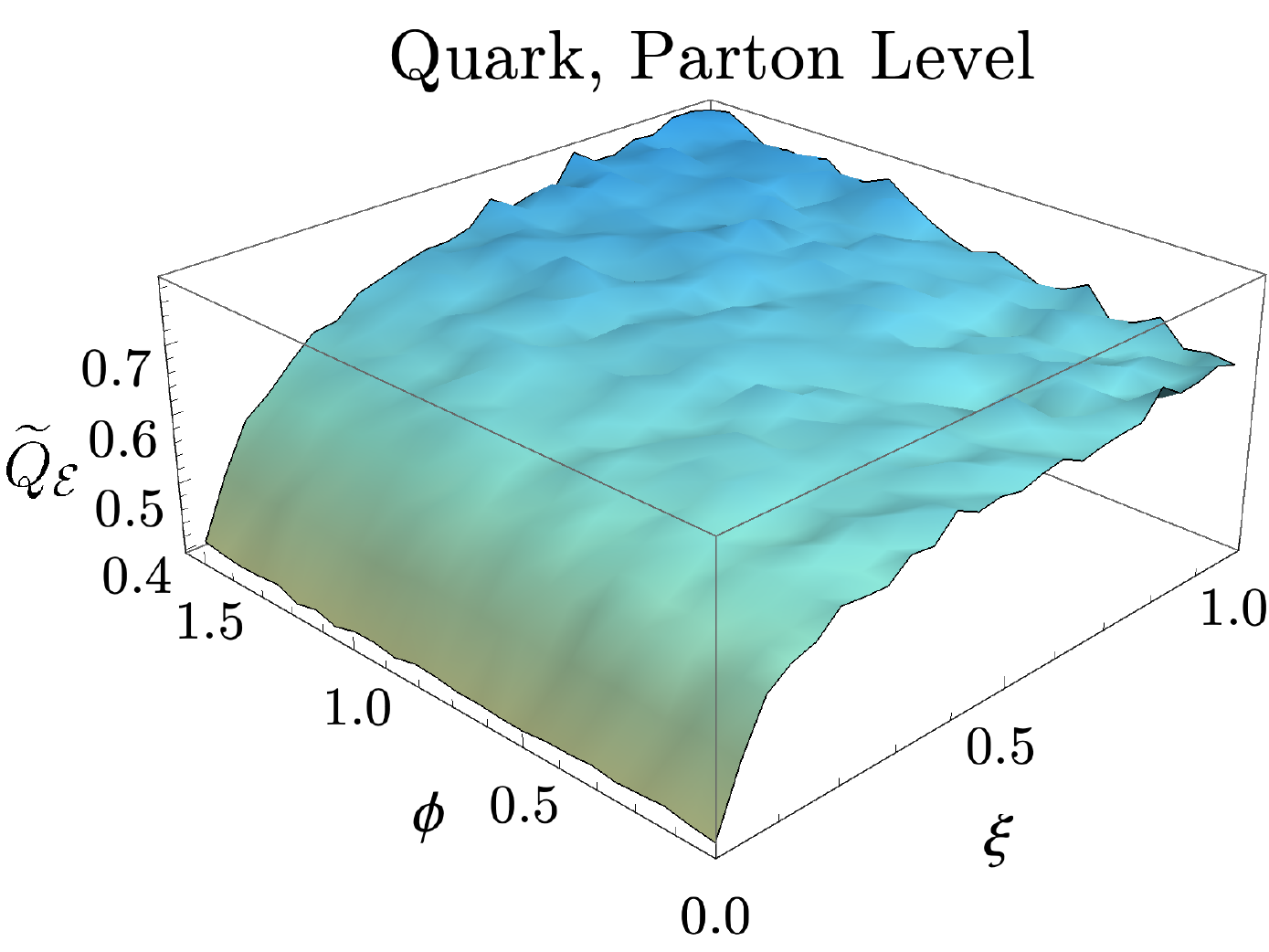}\label{fig:new_parton_quark}
}\qquad 
\subfloat[]{
\includegraphics[width=0.35\textwidth]{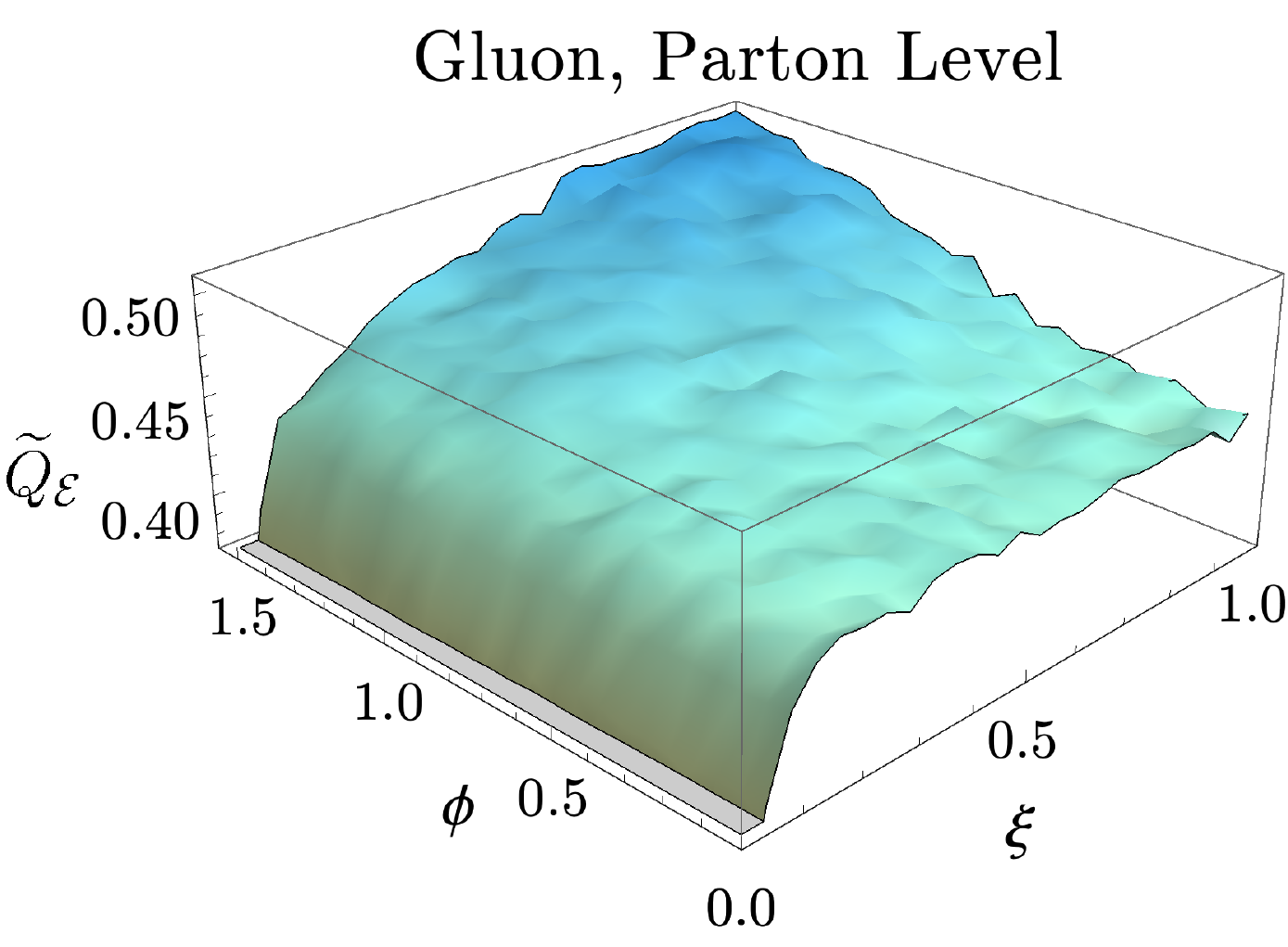}\label{fig:new_parton_gluon}
}\\
\subfloat[]{
\includegraphics[width=0.35\textwidth]{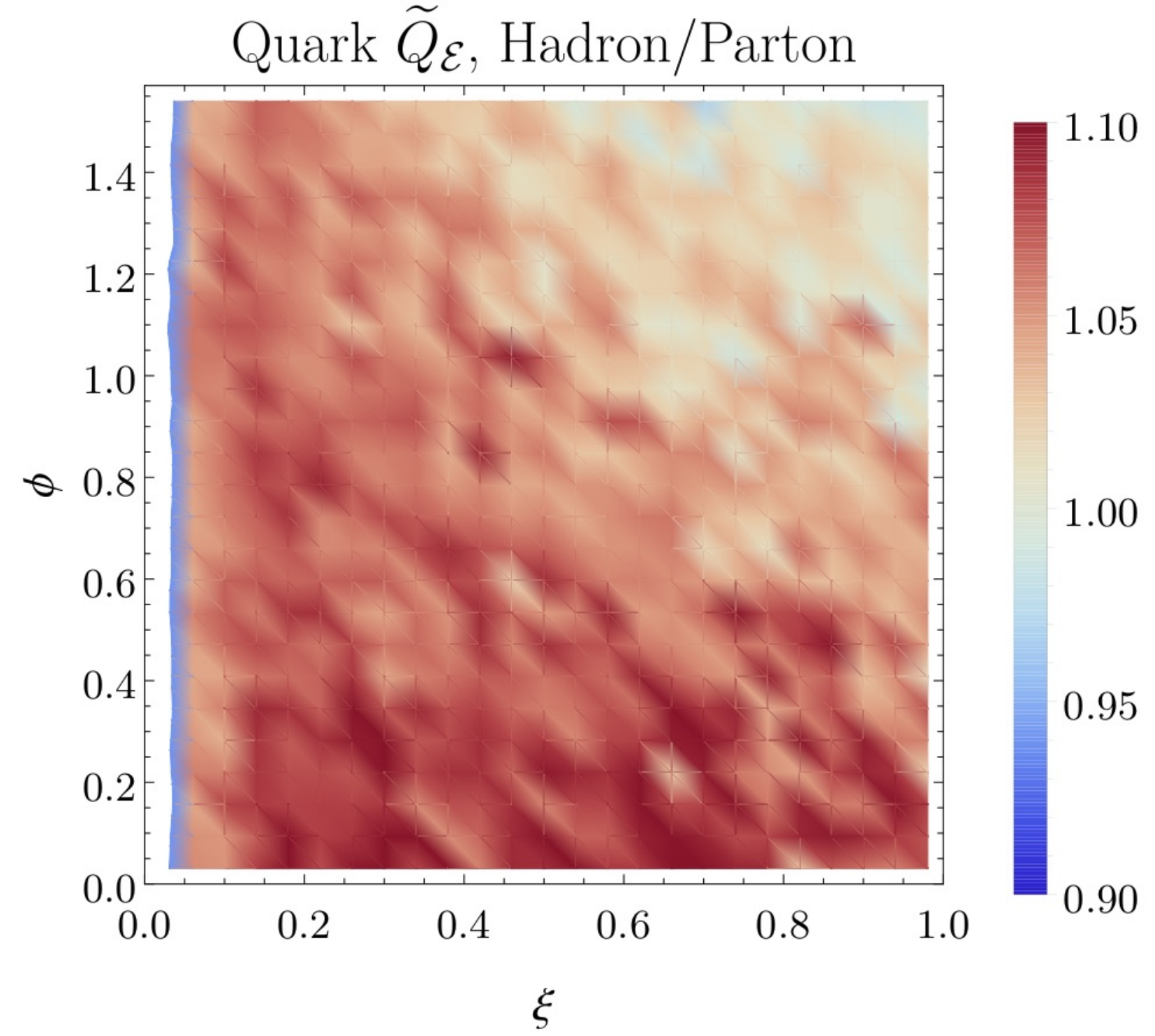}\label{fig:new_quark_ratio}
}\qquad 
\subfloat[]{
\includegraphics[width=0.35\textwidth]{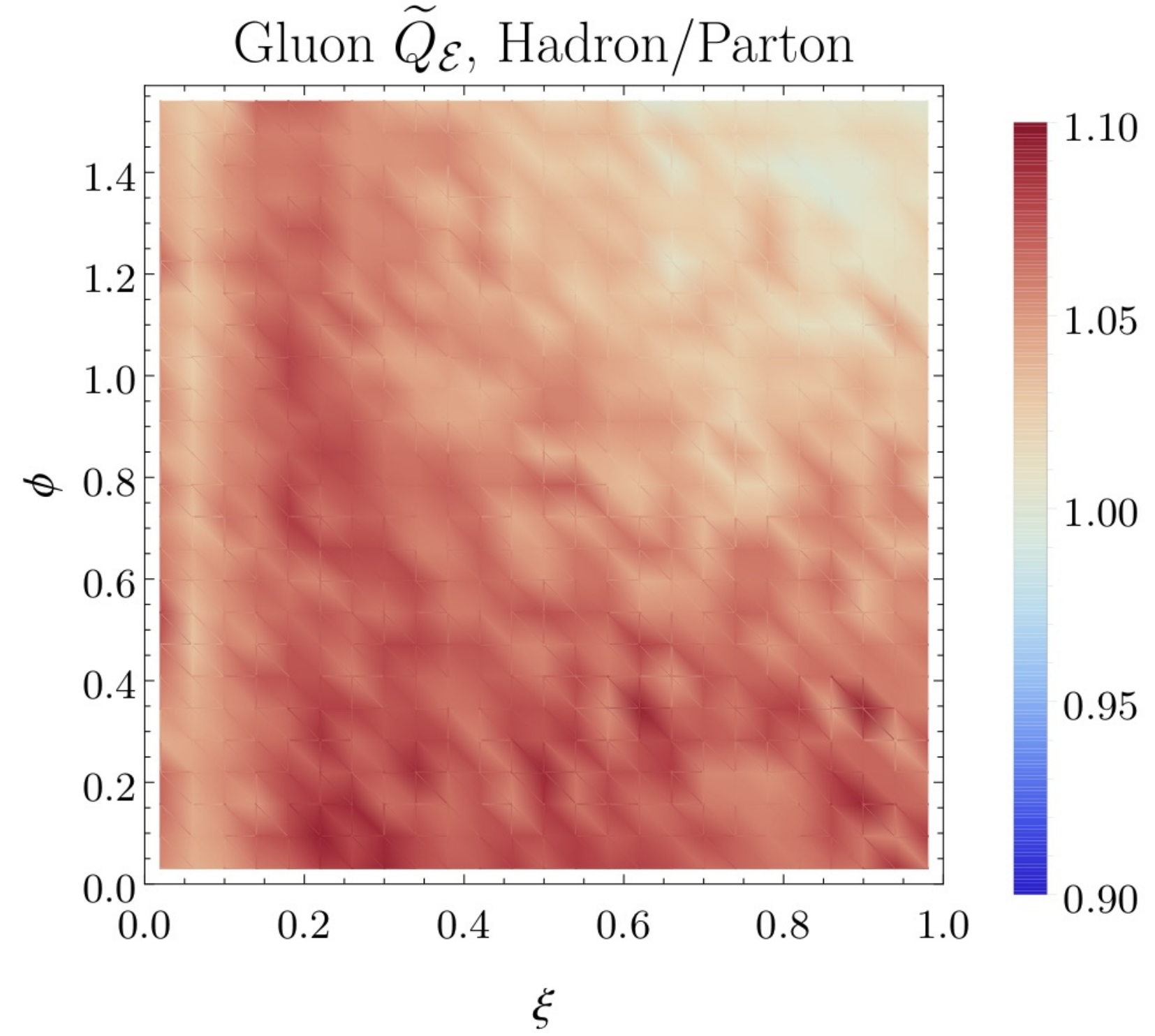}\label{fig:new_gluon_ratio}
}
\end{center}
\caption{The symmetric celestial non-gaussianity, $\widetilde{Q}_{\mathcal{E}}$, for both quark and gluon jets in Pythia, both with and without hadronization. We also show the ratio between parton level and hadron level. Much like $Q_{\mathcal{E}}$, hadronization effects are minimal, showing that the observable is under perturbative control.  The shape dependence is qualitatively different as compared to $Q_{\mathcal{E}}$, exhibiting a much flatter behavior in the perturbative region.
}
\label{fig:new_ratio_pythia}
\end{figure}

\begin{figure}[t]
\begin{center}
\includegraphics[scale=0.9]{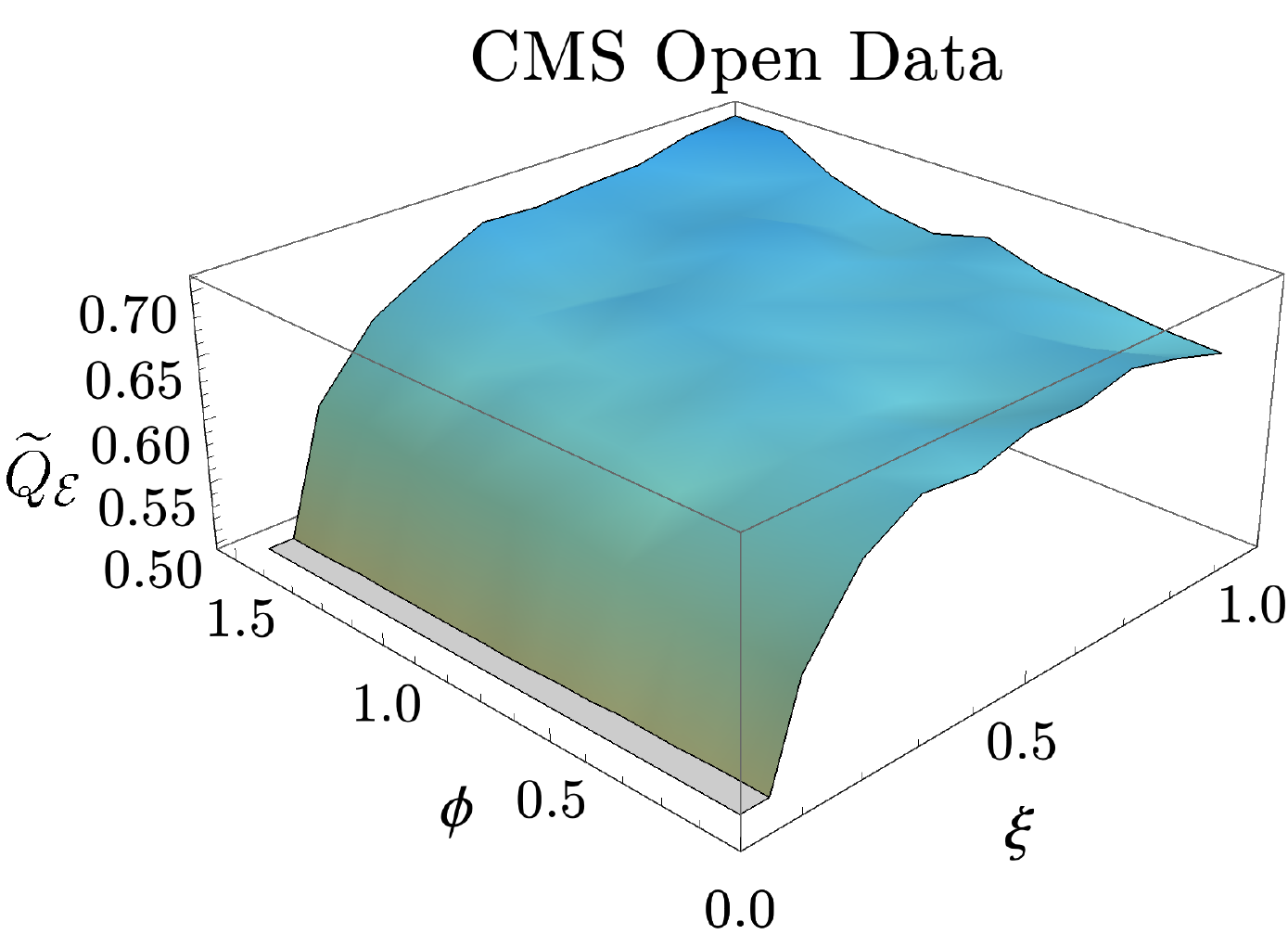}
\caption{The symmetric celestial non-gaussianity in CMS Open Data.
} 
\label{fig:new_ratior_cms}
\end{center}
\end{figure}

\begin{figure}[t]
\begin{center}
\includegraphics[scale=0.5]{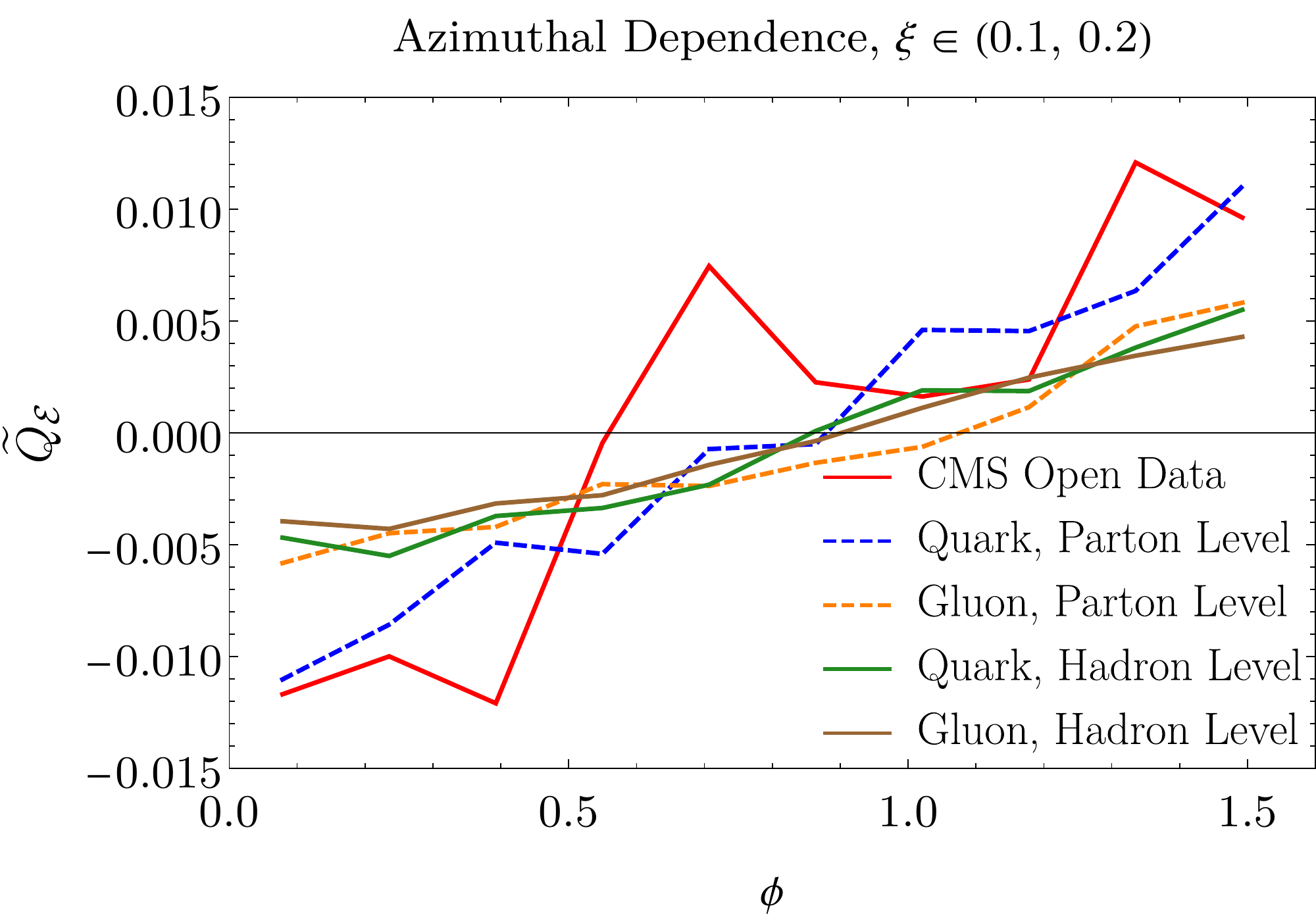}
\caption{
The azimuthal dependence of $\widetilde{Q}_{\mathcal{E}}$ at small $\xi \in (0.1,\, 0.2)$ for CMS Open Data (red), parton-level quark jet (dashed blue), parton-level gluon jet (dashed orange), hadron-level quark jet (green), hadron-level gluon jet (brown). To be shown in the same range, all curves have been shift down with values $0.584$ (CMS Open Data), $0.588$ (parton-level quark jet), $0.434$ (parton-level gluon jet), $0.629$ (hadron-level quark jet), $0.460$ (hadron-level gluon jet).
} 
\label{fig:new_ratior_azimuthal}
\end{center}
\end{figure}

In \Fig{fig:new_ratio_pythia}, we show $\widetilde{Q}_{\mathcal{E}}$ for both quark and gluon jets in Pythia, both with and without hadronization.
We also show the ratio between parton level and hadron level.
As with $Q_{\mathcal{E}}$, we find that the effects of hadronization are small, showing that the observable is under perturbative control.
This alone makes it an interesting observable for studies of perturbative  multi-point correlations inside high energy jets at the LHC.
Corresponding results in the CMS Open Data are shown in \Fig{fig:new_ratior_cms}.

Interestingly, we find that $\widetilde{Q}_{\mathcal{E}}$ varies slowly away from squeezed region $\xi \to 0$, which means the shape of three-point correlator $\langle\mathcal{E}(\vec{n}_1)\, \mathcal{E}(\vec{n}_2)\, \mathcal{E}(\vec{n}_3)\rangle$ is quite close to the denominator:
\begin{equation}
\langle\mathcal{E}(\vec{n}_1)\, \mathcal{E}(\vec{n}_2)\rangle  \langle\mathcal{E}^2(\vec{n}_1)\, \mathcal{E}(\vec{n}_3)\rangle
+\langle\mathcal{E}(\vec{n}_1)\, \mathcal{E}(\vec{n}_2)\rangle   \langle\mathcal{E}^2(\vec{n}_2)\, \mathcal{E}(\vec{n}_3)\rangle
+\langle\mathcal{E}(\vec{n}_2)\, \mathcal{E}(\vec{n}_3)\rangle  \langle\mathcal{E}^2(\vec{n}_1)\, \mathcal{E}(\vec{n}_3)\rangle \,.
\end{equation}
The different shape of $\widetilde{Q}_{\mathcal{E}}$ suggests that it may be useful for unveiling additional features of the three-point correlator.
While the definition of $Q_{\mathcal{E}}$ made clear the behavior of the propagator structure, the extremely flat behavior of $\widetilde{Q}_{\mathcal{E}}$ suggests that it may be useful for probing spin correlations in the three-point correlator, which would be visible as a modulation in the angle, $\phi$ at fixed values of $\xi$.
Since these effects are expected to be small, having a ratio observable that is under perturbative control where they are potentially visible would be extremely interesting.

In \Fig{fig:new_ratior_azimuthal}, we show such a scan of $\phi$ in a small $\xi$ slice, comparing Pythia results for both quark and gluon jets with CMS Open Data.
We do not expect Pythia to fully reproduce the spin correlations, and would therefore expect to see an enhancement in CMS Open Data.
Unfortunately with our current statistics and limited theoretical understanding of this observable, no conclusions can be drawn at the current time.
Nevertheless, we believe that the modified symmetric definition deserves further study as a perturbatively calculable observable that is potentially sensitive to spin correlations in the parton shower.

%%%%%%%%%%%%%%%%%%%%%%%%%%%%%%%%%%%%%%%%%%%
%\section{Additional Plots \label{sec:plots}}
%%%%%%%%%%%%%%%%%%%%%%%%%%%%%%%%%%%%%%%%%%%
%
%In this Appendix we include several additional plots not included in the main text.
%
%\begin{figure}
%\begin{center}
%\includegraphics[scale=0.22]{figures/plots/quark_jet_LL_log}\label{fig:non_g}
%%\captionsetup{font={footnotesize}}
%\end{center}
%\caption{Extra plot on quarks, not sure if need.}
%\label{fig:eflow}
%\end{figure}

\bibliographystyle{JHEP}
\bibliography{spinning_gluon.bib}

\end{document}